\RequirePackage{fix-cm}
\documentclass[natbib]{svjour3}                     
\smartqed  
\pdfoutput=1
\usepackage{graphicx}
\usepackage{mathptmx}      
\usepackage{latexsym}
\usepackage{lineno}
\usepackage{caption}
\usepackage{textcomp}
\usepackage{color}
\usepackage{xcolor}
\usepackage{setspace}

\journalname{Climate Dynamics}
\begin{document}

\title{Co-existing climate attractors in a coupled aquaplanet}


\author{M. Brunetti \and J. Kasparian \and  C. V\'erard}


\institute{M. Brunetti \at
              Institute for Environmental Sciences and Group of Applied Physics, \\ 
              University of Geneva, \\ 
              66 Bd Carl-Vogt, 1205 Geneva, Switzerland \\
              Tel.: +41-22-379 06 25\\
              Fax: +41-22-379 07 44\\
              \email{maura.brunetti@unige.ch}           
            \and 
              J. Kasparian \at
              Institute for Environmental Sciences and Group of Applied Physics, \\ 
              University of Geneva, \\ 
              66 Bd Carl-Vogt, 1205 Geneva, Switzerland \\
           \and
           C. V\'erard \at
              Section of Earth and Environmental Sciences, \\ 
              University of Geneva, \\ 
             13 Rue des Mara\^ichers, 1205 Geneva, Switzerland \\
}

\date{Received: date / Accepted: date}

\maketitle

\begin{abstract}
The first step in exploring the properties of dynamical systems like the Earth climate is to identify the different phase space regions 
where the trajectories asymptotically evolve, called `attractors'. In a given system, multiple attractors can co-exist under the effect of the same forcing. At the boundaries of their basins of attraction, small changes produce large effects. Therefore, they are key regions for understanding the system response to perturbations. Here we prove the existence of up to five attractors in a simplified climate system  where the planet is entirely covered by the ocean (aquaplanet). These attractors range from a snowball to a hot state without sea ice, and their exact number depends on the details of the coupled 
atmosphere--ocean--sea ice configuration. We characterise each attractor by describing the associated climate feedbacks, by using the principal component analysis, and by measuring quantities borrowed from the study of dynamical systems, namely instantaneous dimension and persistence. 
\end{abstract}
\section{Introduction}
\label{intro}

\label{sec:intro}

The Earth climate is an out-of-equilibrium system~\citep{Gallavotti2006} that, under the effect of a inhomogeneous distribution of solar radiation at its boundary, evolves toward statistically stationary states~\citep{1983AdGeo..25..173S,LucariniEtAl2014}. 
Even neglecting possible additional non-steady forcing (for example of anthropic origin), nonlinear interactions between the main components of the climate system, {\it i.~e.} atmosphere, ocean, cryosphere, biosphere, make its study highly difficult. To address the complexity of the climate, modellers have historically developed a hierarchy of models with increasing level of comprehensiveness, ranging from energy balance models to the most realistic general circulation models~\citep{Held2005,2017JAMES...9.1760J}. Here we apply the MIT general circulation model (MITgcm)~\citep{1997JGR...102.5753M,1997JGR...102.5733M,2004MWRv..132.2845A}  to the numerical study of a coupled aquaplanet~\citep{2007JAtS...64.4270M}, that is a system where the full coupling between atmosphere, ocean and sea ice is taken into account, but their motion is simplified by the absence of land, the planet being entirely covered by the ocean. In this way, even if the complexity of the system is reduced with respect to the real climate system, the main nonlinear interactions between its components are still  taken into account. 

Under the action of a constant external forcing and at fixed values of internal parameters, the solutions of a dynamical system are attracted toward stable regions of phase space, named {\it attractors}~\citep{Milnor1985}. The attractor $A$ is the minimal invariant closed set that attracts an open set $U$ of initial conditions converging toward $A$ as time tends to infinity. The largest such $U$ is called {\it basin of 
attraction}~\citep[p. 324]{Strogatz}. Finding the attractors and their basin of attraction is the first step in the analysis of the unperturbed dynamics. 
At the basin boundaries,  also denoted as unstable 
manifolds or edge/melancholia states~\citep{2017Nonli..30R..32L}, the dynamics is highly nonlinear, 
small perturbations giving rise to abrupt and potentially irreversible changes that correspond to the passage from an attractor to the other. Such endogenous crises are generally called `tipping points' in climate dynamics or `critical transitions' in statistical physics. 
Examples of tipping elements in the present-day climate are the shut-down of the overturning circulation in the Atlantic ocean, the methane release from  the melting of the permafrost or the dieback of the Amazon forest~\citep{Lenton2008}.

Thus the attractors characterisation helps in predicting the system response to perturbations.  
A huge effort is presently under way 
to understand properties of attractors and tipping points~\citep{Bathiany2016}, to develop early warning systems 
for detecting the approach to such thresholds~\citep{Dakos14308} and to clarify the nature of perturbations (internal variability and self-reinforcing feedbacks) or external forcing (for example of astronomical or anthropic origin) that could give rise to critical transitions~\citep{2013JCli...26.2862R}.  A treatment of noise-induced transitions is given in~\citet{2019PhRvL.122o8701L}.   
  
Here we systematically search for attractors in an aquaplanet, that is the simplest configuration in coupled atmosphere-ocean-sea ice models. 
We show how the large-scale structure of the unperturbed system changes when we improve the physical description of the MITgcm used to run the simulations. Namely, we compare the more accurate model set-up where dissipated kinetic energy is re-injected within the system as thermal energy, and the bulk cloud albedo varies with latitude ({\tt setUp1}), with the one where dissipated kinetic energy is lost and bulk cloud albedo is constant ({\tt setUp2}). 
While previous studies identified two~\citep{1969TellA..21..611B,1969JApMe...8..392S,1976JAtS...33....3G,2010QJRMS.136....2L,Abbot2011,2013Icar..226.1724B}, three~\citep{2011JCli...24..992F,2017Nonli..30R..32L}, or four co-existing attractors \citep{Rose2015}, we obtain up to five attractors that co-exist under the same external forcing, represented in our simulations by fixed values of solar irradiation and atmospheric CO$_2$ content. This demonstrates that a simplified system such as a coupled aquaplanet is sufficiently rich to produce a complex dynamical structure. 

Each attractor is associated to a different climate, ranging from snowball conditions \citep{Kirschvink1992} 
to a hot state where the sea ice completely disappears.  
We characterise each climate by describing ocean overturning circulation, heat transport, cloud cover and surface air temperature distribution, and by estimating a list of averaged global quantities and energy budget components to reveal the dominant nonlinear feedbacks at play in each attractor. 
We take advantage of some powerful tools borrowed from the theory of dynamical systems and statistics~\citep{1983AdGeo..25..173S}, such as instantaneous dimension and persistence~\citep{lucarini2016extremes,2017NatSR...741278F,2017NPGeo..24..713F}, and principal component analysis, to further explore the dynamics of the attractors and differentiate them.  

The paper is organised as follows: in Section~2 we present the MITgcm configurations and the techniques for testing the complexity of time series, in Section~3 we describe the results of simulations, and finally in Section~4 we discuss the results in the light of previous findings and we suggest 
further developments. 
 
\section{Methods}
\label{sec:meth}

\subsection{MITgcm}
\label{sec:meth1}

We use the MIT general circulation model (MITgcm, version c65q)~\citep{1997JGR...102.5753M,1997JGR...102.5733M,2004MWRv..132.2845A} to simulate a coupled aquaplanet, {\it i.e.} a planet homogeneously covered by a 3000~m deep ocean. It offers a simpler system than the actual Earth configuration, as no continents perturb the ocean currents and the atmosphere circulation. 

The MITgcm 
is a coupled atmosphere-ocean-sea ice model, where the same 
kernel is employed for representing atmosphere and ocean
dynamics \citep{MarshallAdcroft2004} on a common cubed-sphere grid. 
The physics packages activated for the oceanic component are the Gent and McWilliams scheme \citep{1990JPO....20..150G}, used to parameterise mesoscale eddies, and
the KPP scheme \citep{LargeYeager2004} for accounting of vertical
mixing processes in the ocean surface boundary layer and
the interior. 
The thermodynamic properties of sea ice are implemented through the Winton model \citep{2000JAtOT..17..525W}, while sea ice dynamics is neglected. 
Concerning the atmospheric component, the physics module called 5-layer SPEEDY \citep{Molteni2003} comprises a four-band radiation scheme, boundary layer and moist
convection schemes, resolved baroclinic eddies and diagnostic
clouds. Orbital forcing is prescribed at present-day
values and the atmospheric $CO_2$ content is fixed to a constant
value of 326 parts per million. 

\begin{figure}[ht!]
\includegraphics[width=0.5\linewidth]{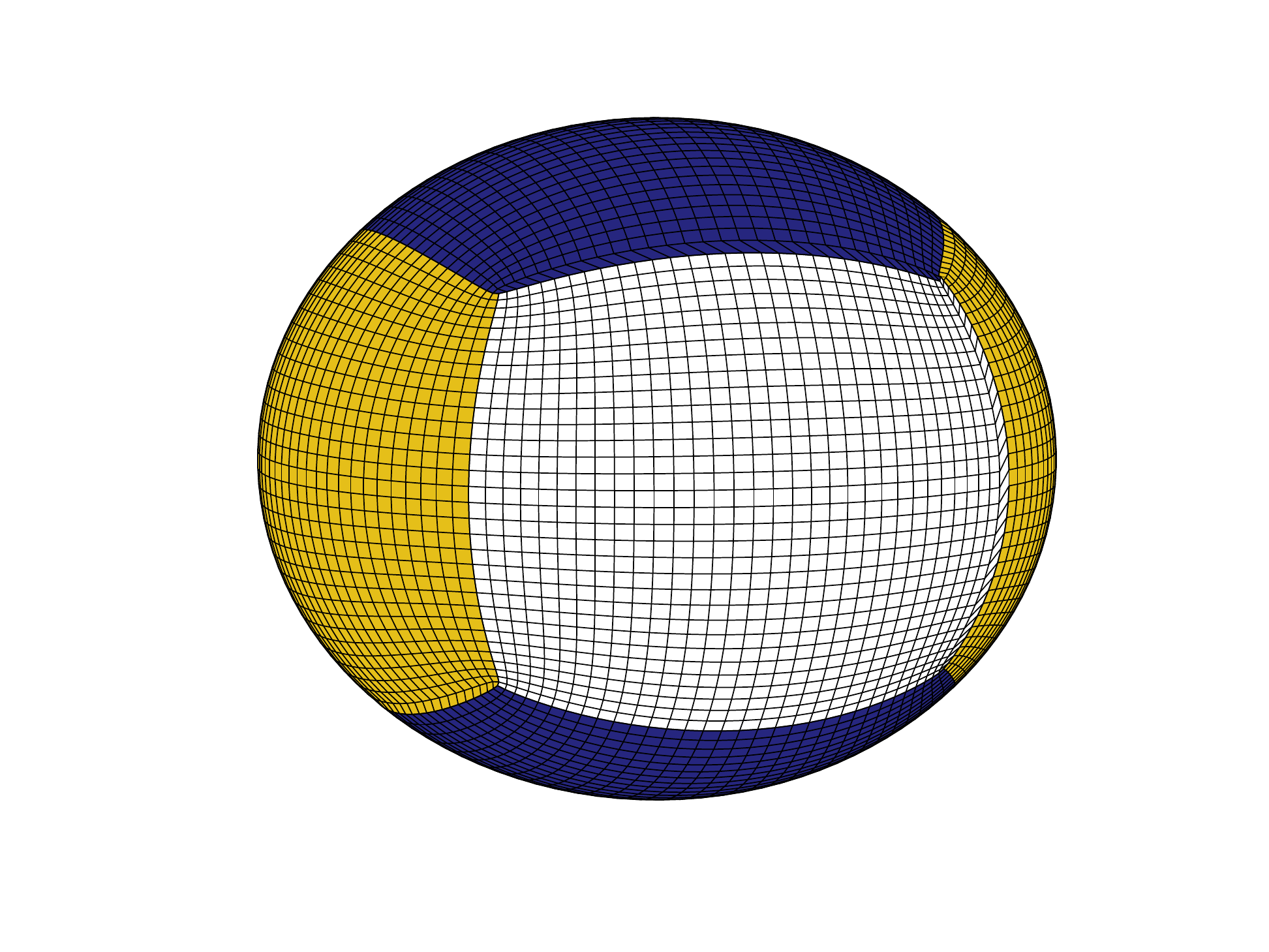}
\includegraphics[width=0.5\linewidth]{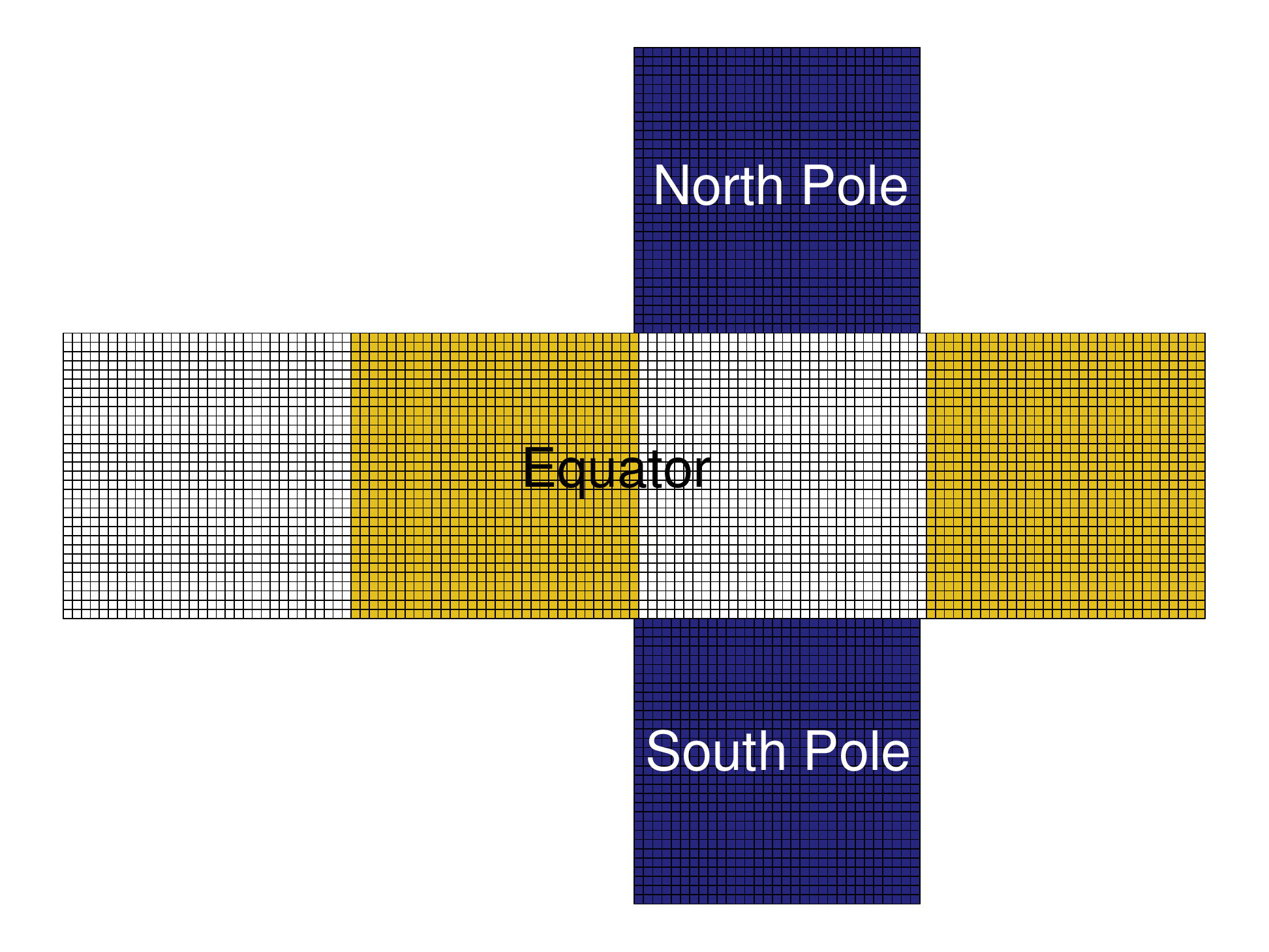} 
\caption{Cubed sphere used in MITgcm and crossmap representation.}
\label{fig:zero}
\end{figure}

We consider a low resolution cubed-sphere (CS) grid (Fig.~\ref{fig:zero}),
where each face of the cube has 32 $\times$ 32 cells (CS32), giving a horizontal
resolution of 2.8$^\circ$. The ocean grid has 15 vertical levels
with different thickness, from 50 m near the surface to 690
m in the abyss, for a total depth of $h=3$~km. In this way, simulations
over thousands years can be run in a reasonable amount of
CPU time (namely, 100 yr in 1 day on a typical workstation).

Two different configurations are considered. In the first, more comprehensive set-up,  denoted as {\tt setUp1},
the bulk cloud albedo varies with latitude, like in version 40 of SPEEDY \citep{2006ClDy...26...79K,kucharski2013}, where this modification has been implemented in order to correct a too strong high-latitude solar radiation flux. 
Furthermore, dissipated kinetic energy is re-injected within the system as thermal energy, allowing for an improved budget at Top-Of-Atmosphere (TOA)~\citep{2011RvGeo..49.1001L,2014ClDy...43..981L}. 
This configuration is the same as that used for {\tt Run4} in \citet{2018ClDy...50.4425B}. 
In the second set-up, called {\tt setUp2}, the bulk cloud albedo is constant, as implemented in the 5-layer SPEEDY~\citep{Molteni2003} and friction energy is lost, like in the configuration used for {\tt Run3}  in \citet{2018ClDy...50.4425B}.

The values of input numerical parameters, such as vertical diffusivity
within the ocean and snow/ice/ocean albedos, are the same as those used in~\citet{2018ClDy...50.4425B} and 
are listed in Table~\ref{tab:1}. Their choice minimises the drift in the global ocean
temperature, and are within the observed range
reported, for example, in \citet{2011JGRC..116.4025N}. Afterwards,
following the tuning procedure used by other groups \citep{2011JCli...24.4973G,2016Tang}, only the relative humidity threshold for low clouds (a parameter referred as $RHCL2$ in MITgcm atmospheric module)
has been adjusted in such a way that, starting from an initial condition with average ocean temperature of 8.9$^\circ$C, the simulation ends up to a warm state.  
Then, all the simulations within a given set-up have been performed using the same value of $RHCL2$, denoted by $RHCL2^*$ (namely, 
$RHCL2^* = 0.7645$ for {\tt setUp1} and $RHCL2^* = 0.7239$ for {\tt setUp2}).

\begin{table}[ht!]
\caption{Input parameters used in the simulations.}
\label{tab:1}
\centering
\begin{tabular}{l c}
\hline
Ocean albedo & 0.07 \\
Bulk cloud albedo & 0.38 \\ 
Max sea ice albedo & 0.64 \\
Min sea ice albedo & 0.20 \\
Cold snow albedo &  0.85 \\
Warm snow albedo &  0.70 \\
Old snow albedo & 0.53 \\
Vertical diffusivity [m$^2$/s] & $3\cdot 10^{-5}$ \\
Horizontal diffusivity [m$^2$/s]  & 0  \\
\hline
\end{tabular}
\end{table}

\subsection{Complexity assessment of time series}
 \label{sec:meth2}

When a dynamical system is multistable, the solutions for a given value of internal parameters can be attracted to different 
regions in phase space, called {\it attractors}. The properties of such regions can be reconstructed from the time series of given observables. 
Several techniques exist for testing the complexity of time 
series~\citep{2015CSF....81..117T}. Here we consider two methods. The first is derived from nonlinear dynamics to describe attractor properties in phase space, namely instantaneous dimension and persistence~\citep{lucarini2016extremes,2017NatSR...741278F,2017NPGeo..24..713F}. The second,  
the principal component analysis, is used in statistical studies and provides information on the spatial modes of variability of the climate dynamics on each attractor.  
We apply these techniques to the yearly averaged SAT series, at each horizontal grid point, after the transients die out and the state is stabilised on a given attractor. For each attractor, the length of the time series is at least 1500 yr.

\subsubsection{Instantaneous dimension and persistence of trajectories} 

The mean dimension of the attractor is the number of degree of freedom sufficient to describe the dynamics when the system has settled down into the attractor. Considering the large number of degree of freedom, the so called embedding method~\citep{1983PhRvL..50..346G} cannot be used. Rather, 
we rely on a more robust approach that allows to estimate the distribution of instantaneous dimensions of the attractor~\citep{lucarini2016extremes,2017NatSR...741278F,2017NPGeo..24..713F}. The idea is to count the number of times a trajectory in phase space returns within a sphere centered on an arbitrary point $\zeta$ on the attractor. Taking the logarithm of the distance $\delta$ between $\zeta$ and all the other observations $x(t)$, and defining
\begin{equation}
g(x(t)) = -\log(\delta(x(t),\zeta))
\end{equation}
the problem of finding the number of returns within the small sphere becomes equivalent to find the probability of exceeding 
a certain threshold $s(q)$ associated to the quantile $q$ of the series $g(x(t))$. The theorem presented in~\citet{Freitas2010} and modified by~\citet{2012JSP...147...63L}  guarantees that in the case of large number of degrees of freedom such probability is the exponential member of the generalised Pareto distribution:
\begin{equation}
P[g(x(t))>s(q),\zeta] \simeq \exp\left(-\frac{x-\mu(\zeta)}{\sigma(\zeta)}\right)
\label{GPD}
\end{equation}
where $\mu$ is the location parameter and $\sigma$ is the scale parameter, both depending on the chosen point $\zeta$. Interestingly, $\sigma$ provides the dimension around $\zeta$, $d(\zeta)=1/\sigma(\zeta)$.  The average over a sufficiently large sample of points 
gives the attractor dimension, 
$D = \overline{d(\zeta)}$.    
We have checked the stability of the results against changes in
the quantile within the range $0.90<q<0.975$, and we have used the Anderson-Darling test for testing the hypothesis that 
the extreme values of the distribution $g(x(t))>s(q)$ follow a generalised Pareto distribution.

One can further develop the analysis of the attractor by considering the extremal index $\theta$, a parameter that measures the 
degree of clustering of extreme values in a stationary process~\citep{Smith1994,2019Chaos..29b2101M}. 
The extremal index can take values in the interval  $[0,1]$. It has been shown that it is related to the reciprocal of the mean cluster size of exceedances over a high threshold~\citep{Leadbetter1983}, or equivalently
to the reciprocal of the averaged residence time of trajectories in the vicinity of $\zeta$~\citep{2017NatSR...741278F,2017NPGeo..24..713F}: low (respectively high) values of $\theta$ correspond to long (short) persistence of trajectories near $\zeta$. The value of $\theta$ is obtained through the estimator defined in \citet{Ferro2003}.

\subsubsection{Principal component analysis}

We sought for spatial modes in the dynamics of each attractor, by performing a principal component analysis (PCA) on the $32\times 32 \times 6=6144$ temporal series of yearly averaged SAT after the system was stabilised. The PCA was performed with Matlab's function {\it pca}. 
The temporal series at each point are projected on the successive principal components. Their mapping defines spatial modes displaying regions that are spatially correlated, anti-correlated or independent at inter-decadal time scale.

\section{Results} 
\label{sec:res}

\subsection{Characterisation of the climate in each attractor}

Under the same forcing, represented by a mean annual incoming solar radiation at the top of the Earth's atmosphere of 1368~W m$^{-2}$
and a fixed level of atmospheric CO$_2$ of
326 parts per million, the ensemble of considered initial conditions relaxes to different final statistically stationary states, {\it i.e.} attractors in the phase space. We have obtained five different attractors, respectively denoted as {\it hot state, warm state, cold state, waterbelt and snowball} in  {\tt setUp2} and four attractors (the same as before except the hot one) in {\tt setUp1}. This is remarkable because, to our knowledge, three 
(hot\footnote{The hot state is called `warm' in~\citet{2011JCli...24..992F,Rose2015}. We prefer to call `hot' the attractor without ice cover and `warm' the temperate state that is also obtained in energy balance models.}, cold state and snowball in \citet{2011JCli...24..992F}) and four attractors (the previous ones plus the waterbelt in \citet{Rose2015}) 
had been obtained to date using MITgcm with lower 
resolution\footnote{The attractors in~\citet{2011JCli...24..992F,Rose2015} are obtained 
using CS24 ($24\times 24$ cells per cube face), slightly different ice/snow albedos with respect to the ones used here and listed in Table~\ref{tab:1}, and a solar constant of $S_0 = 1366$~W/m$^2$.
Vertical diffusivity and CO$_2$ content are the same as here. Ocean and bulk cloud albedos, the parameter $RHCL2$ and the type of cloud albedo parameterisation are not specified nor if thermal energy has been re-inserted within the system. 
Ice thickness diffusion is included to mimic ice dynamics. We have verified that the waterbelt can be obtained in CS24 without ice thickness diffusion, while preliminary results show that the warm state is not stable in CS24 with the considered parameters.}, 
and only two~\citep{1969TellA..21..611B,1969JApMe...8..392S,1976JAtS...33....3G,Abbot2011,2010QJRMS.136....2L,2013Icar..226.1724B} or three 
attractors~\citep{2009JAtS...66.2828R,2017Nonli..30R..32L} 
using energy balance models or intermediate complexity models. 

The initial conditions (ICs) are constructed in two ways. In the first, the ocean is initialised from a state of rest, without 
sea ice and with a homogeneous salinity at 35~psu; we take a zonal mean temperature given by $(N+1-k)(22 \cos\phi +10)/N$, where $N=15$ is the number of vertical levels, $k$ is the vertical index, $k=1,\ldots N$, and $\phi$ is the latitude; this IC has an average ocean temperature of 8.9$^\circ$C and is denoted by IC8.9; we homogeneously change the ocean temperature distribution by varying its mean value. In the second, we run simulations with different values of the $RHCL2$ parameter, so to obtain extremely cold or warm conditions starting from the same IC8.9; we thus use the final pickup files as ICs, this time with the value of $RHCL2^*$ that is used for all the simulations within a given set-up (see Section 2.1). These two methods allow us to construct an ensemble of different ICs that span initial global averages of surface air temperature (SAT) ranging from -40$^\circ$C to 35$^\circ$C.

\begin{figure}[ht!]
\begin{center}
\includegraphics[width=0.7\linewidth]{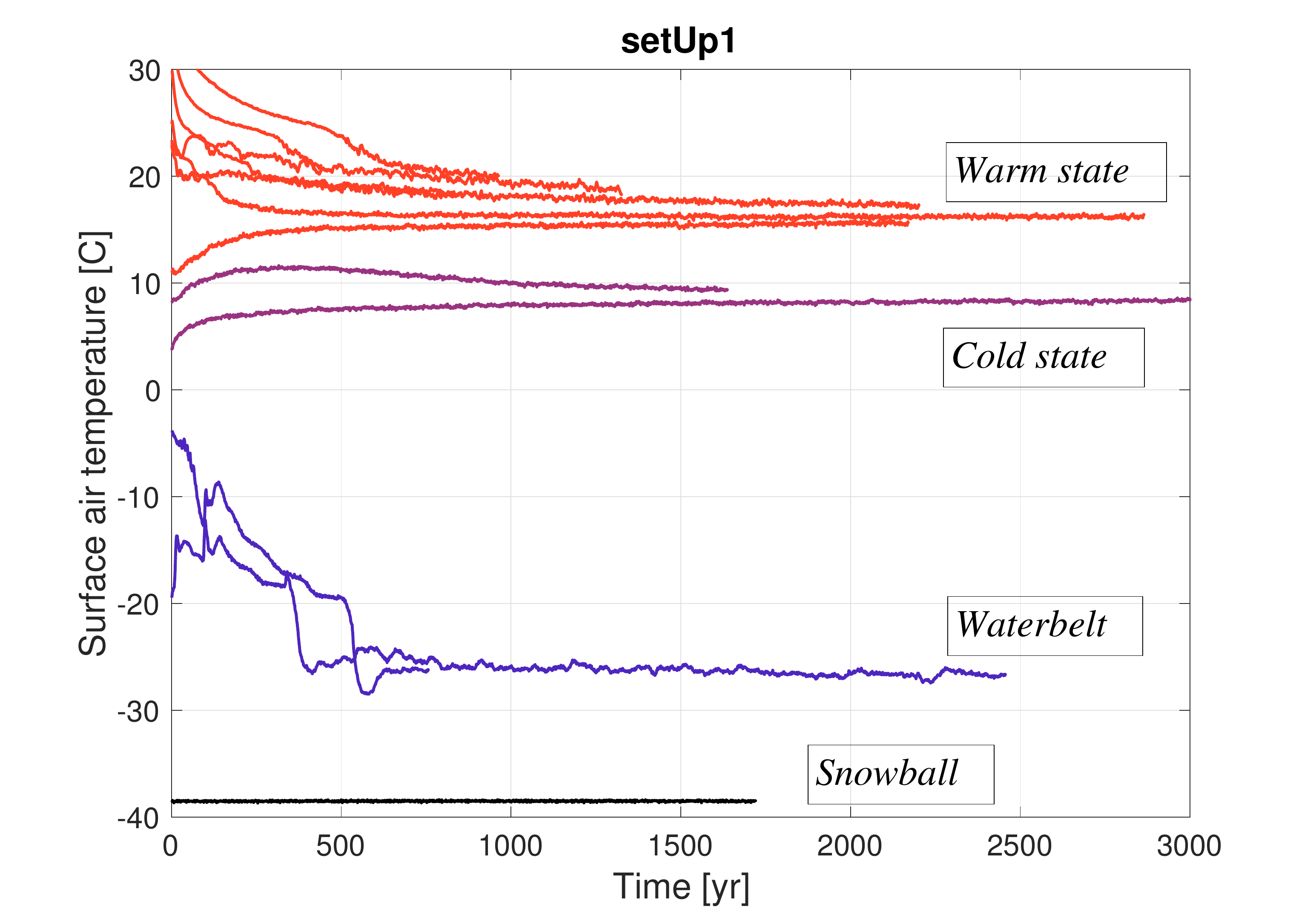}
\includegraphics[width=0.7\linewidth]{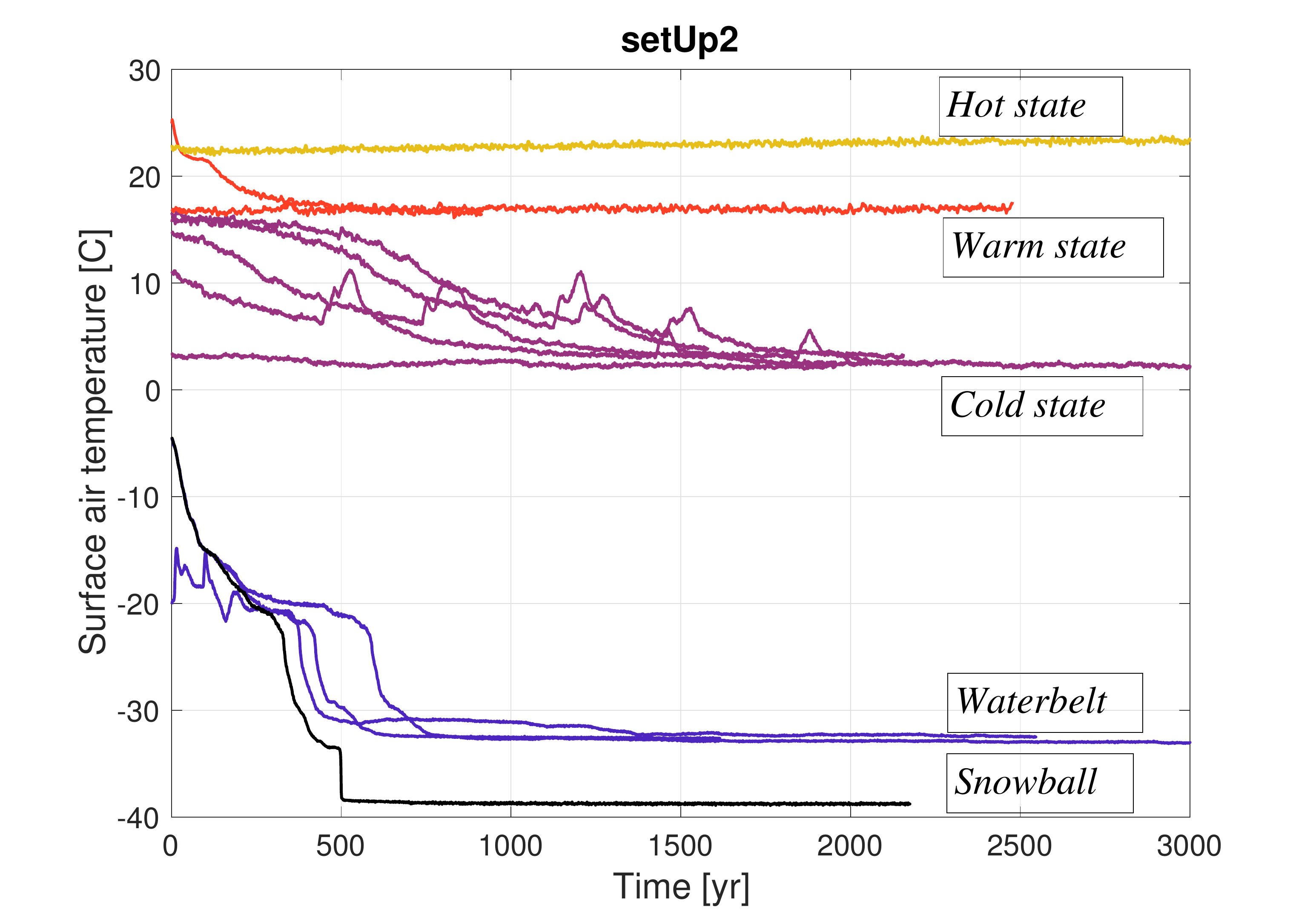}  
\caption{Evolution of surface air temperature and attractors for both set-ups.}
\label{fig:one}
\end{center}
\end{figure}

Fig.~\ref{fig:one} shows the time evolution of SAT  and how different initial conditions converge toward the attractors. 
Note that not only the number, but also the position of the attractors depends on the considered set-up. In order to obtain the snowball state, where ice entirely covers the planet,  and the waterbelt, where there is an equatorial oasis of open water, we have varied  
the maximal allowed thickness of sea ice in the range $h_{\rm{max}} = [10, 1000]$~m (not all the attempts are shown in Fig.~\ref{fig:one}). Among all, 
only one single IC converged to the snowball attractor in {\tt setUp2}, namely the one with $h_{\rm{max}} = 60$~m, all the others being attracted by the waterbelt state. This means that obtaining a snowball starting from non-snowball ICs is unlikely in the considered set-ups of MITgcm. Moreover, the dependence on $h_{\rm{max}}$ may be a spurious effect of the absence of sea ice dynamics in our simulations.  
The waterbelt~\citep{Pierrehumbert2011}\footnote{Also called `slushball', `soft' snowball or Jormungand state~\citep{Abbot2011}, 
depending on the mechanism that determines such state.} and the snowball 
had already been shown to strongly depend 
on ocean~\citep{2001GeoRL..28.1575P} and sea ice dynamics~\citep{2007JGRC..11211014L,cp-8-2079-2012}, sea ice/snow albedo parameterisations, cloud radiative forcing and ocean/atmosphere heat 
transports~\citep{2012CliPa...8..907Y}. In particular,  we confirm that is harder to enter into the snowball state when sea ice dynamics is excluded as in our simulations~\citep{2007JGRC..11211014L,cp-8-2079-2012,Rose2015}. 

Some of the ICs ending up to the cold state in {\tt setUp2} (see Fig.~\ref{fig:one}, right panel, purple curves) show fast and transient increases in SAT. We have verified that such abrupt changes are related to the sudden reduction of sea ice in small parts of the polar regions, the SAT responding through the ice-albedo feedback on a short timescale of the order of some decades. Such abrupt changes are coupled to variations in salinity, as also discussed by~\citet{2013JCli...26.2862R} where they were shown to be related to the abrupt loss of salt stratification in the polar ocean. However, this internal variability is not sufficient to attract the solution to the warm state, and the solution finally evolves toward the cold attractor.       
 
In Table~\ref{tab:2} global mean values of SAT, ocean temperature, sea ice extent and corresponding latitude of sea ice boundary, calculated from the last 100 yr of simulations, are listed in order to illustrate their dependence on the chosen set-up ({\tt setUp1}  in black, {\tt setUp2} in gray). The two main indicators of simulation quality discussed in~\citet{2018ClDy...50.4425B}, namely the energy imbalance at Top-Of-Atmosphere (TOA)  and at the ocean surface, are also listed in Table~\ref{tab:2}. 

The former is an indicator of the limitations existing within a given climate model, at the level of both numerical
implementation and physical parameterisations~\citep{2011RvGeo..49.1001L,2014ClDy...43..981L,Lembo2019}. 
In our case it is in general larger in {\tt setUp2} than in {\tt setUp1}, confirming that the latter configuration is more accurate than the former, since it properly includes friction 
heating\footnote{We have verified that the TOA imbalance does not depend on the considered cloud albedo parameterisation.}. 
Note that the values of TOA imbalance in {\tt setUp2} are of the same order of magnitude of some IPCC climate models in the preindustrial scenario (see Fig.~2 in~\citet{2011RvGeo..49.1001L}).   

The ocean surface imbalance $Q_{\rm{net}}$ of radiation flux is related to the ocean temperature drift, ${\rm{d}}T/{\rm{d}}t$, through $Q_{\rm{net}} = c_{\rm{p}}\, \rho\, h\, {\rm{d}}T/{\rm{d}}t$ (where $\rho = 1023$~kg m$^{-3}$ is the sea water density, $c_{\rm{p}} = 4000$~J(K kg)$^{-1}$ is the specific heat capacity, and $h = 3000$~m is the ocean depth). We have verified 
that all our simulations have $Q_{\rm{net}} \simeq 0$ at the equilibrium. 
From the values in Table~\ref{tab:2}, it can be seen that in all the simulations the imbalance is indeed much smaller than the standard deviation.

\begin{table}
\caption{Global mean values for the different attractors in the two configurations, {\tt setUp1} in black and {\tt setUp2} in gray, averaged over the last 100 yr of simulations. 
}
\label{tab:2}
\centering
\small
\begin{tabular}{llllll}
\hline
Attractor & Snowball & Waterbelt & Cold state & Warm state & Hot state \\
SAT  [$^\circ$C]  &  $-38.50 \pm 0.06$ &  $-26.69 \pm 0.13$ &   $8.3 \pm 0.1$ &   $16.23 \pm 0.09$ &   ---  \\
      & {\color{gray} $-38.75 \pm 0.05$} & {\color{gray} $-33.00 \pm 0.03$} & {\color{gray} $2.2 \pm 0.1$} & {\color{gray} $17.0 \pm 0.2$} & 
      {\color{gray} $23.4 \pm 0.1$} \\
Ocean temperature [$^\circ$C]            & $-1.918011\pm 0.000005$ &  $-1.358 \pm 0.009$ &  $5.241\pm 0.002$ &  $8.643\pm 0.002$ &  ---  \\ 
 & {\color{gray} $-1.918153\pm 0.000006$} & {\color{gray} $-1.6449 \pm 0.0008$} & {\color{gray} $3.225\pm 0.002$} & {\color{gray} $9.877\pm 0.004$} & {\color{gray} $17.418\pm 0.006$} \\ 
 Sea ice extent [$10^6$~km$^2$]     &  509.9  &  $369 \pm 1$ & $126\pm 1$ 
 &  $80.2\pm 0.8$ &   ---   \\
& {\color{gray} 509.9}  & {\color{gray} $429.51 \pm 0.06$} & 
{\color{gray} $160.4 \pm 0.9$} & {\color{gray} $67\pm 1$} 
& {\color{gray} 0}  \\
Latitude of  sea ice boundary & 0 &  16 &  49 &  57 &  ---   \\ 
 & {\color{gray} 0} & {\color{gray} 9} & {\color{gray} 43} & {\color{gray} 60} & {\color{gray} 90}  \\
TOA budget [$W/m^2$]    &   $-0.5\pm0.1$ &  $-1.1 \pm 0.2$ & $-0.3\pm0.2$ 
&  $-0.3\pm 0.2$ &  ---  \\                 
 & {\color{gray} $0.3\pm 0.1$} & {\color{gray} $1.6\pm 0.1$} & {\color{gray} $3.0\pm 0.2$} & {\color{gray} $2.6\pm 0.2$} & {\color{gray} $2.3\pm 0.2$}  \\
 Ocean surface budget  &   $(-1\pm 2)\cdot 10^{-5}$ &  $0.01 \pm 0.23$ &  $0.03\pm0.18$ &  $0.04\pm 0.23$ &  ---   \\
 $Q_{net}$ [$W/m^2$]   &  {\color{gray} $(-2\pm 4)\cdot 10^{-5}$} & {\color{gray} $0.01\pm 0.14$} & {\color{gray} $-0.009\pm 0.217$} & 
 {\color{gray} $0.03\pm 0.29$} & {\color{gray} $0.05\pm 0.27$}  \\
\hline
\end{tabular}
\end{table}

Each attractor corresponds to a qualitatively different climate, which is the result of the competition among several nonlinear mechanisms and interactions between ocean, atmosphere and sea ice. Among them, the main feedbacks influencing 
the climate system include: {\it (i)} the positive ice-albedo feedback (ice is more reflective than water surfaces: the more incoming radiation, the more the ice melts, the more the surface warms); {\it (ii)} the negative Boltzmann radiative feedback (warmer surface emits more radiation back to space); {\it (iii)} the cloud feedback (more clouds reflect more short-wave radiation back to space, and at the same time more long-wave radiation back to the surface). 

\begin{figure}[ht!]

\hfill \phantom{spazio qui} \hfill SAT [$^\circ$C] \phantom{qqq} \hfill  \phantom{d} Cloud cover \% \phantom{qqq}  
\hfill Overturning~circ.~[Sv]  \hfill\hfill Heat transport [PW] \hfill{} 

\hfill 
\phantom{q}
\includegraphics[width=0.24\linewidth]{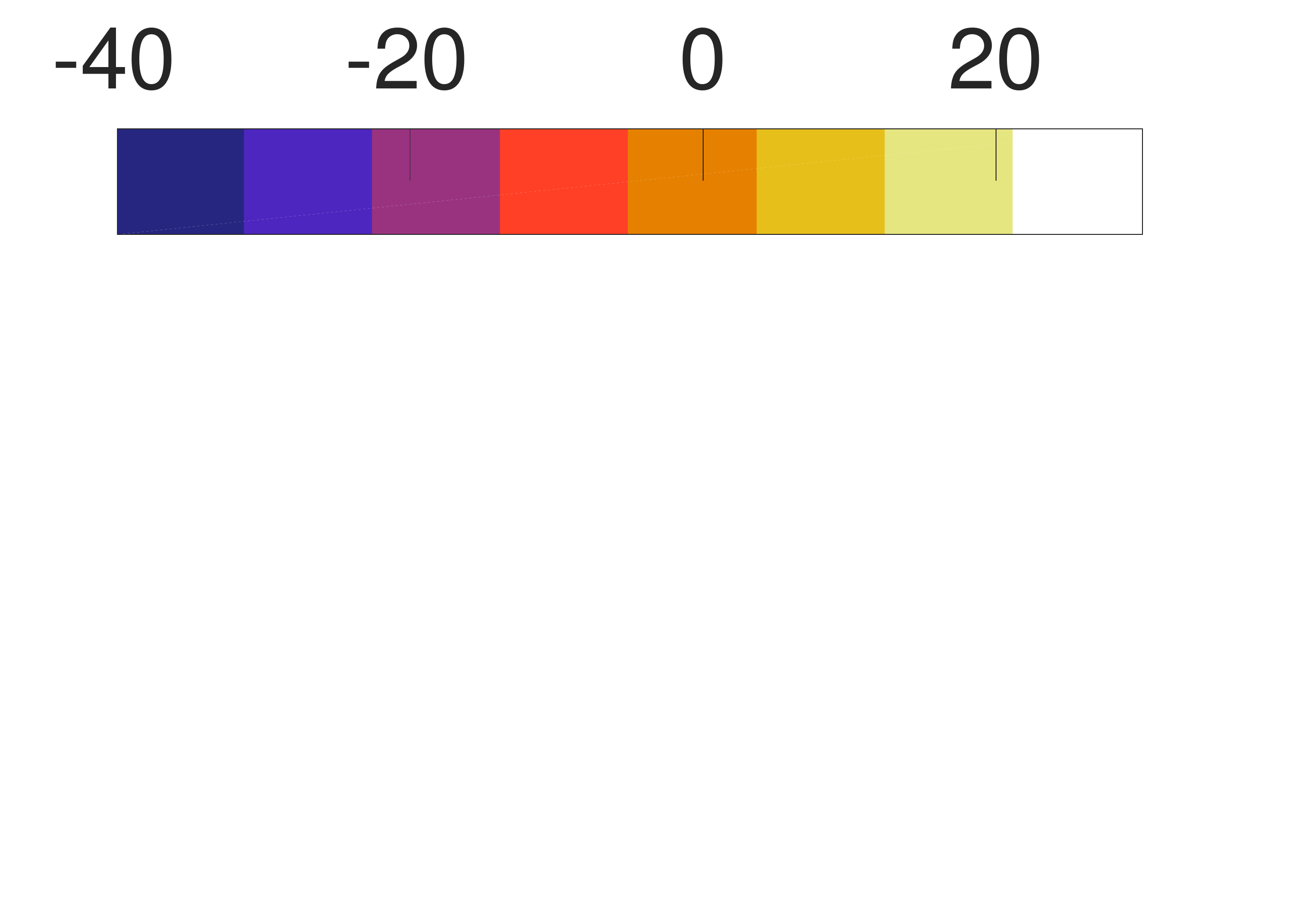} 
\includegraphics[width=0.24\linewidth]{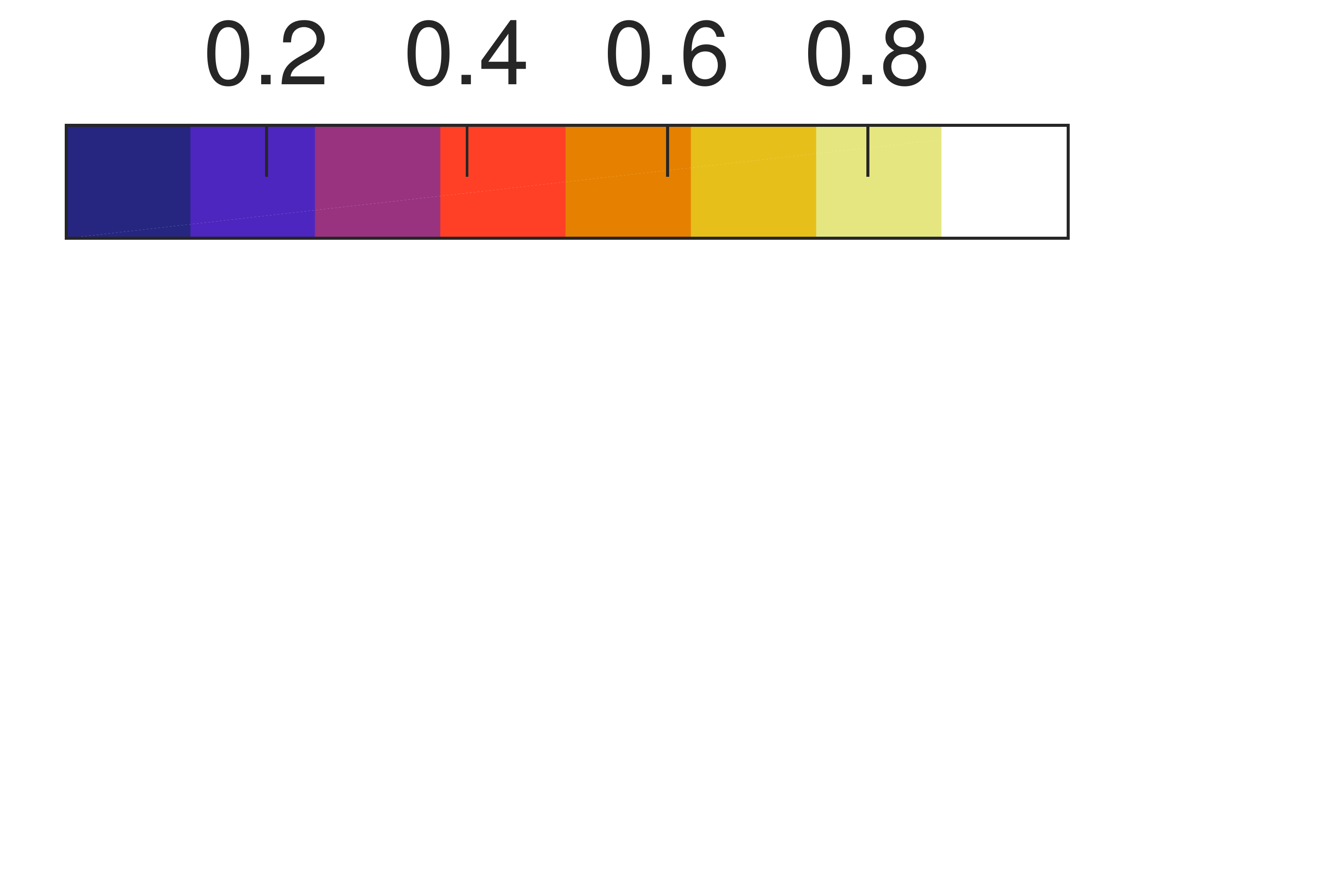} 
\includegraphics[width=0.24\linewidth]{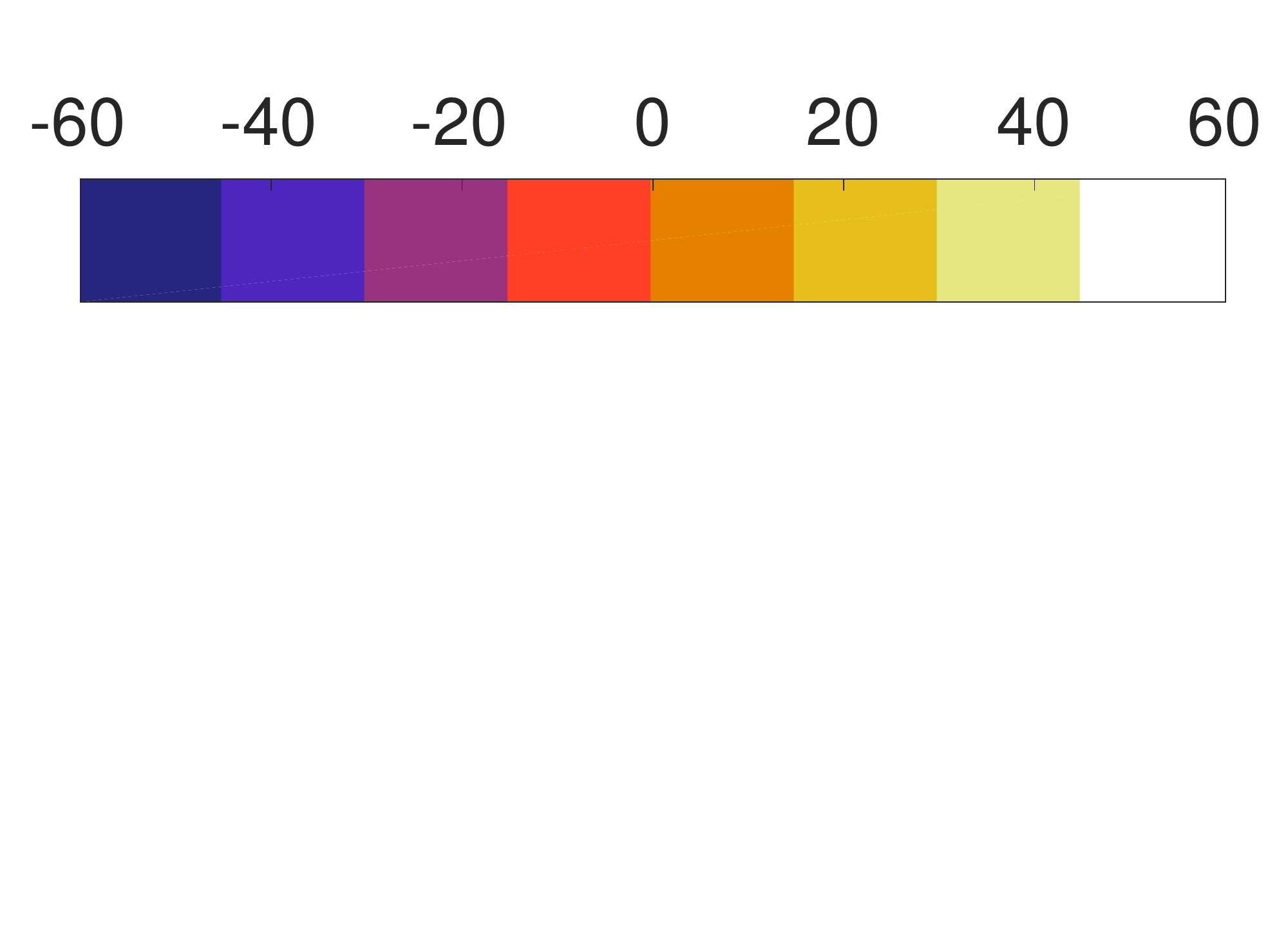} \phantom{mmmmmmmmmm}
\hfill
\vspace{-1.5cm}

\begin{minipage}{0.02\textwidth}
 \vspace{0pt}\raggedright
 \vspace{0.9cm}
{\rotatebox[origin=c]{90}{Snowball}}\vspace{1cm}
{\rotatebox[origin=c]{90}{Waterbelt}}\vspace{1.cm}
{\rotatebox[origin=c]{90}{Cold state}}\vspace{1.2cm}
{\rotatebox[origin=c]{90}{Warm state}}\vspace{1.cm}
\end{minipage}\hfill
\begin{minipage}{0.24\textwidth}
\includegraphics[width=\linewidth]{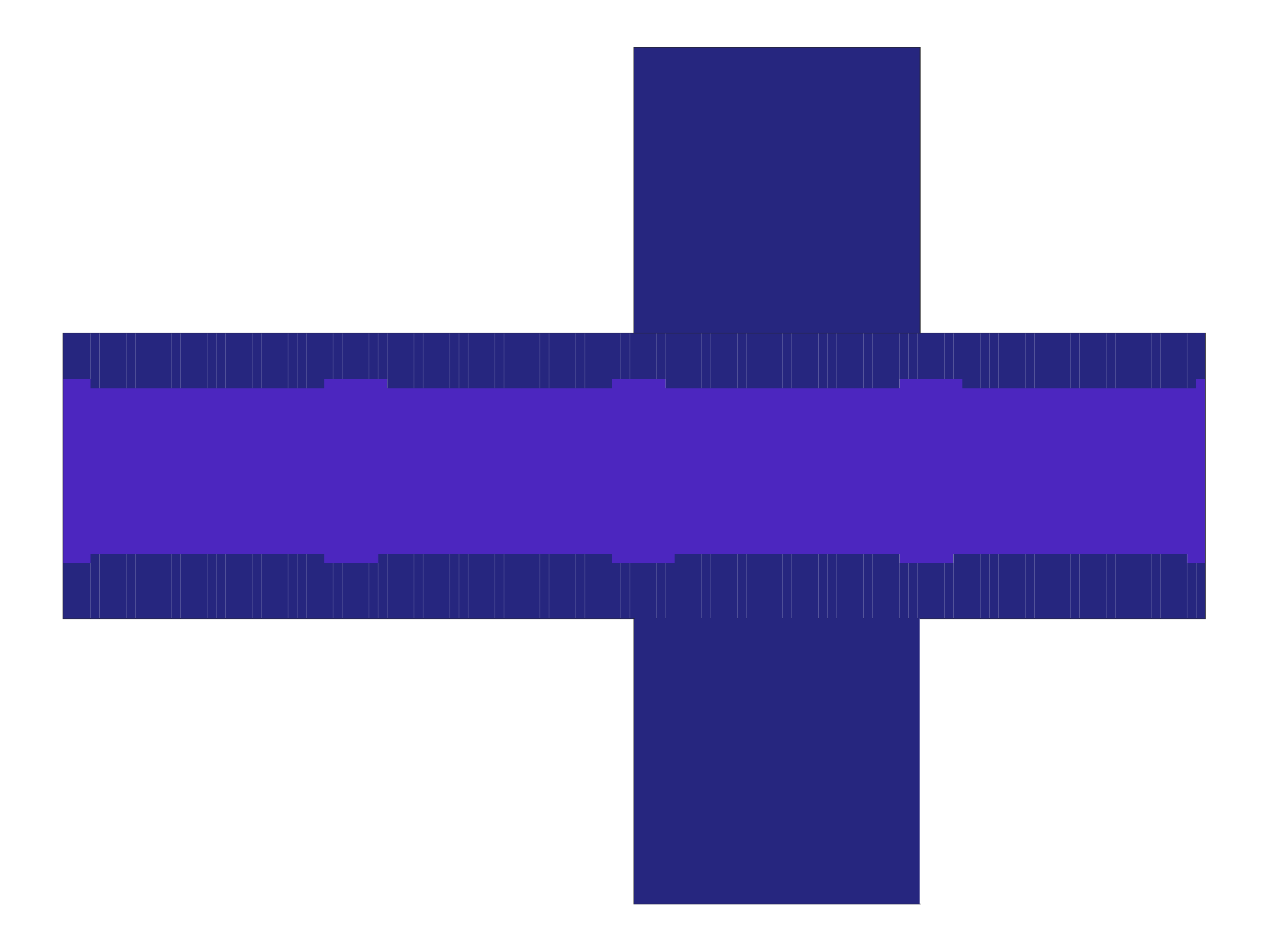}
\includegraphics[width=\linewidth]{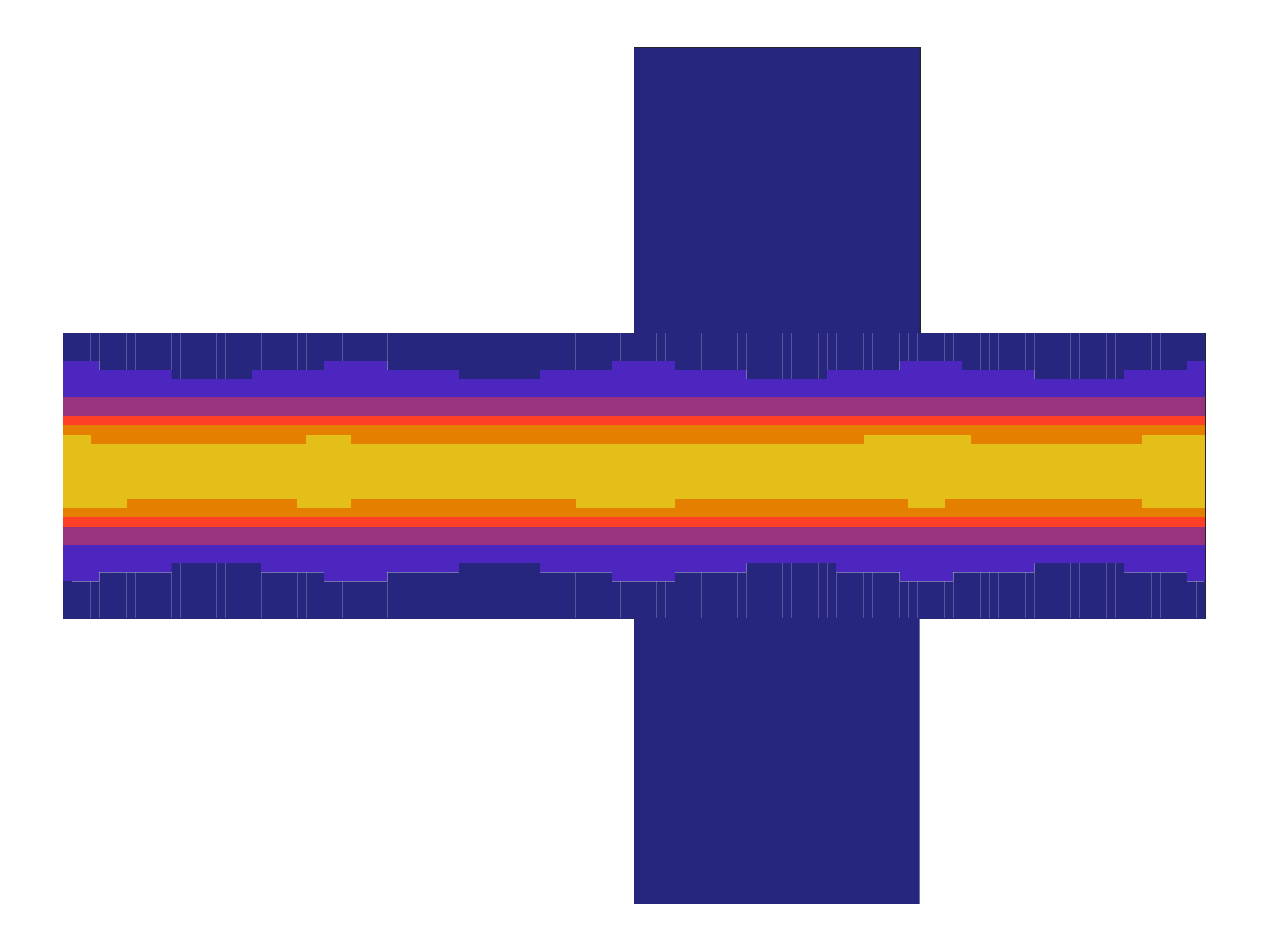}
\includegraphics[width=\linewidth]{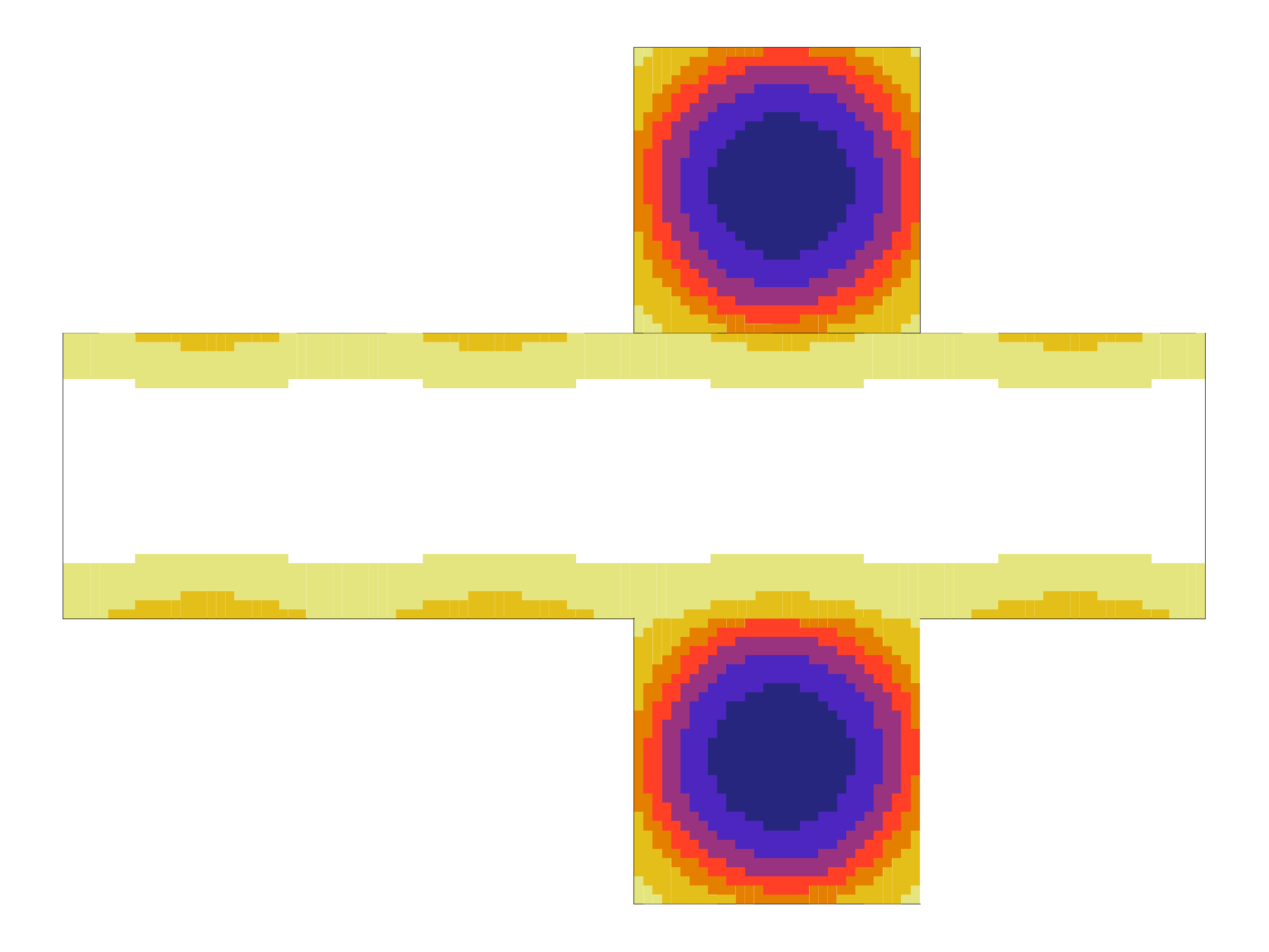}
\includegraphics[width=\linewidth]{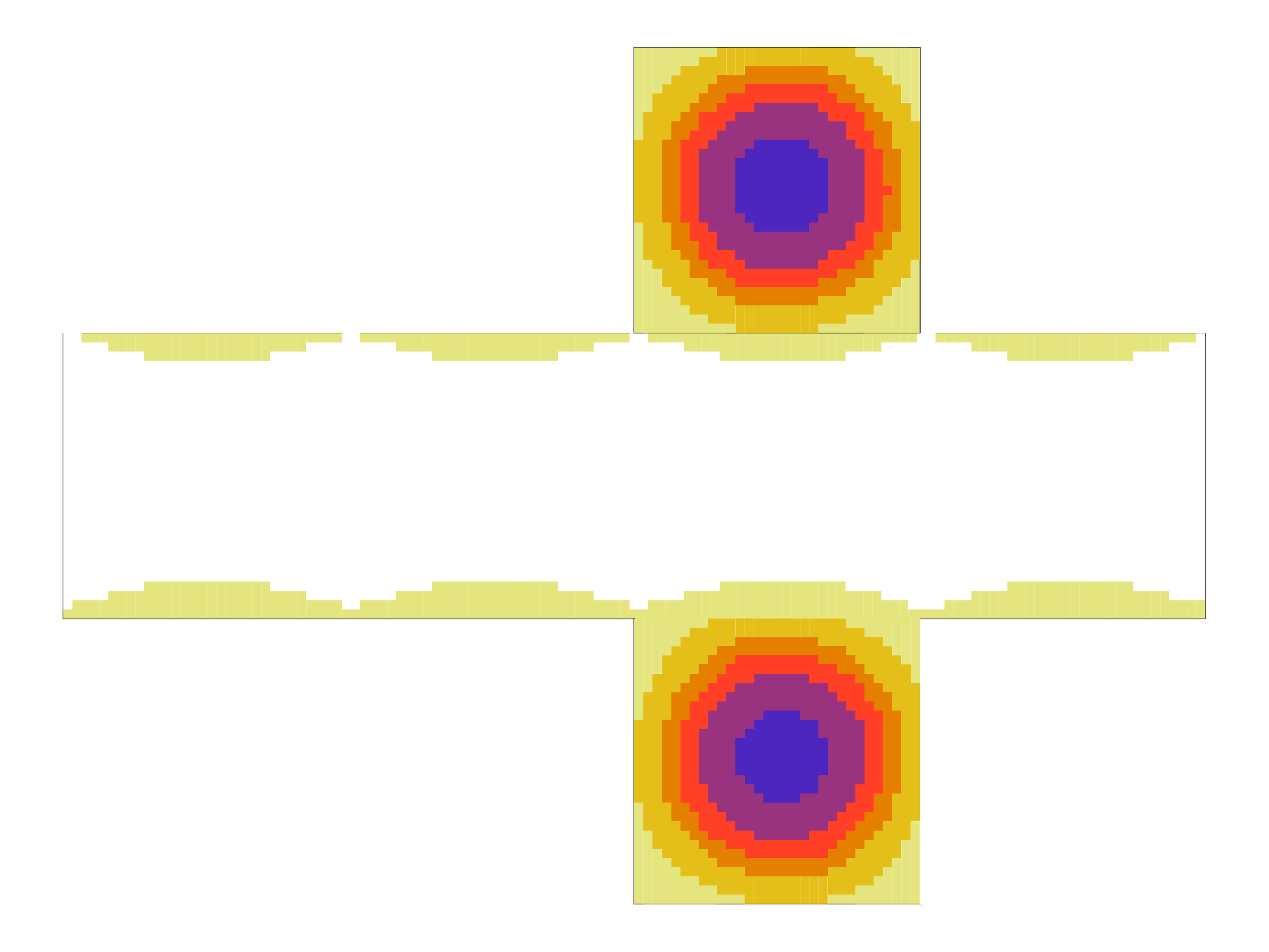}
\end{minipage}\hfill
\begin{minipage}{0.24\textwidth}
\includegraphics[width=\linewidth]{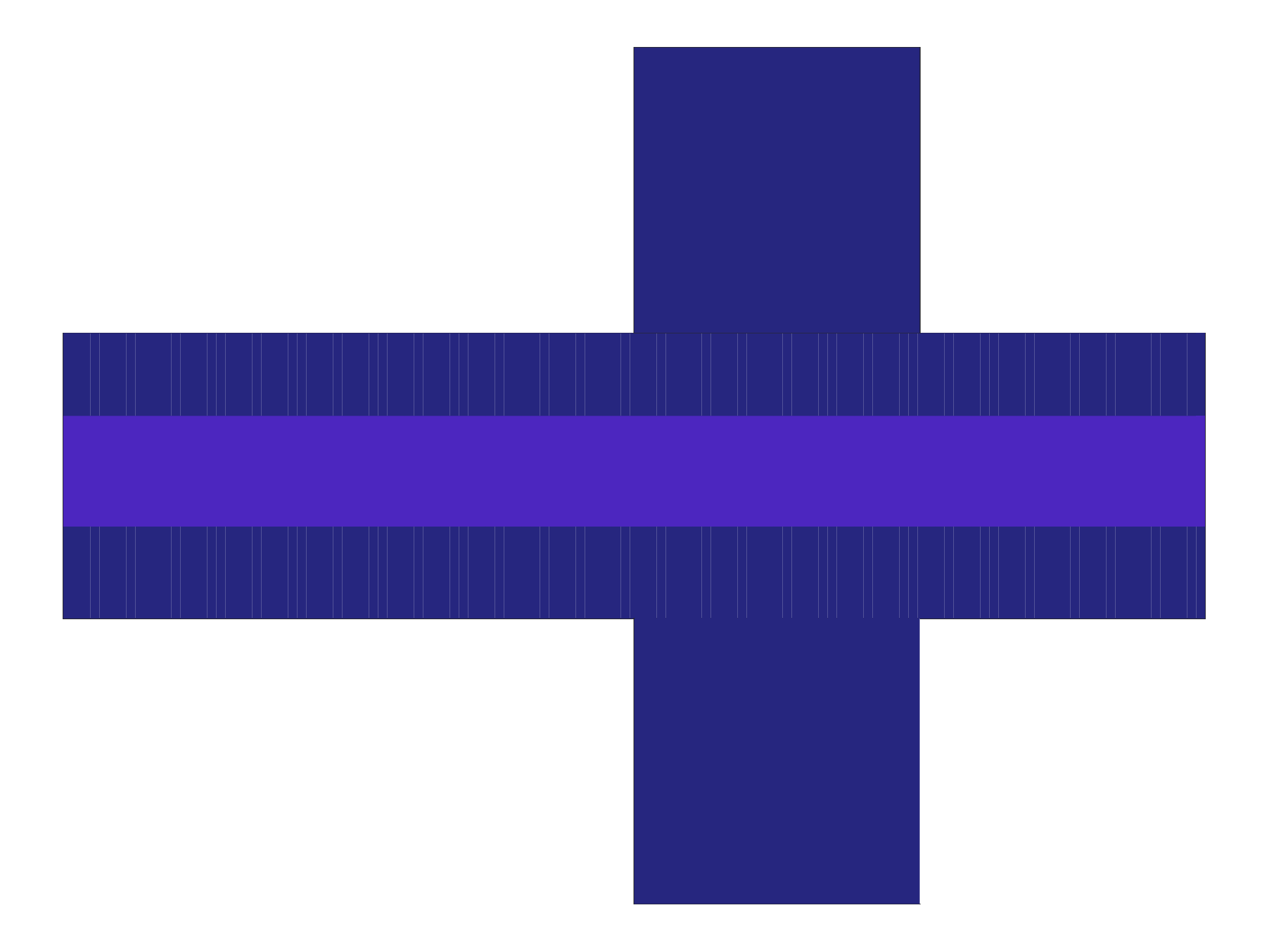}
\includegraphics[width=\linewidth]{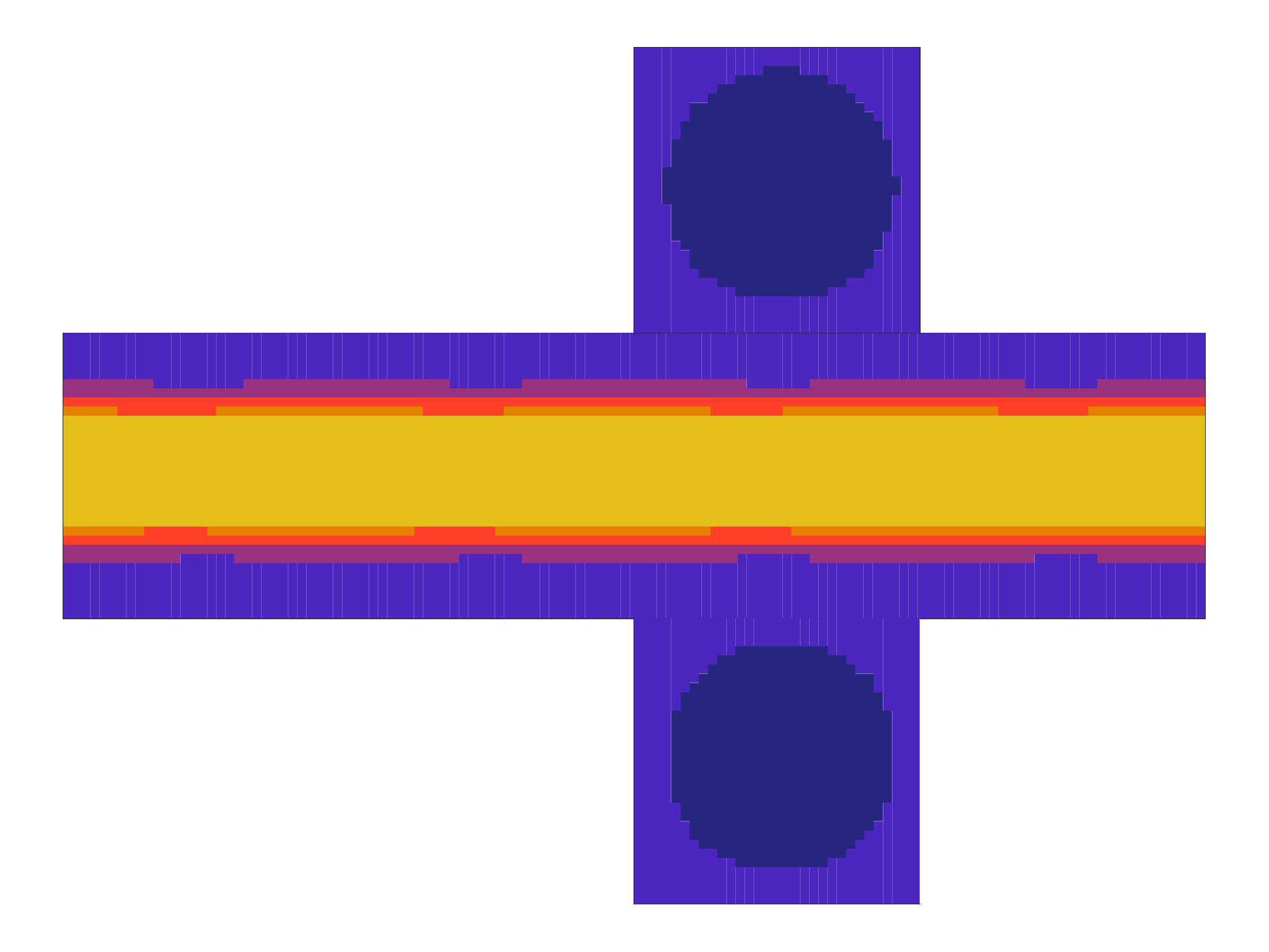} 
\includegraphics[width=\linewidth]{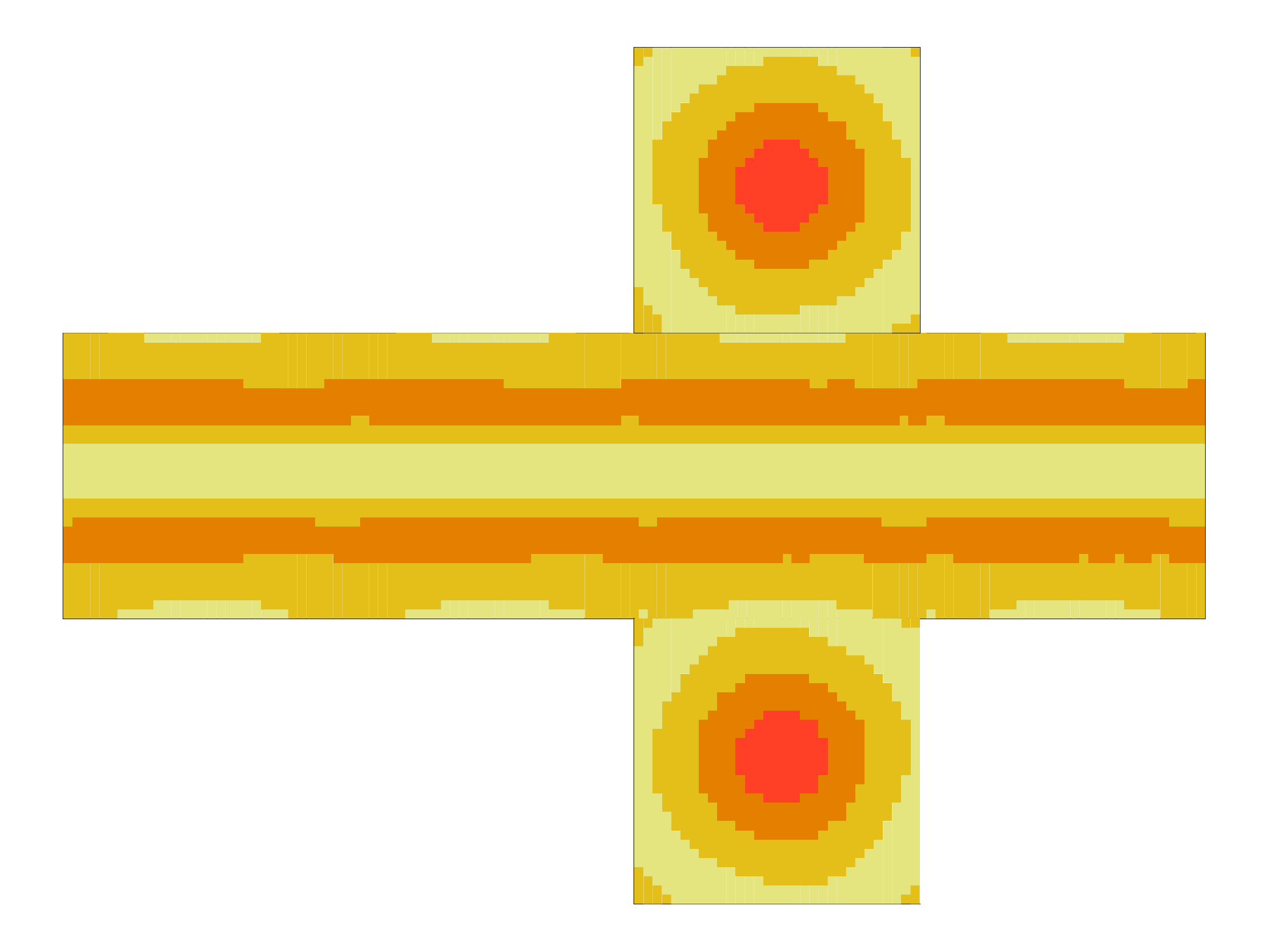}
\includegraphics[width=\linewidth]{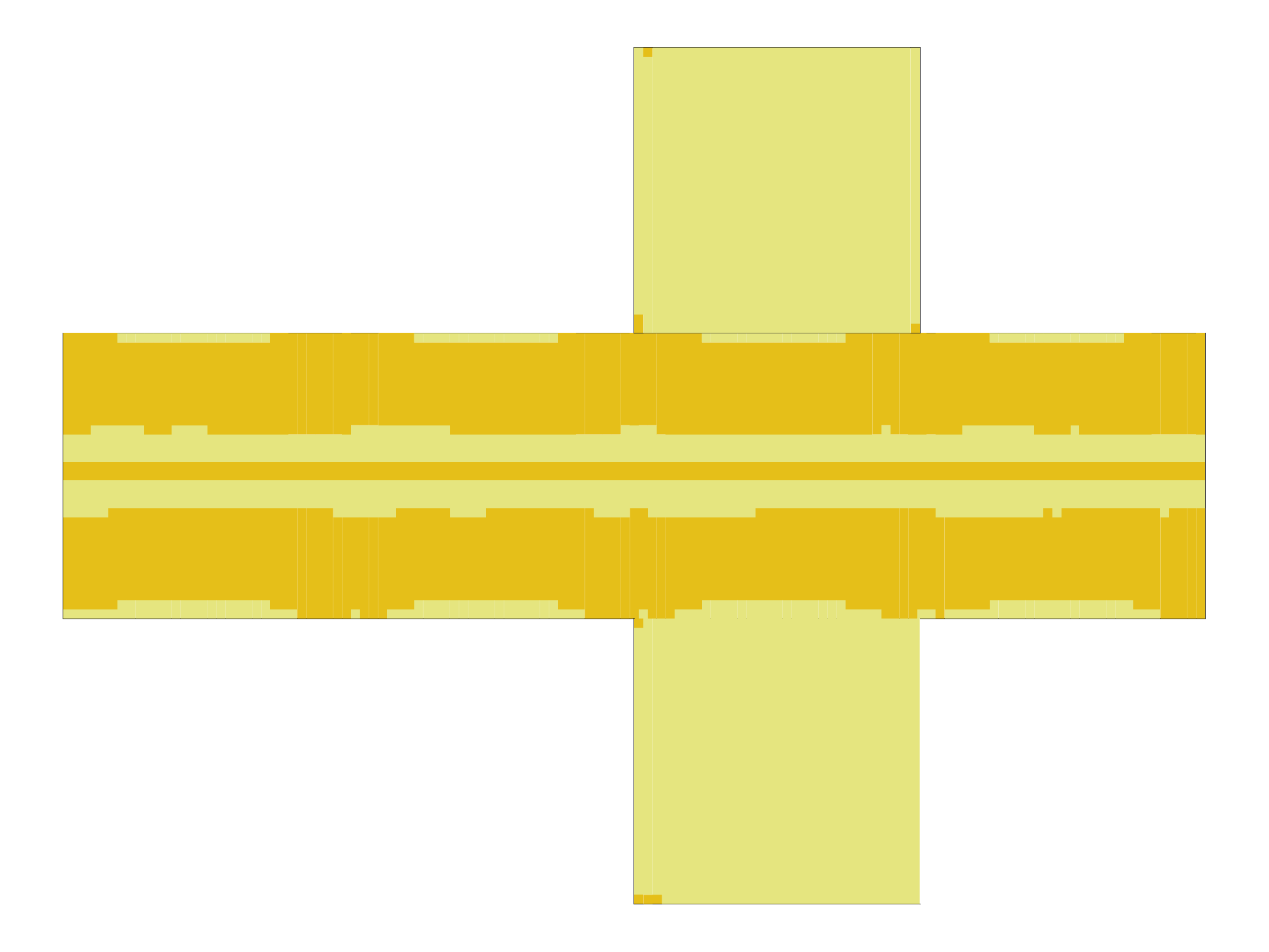}
\end{minipage}\hfill
\begin{minipage}{0.24\textwidth}
\includegraphics[width=\linewidth]{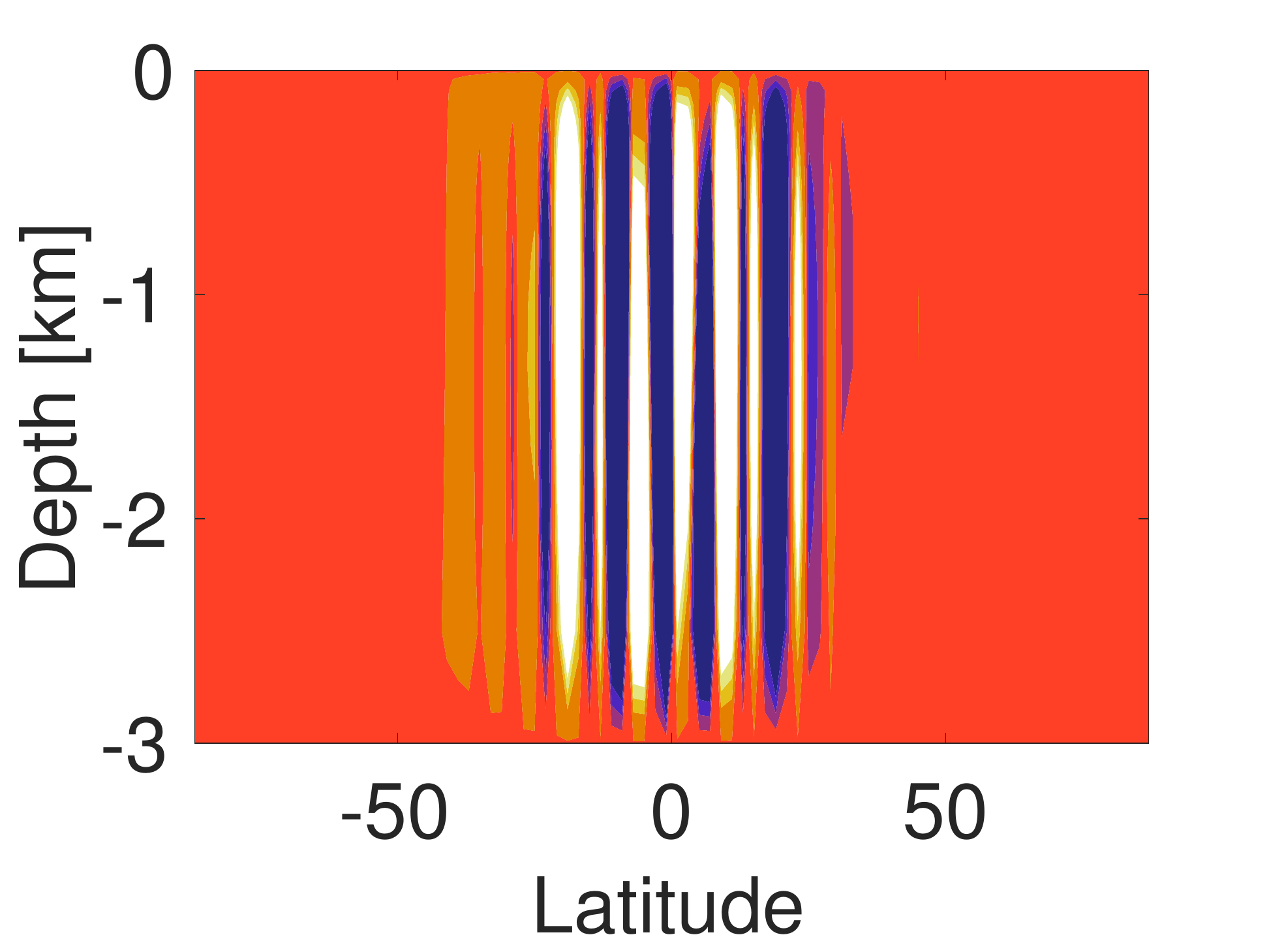}
\includegraphics[width=\linewidth]{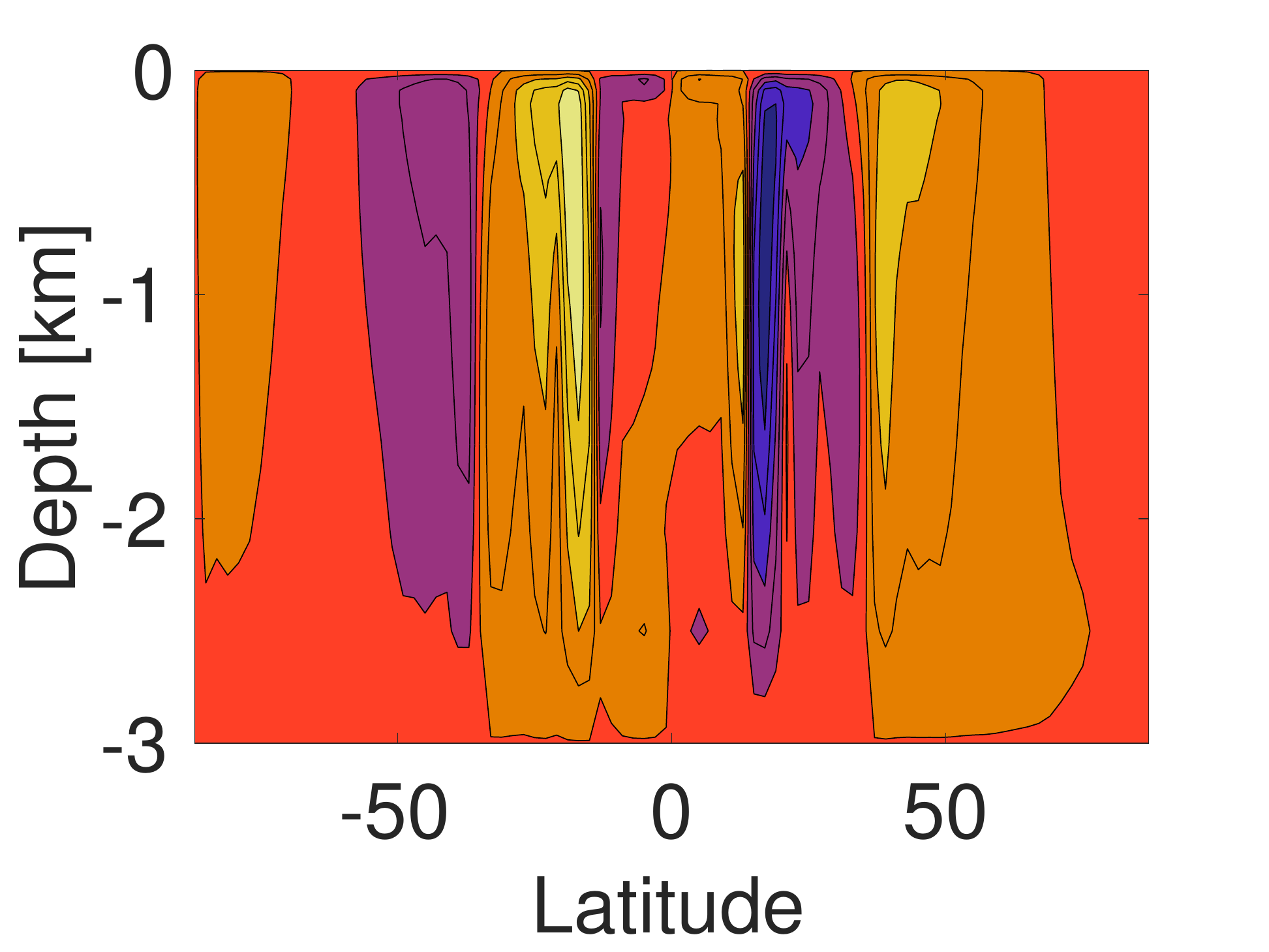} 
\includegraphics[width=\linewidth]{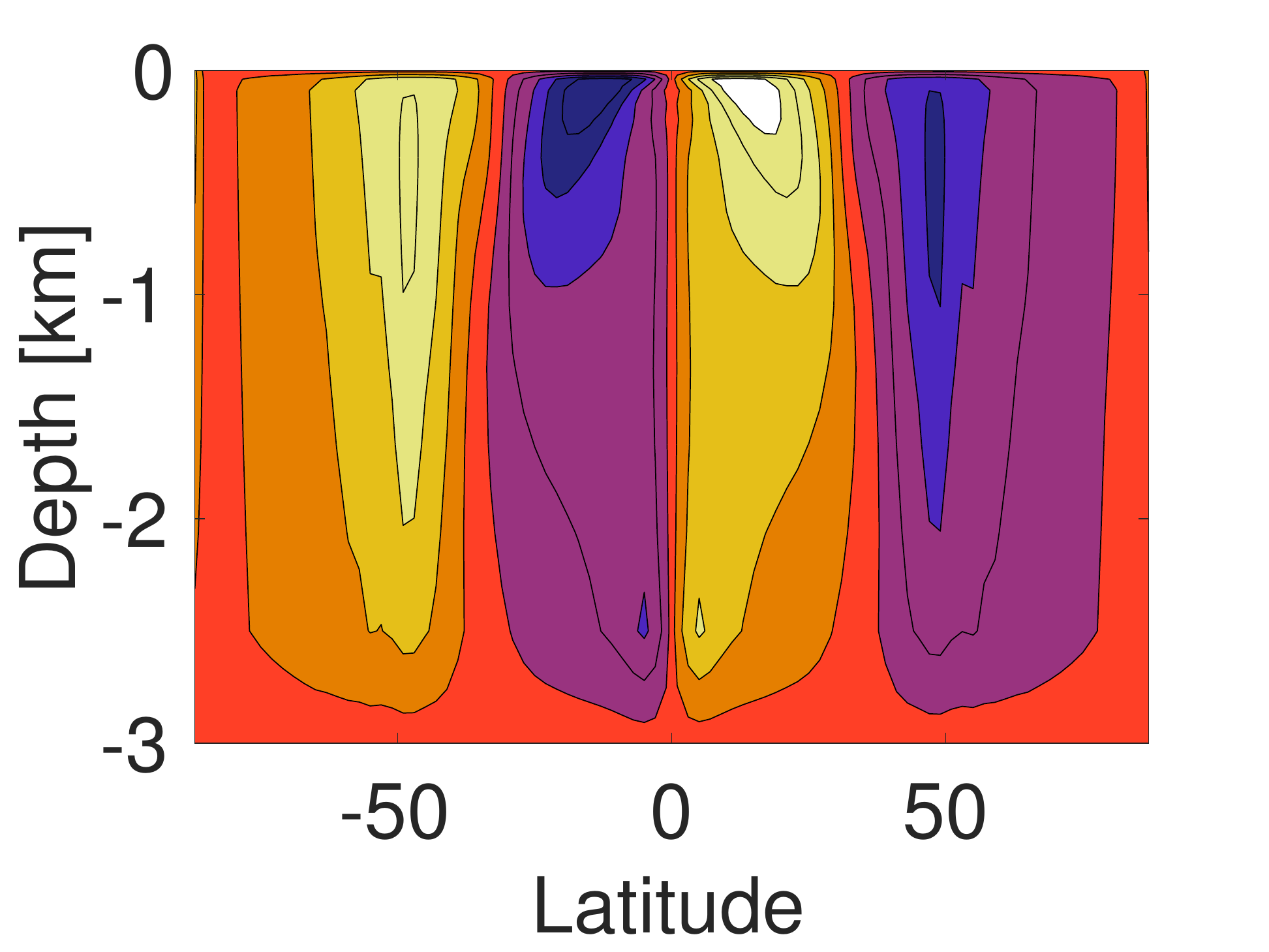}
\includegraphics[width=\linewidth]{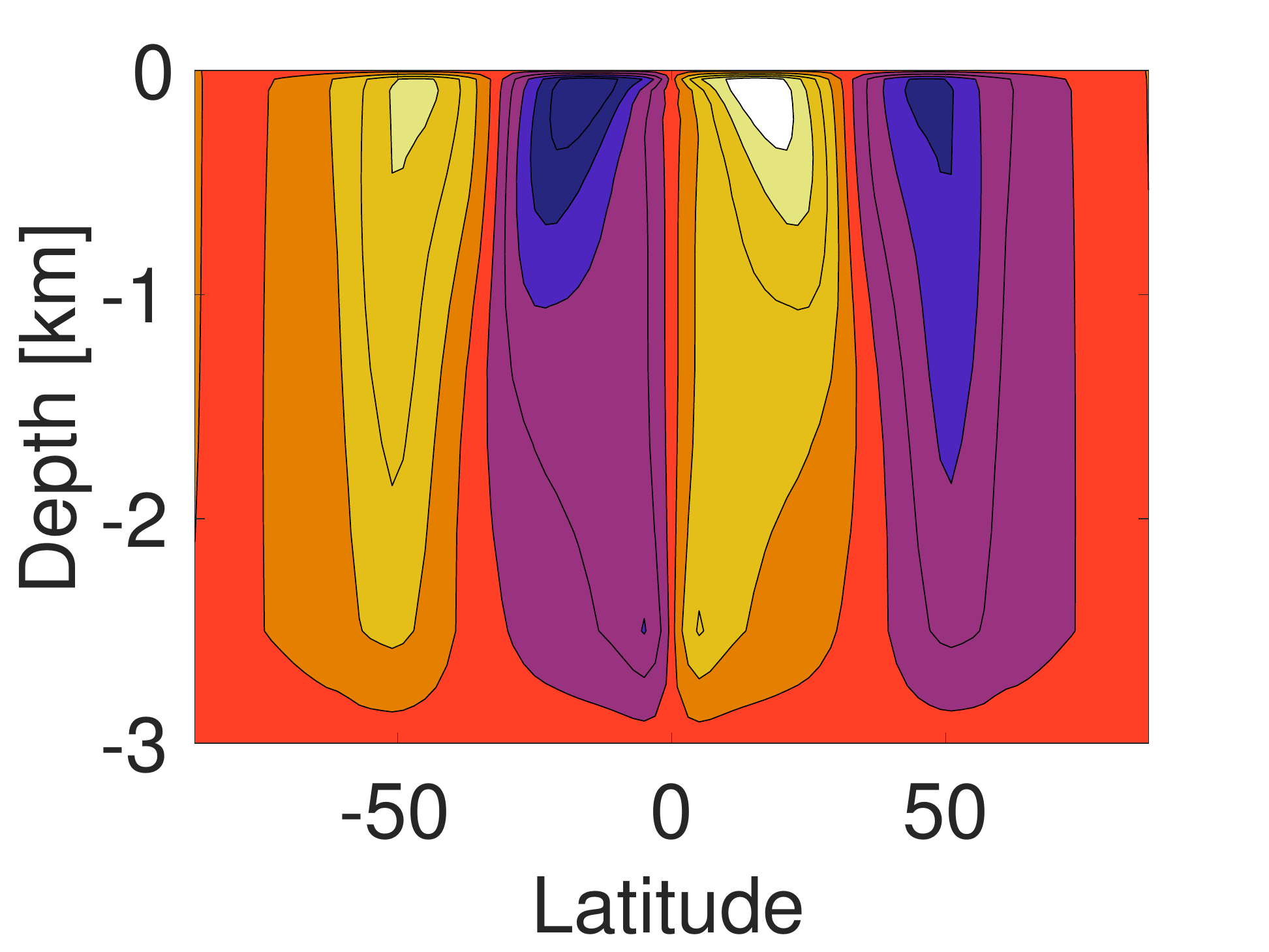}
\end{minipage}\hfill
\begin{minipage}{0.24\textwidth}
\includegraphics[width=\linewidth]{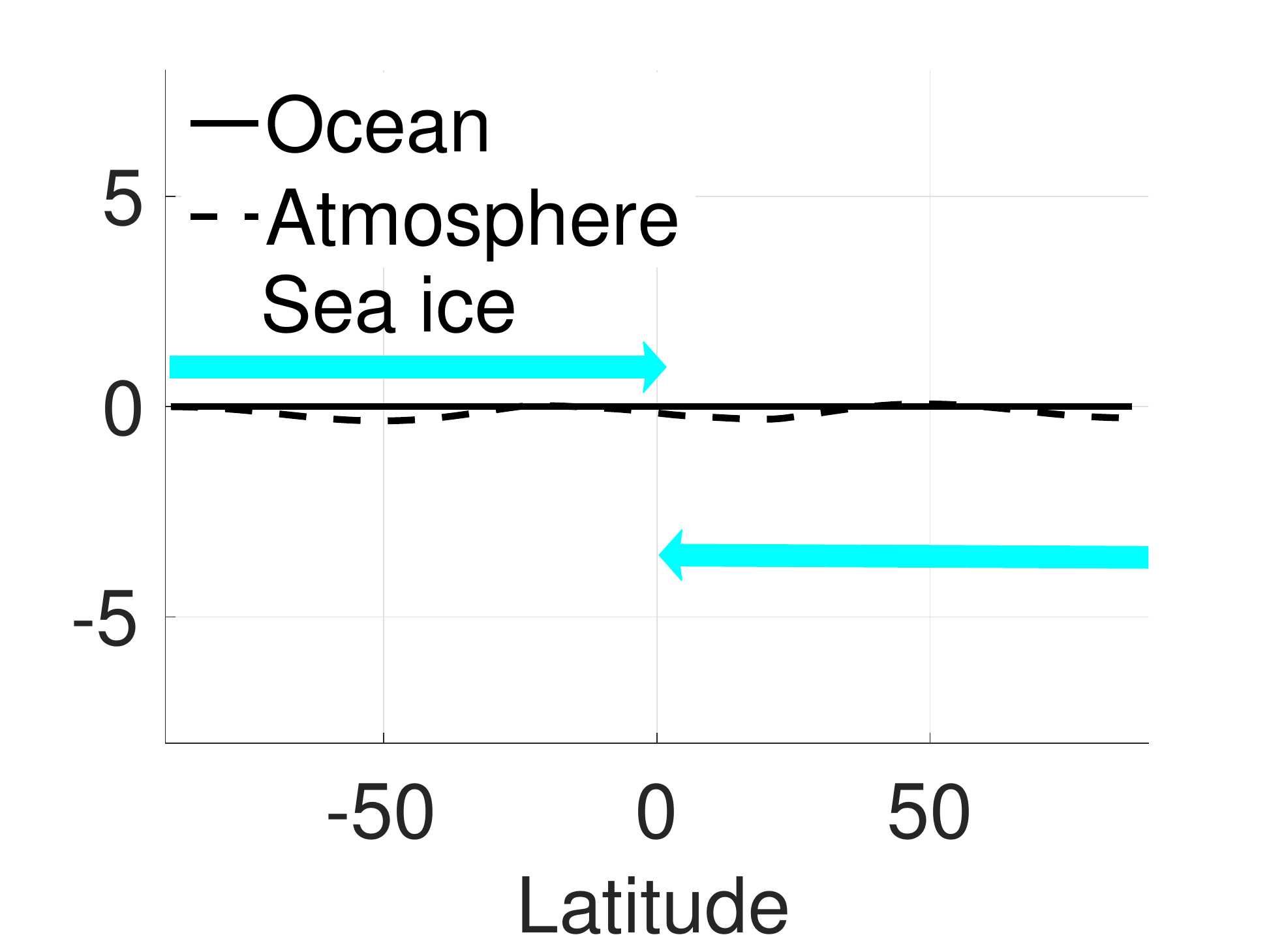}
\includegraphics[width=\linewidth]{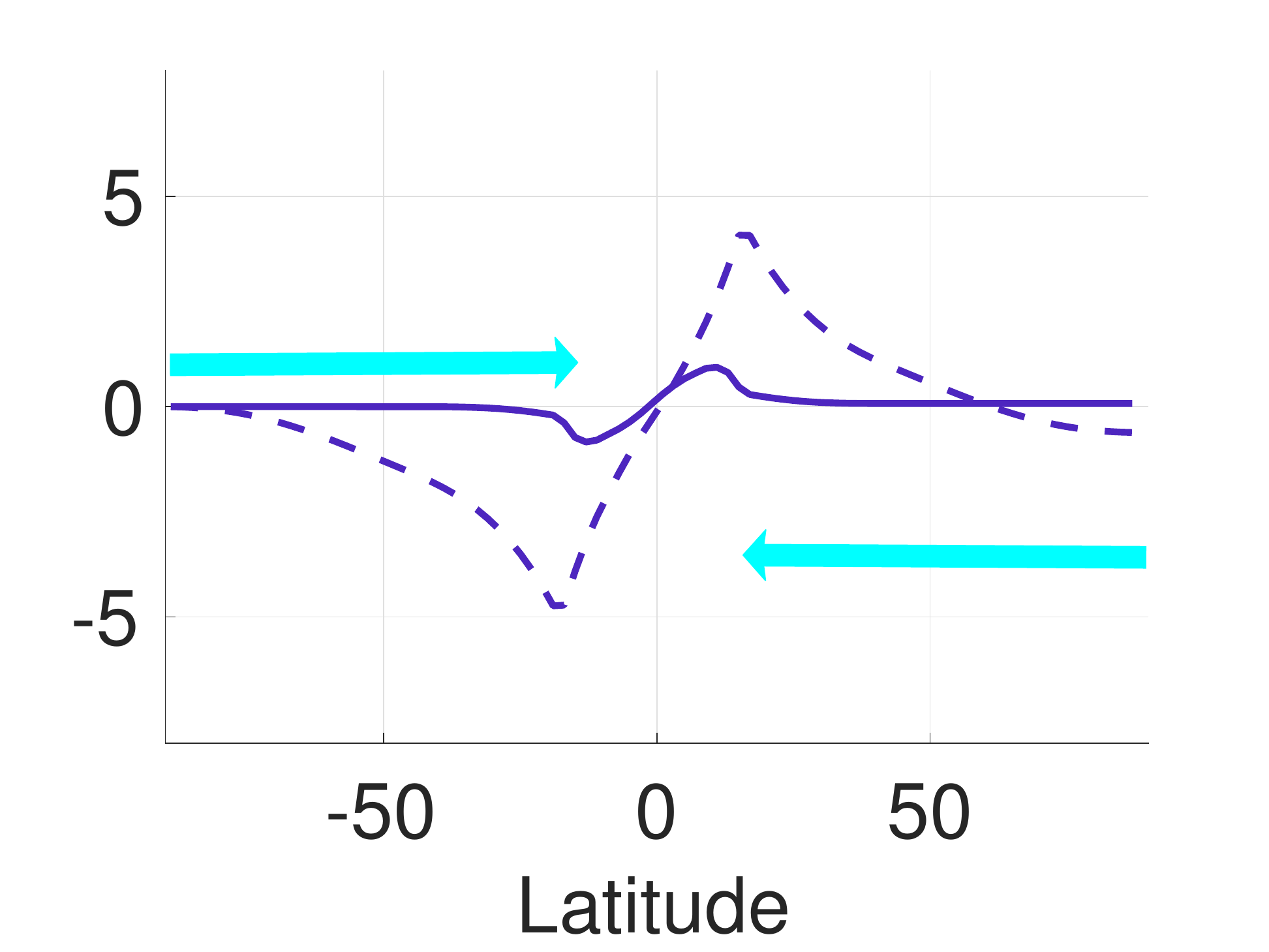}
\includegraphics[width=\linewidth]{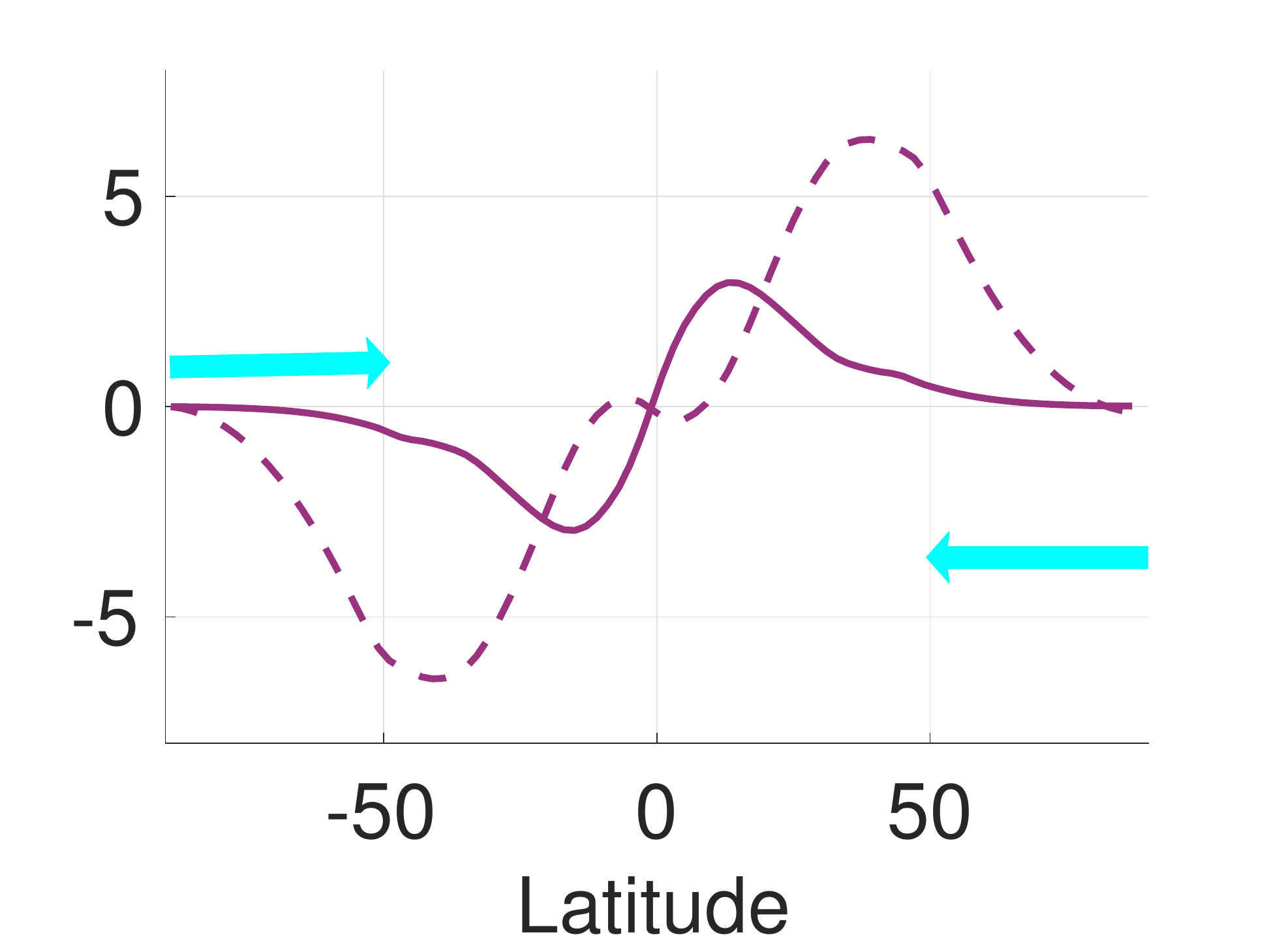}
\includegraphics[width=\linewidth]{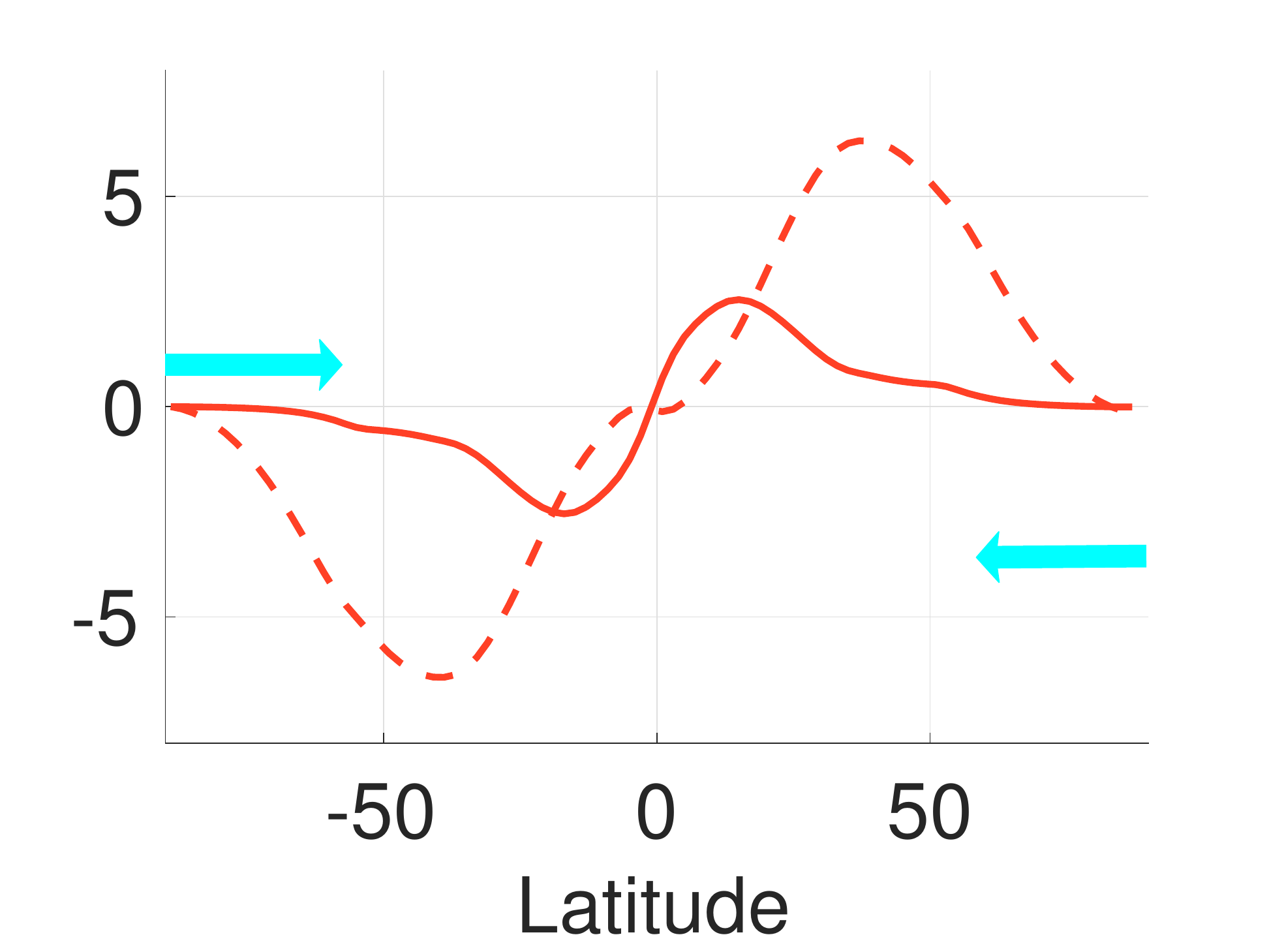}
\end{minipage}\hfill
\caption{Characterisation of the climate in {\tt setUp1} attractors. 
For each  
attractor (along the lines), the corresponding climate is described through the SAT distribution (first column), cloud cover (second column), ocean overturning circulation (third column), atmospheric and ocean heat transport (dashed and solid line, respectively, fourth column), and sea ice covered region (arrow, fourth column). Values are obtained by averaging over the last 100~yr of simulations. In the fourth column, a positive value of the heat transport on the vertical axis corresponds to a northward transport.  } 
\label{fig:two}
\end{figure}

\begin{figure}[ht!]

\hfill \phantom{spazio qui} \hfill SAT [$^\circ$C] \phantom{qqq} \hfill  \phantom{d} Cloud cover \% \phantom{qqq}  
\hfill Overturning~circ.~[Sv]  \hfill\hfill Heat transport [PW] \hfill{} 

\hfill 
\phantom{q}
\includegraphics[width=0.24\linewidth]{FIGS/colorbarSAT-eps-converted-to.pdf} 
\includegraphics[width=0.24\linewidth]{FIGS/tmpClouds2-eps-converted-to.pdf} 
\includegraphics[width=0.24\linewidth]{FIGS/mocColorbar-eps-converted-to.pdf} \phantom{mmmmmmmmmm}
\hfill
\vspace{-1.5cm}

\begin{minipage}{0.02\textwidth}
 \vspace{0pt}\raggedright
 \vspace{0.9cm}
{\rotatebox[origin=c]{90}{Snowball}}\vspace{1cm}
{\rotatebox[origin=c]{90}{Waterbelt}}\vspace{1.cm}
{\rotatebox[origin=c]{90}{Cold state}}\vspace{1.2cm}
{\rotatebox[origin=c]{90}{Warm state}}\vspace{1.cm}
{\rotatebox[origin=c]{90}{Hot state}}\vspace{1.cm}
\end{minipage}\hfill
\begin{minipage}{0.24\textwidth}
\includegraphics[width=\linewidth]{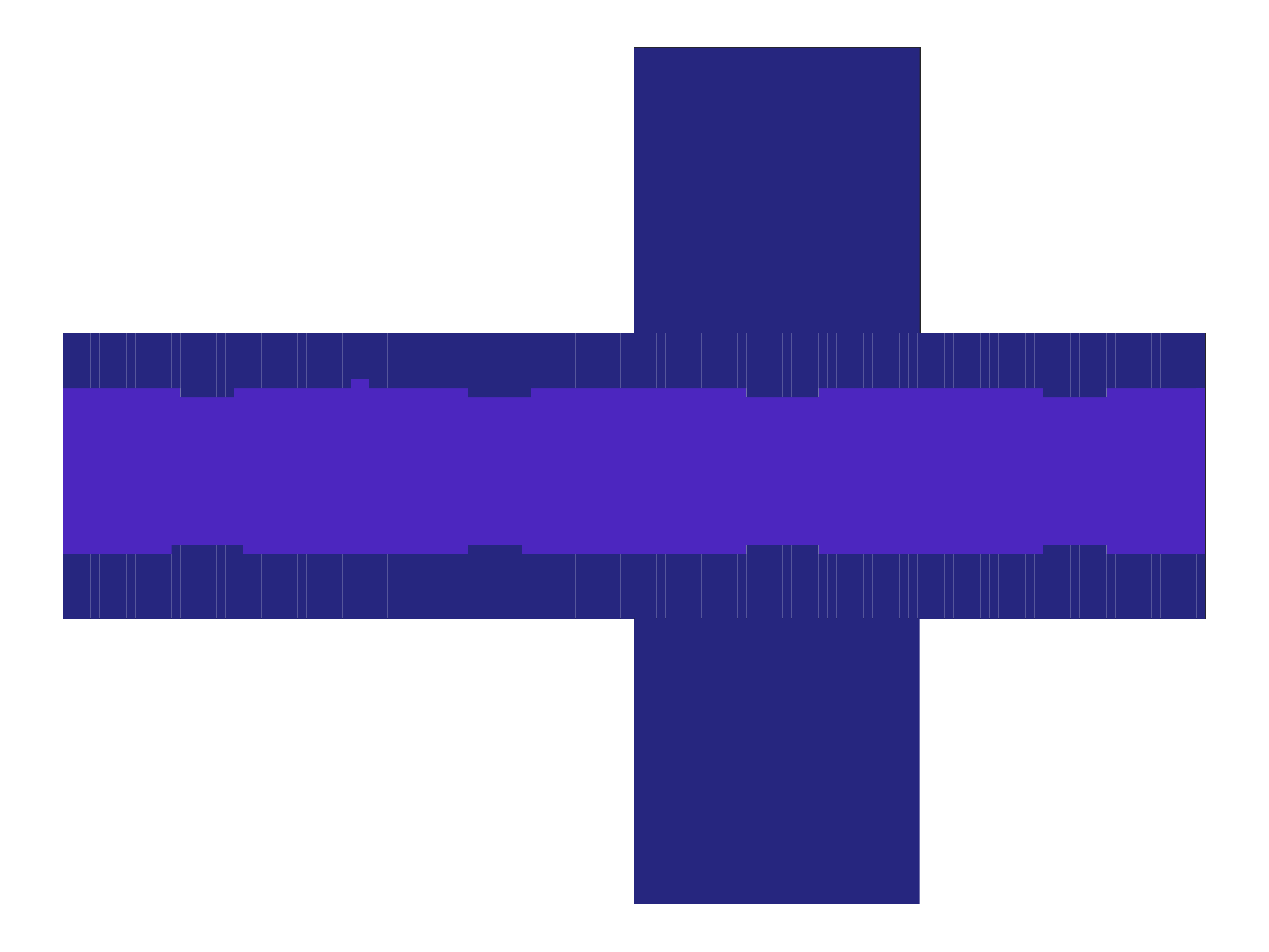}
\includegraphics[width=\linewidth]{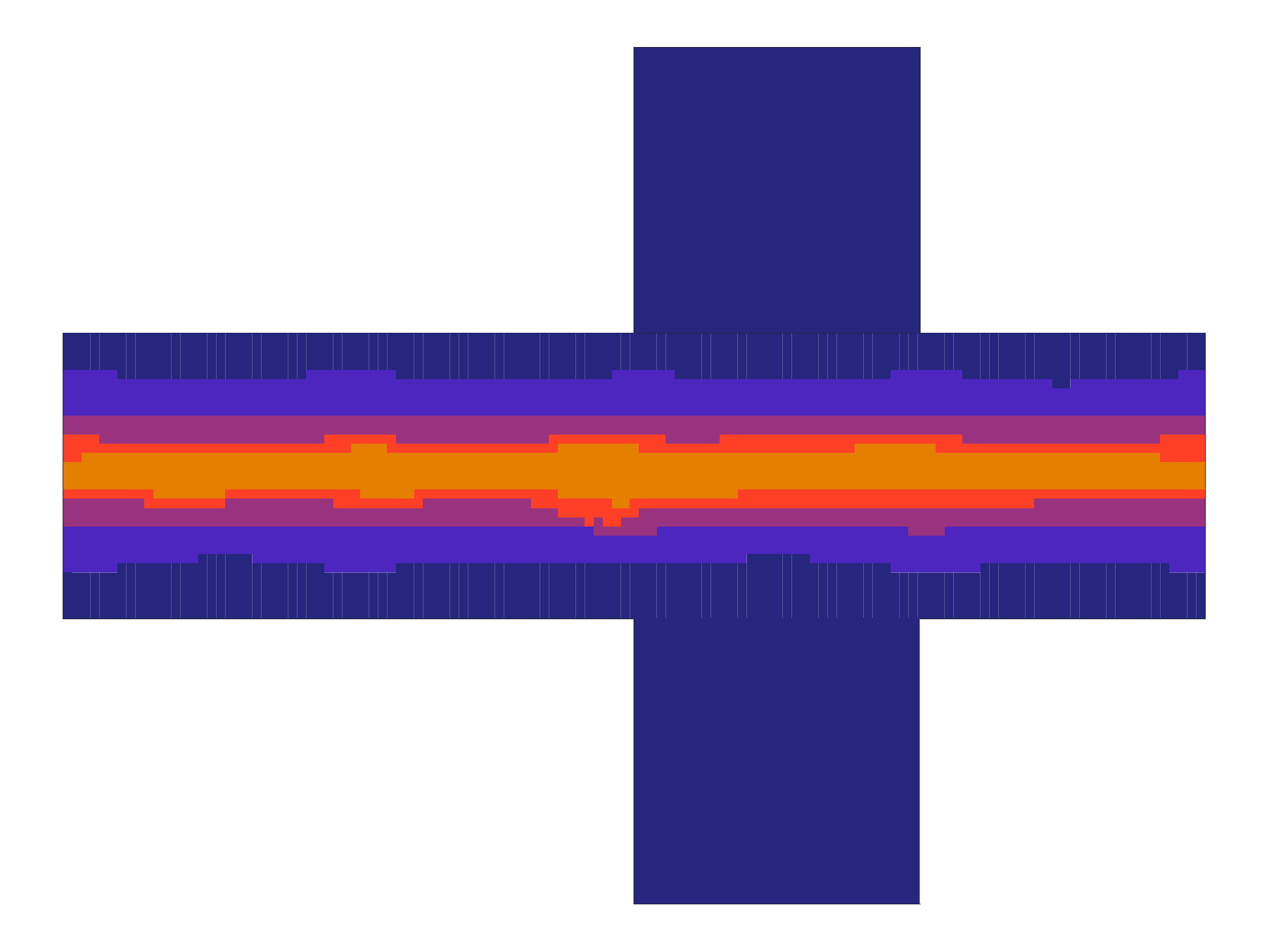}
\includegraphics[width=\linewidth]{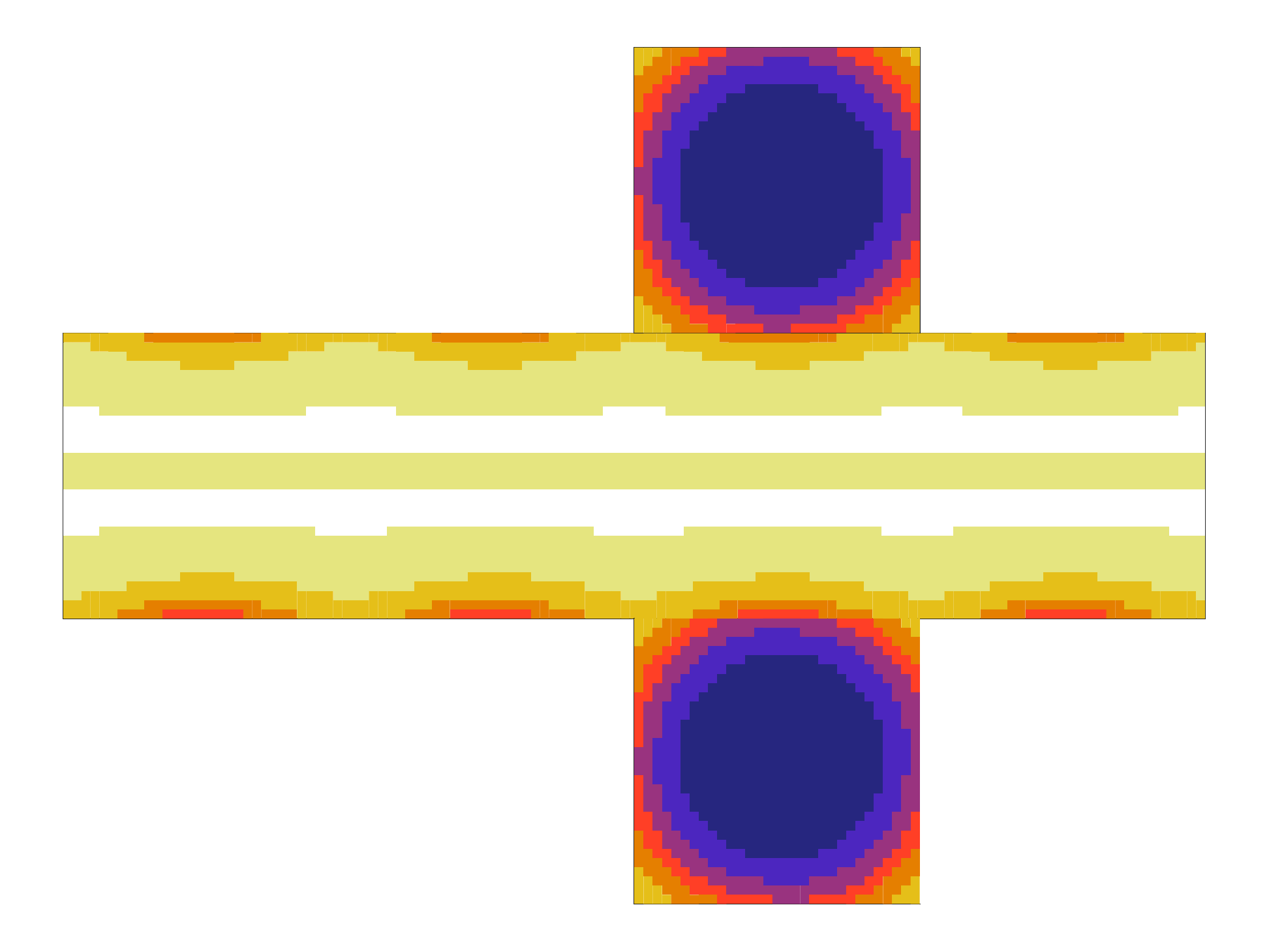}
\includegraphics[width=\linewidth]{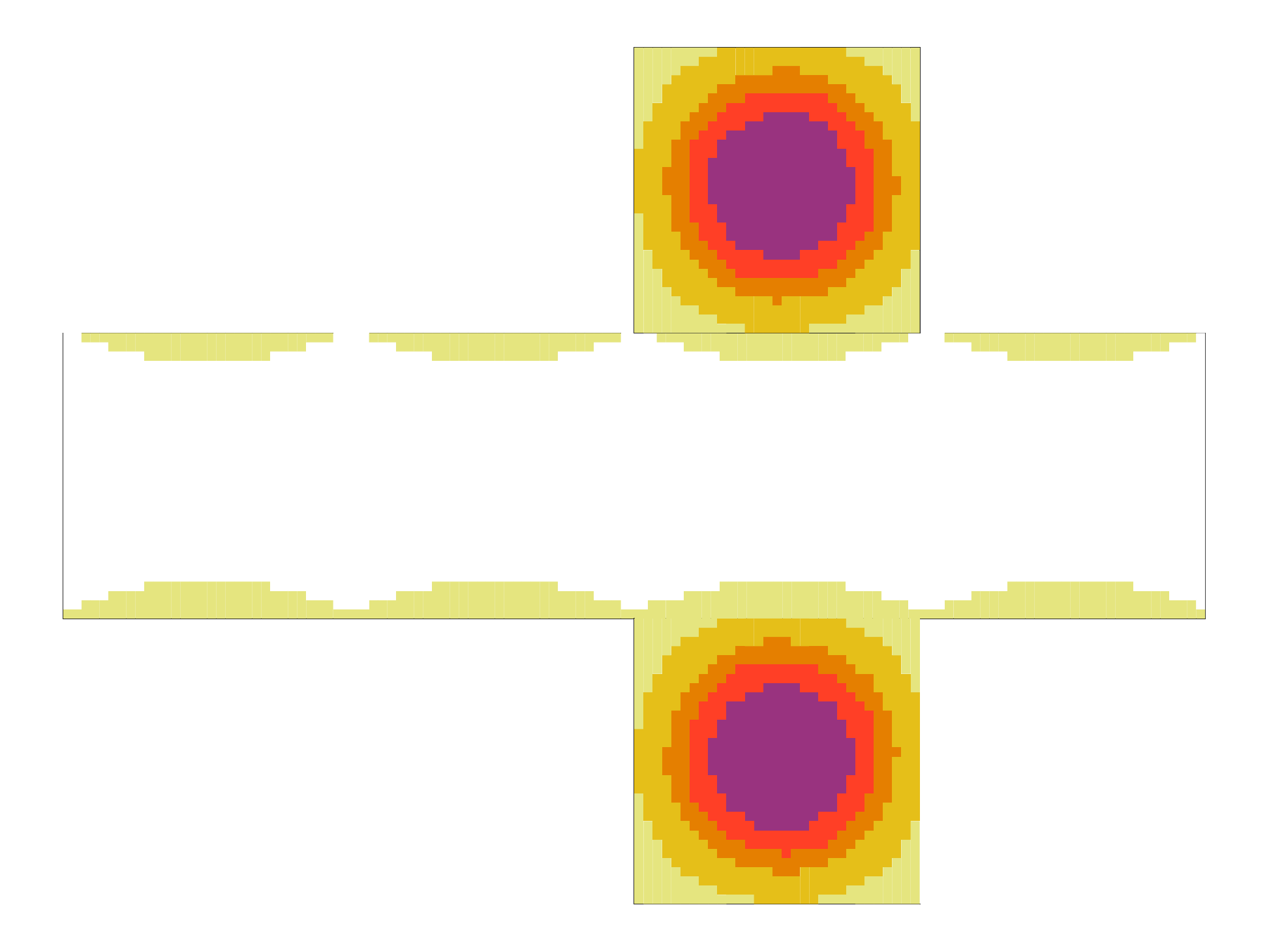}
\includegraphics[width=\linewidth]{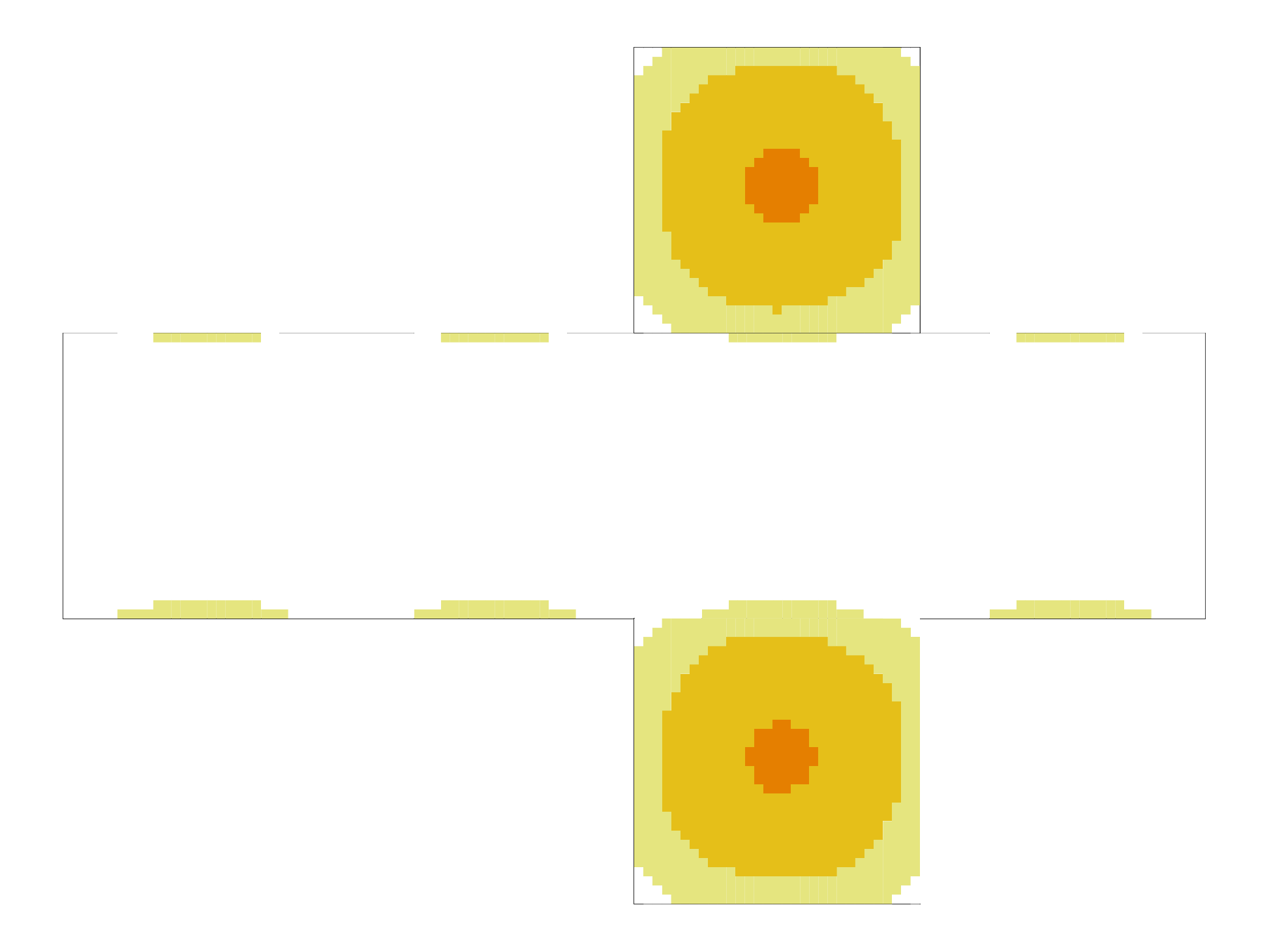}
\end{minipage}\hfill
\begin{minipage}{0.24\textwidth}
\includegraphics[width=\linewidth]{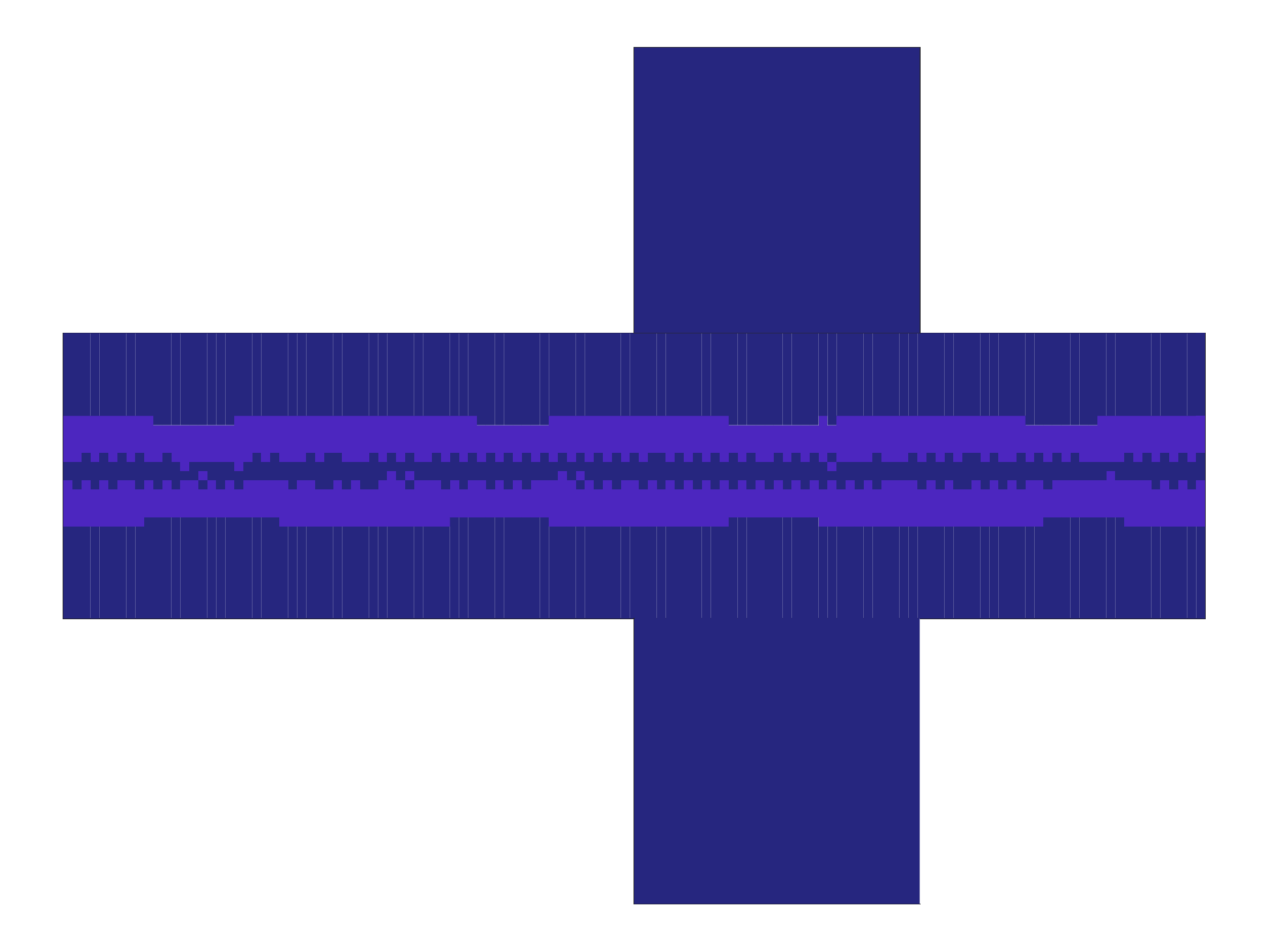}
\includegraphics[width=\linewidth]{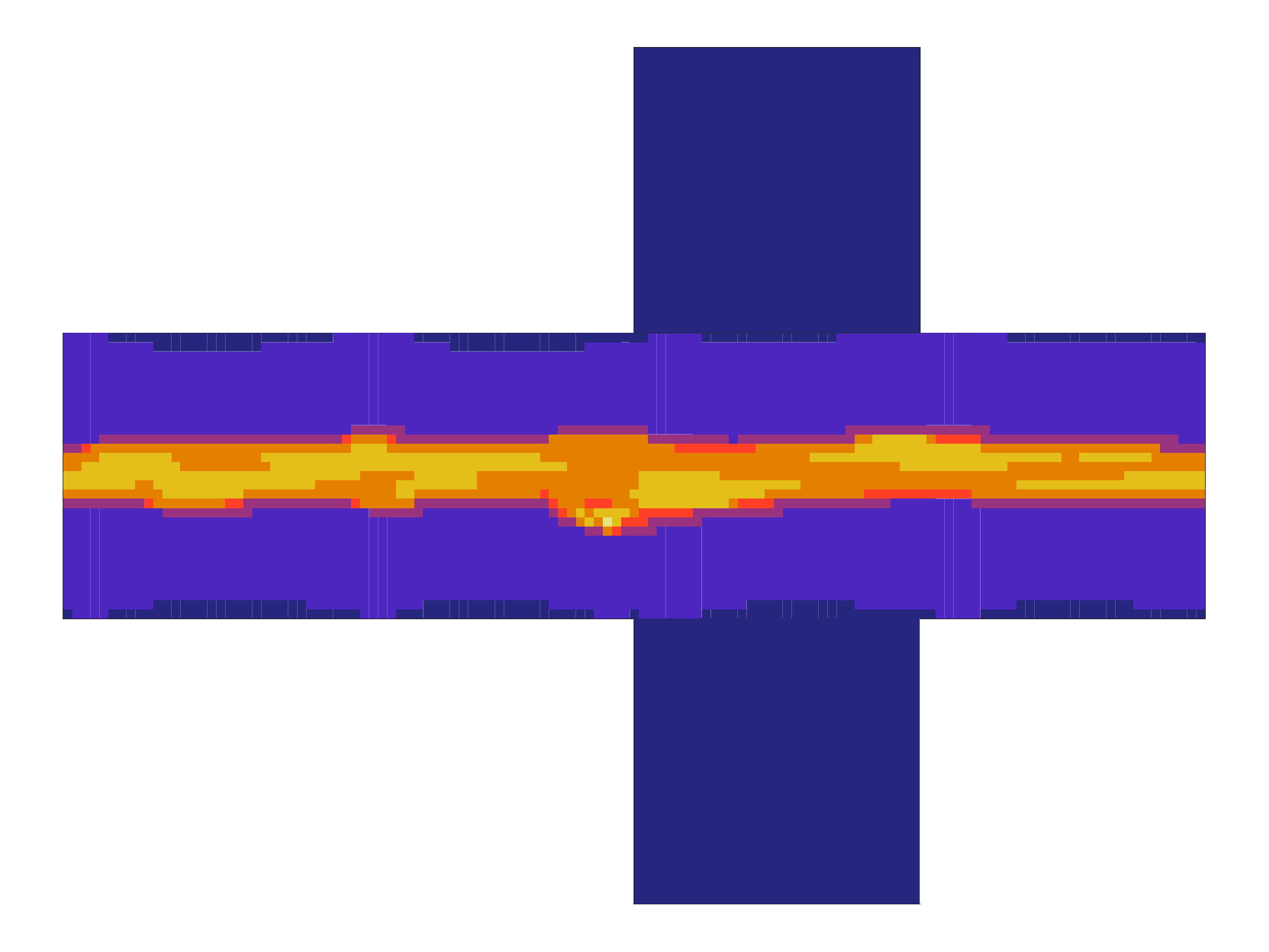} 
\includegraphics[width=\linewidth]{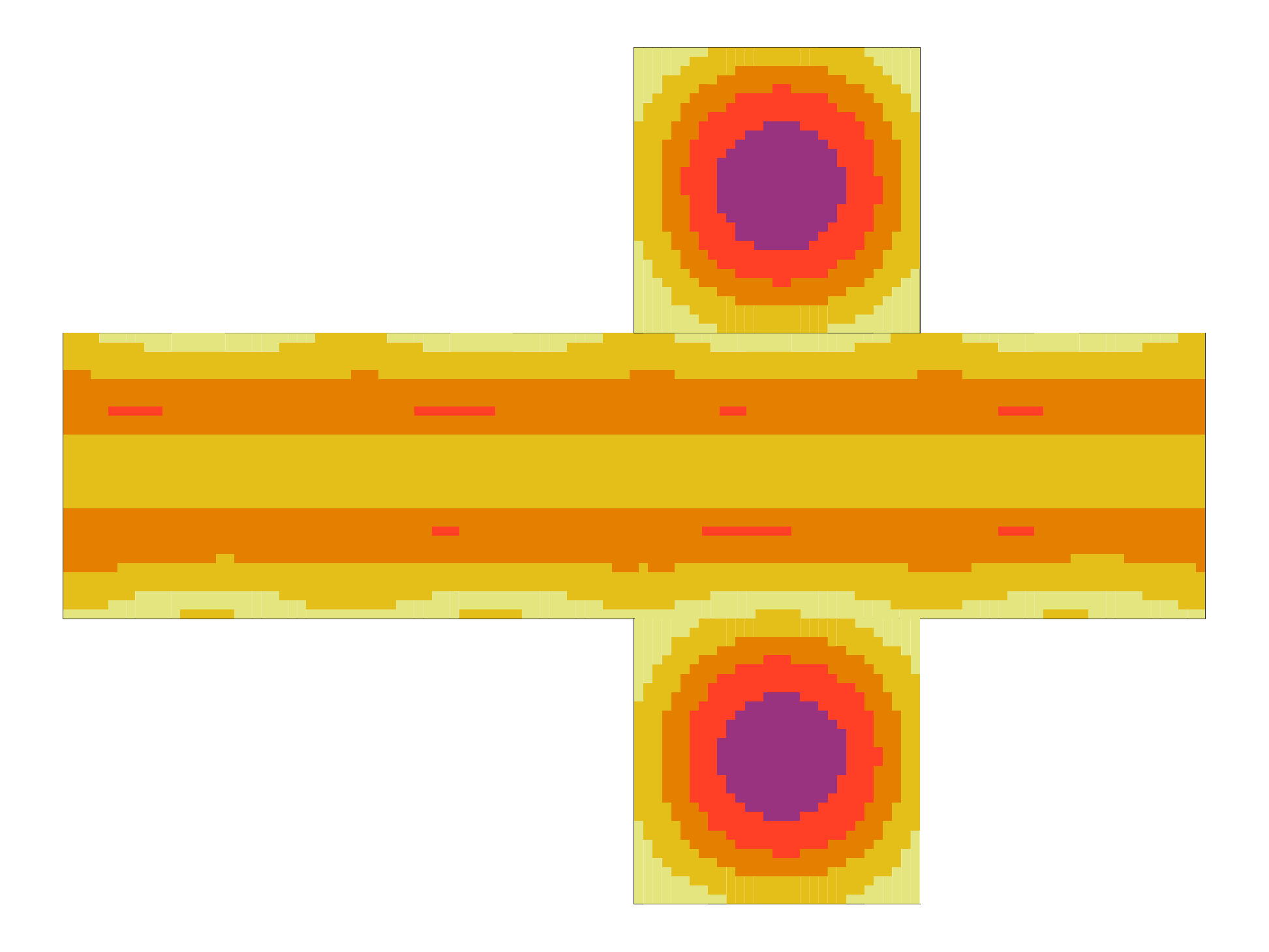}
\includegraphics[width=\linewidth]{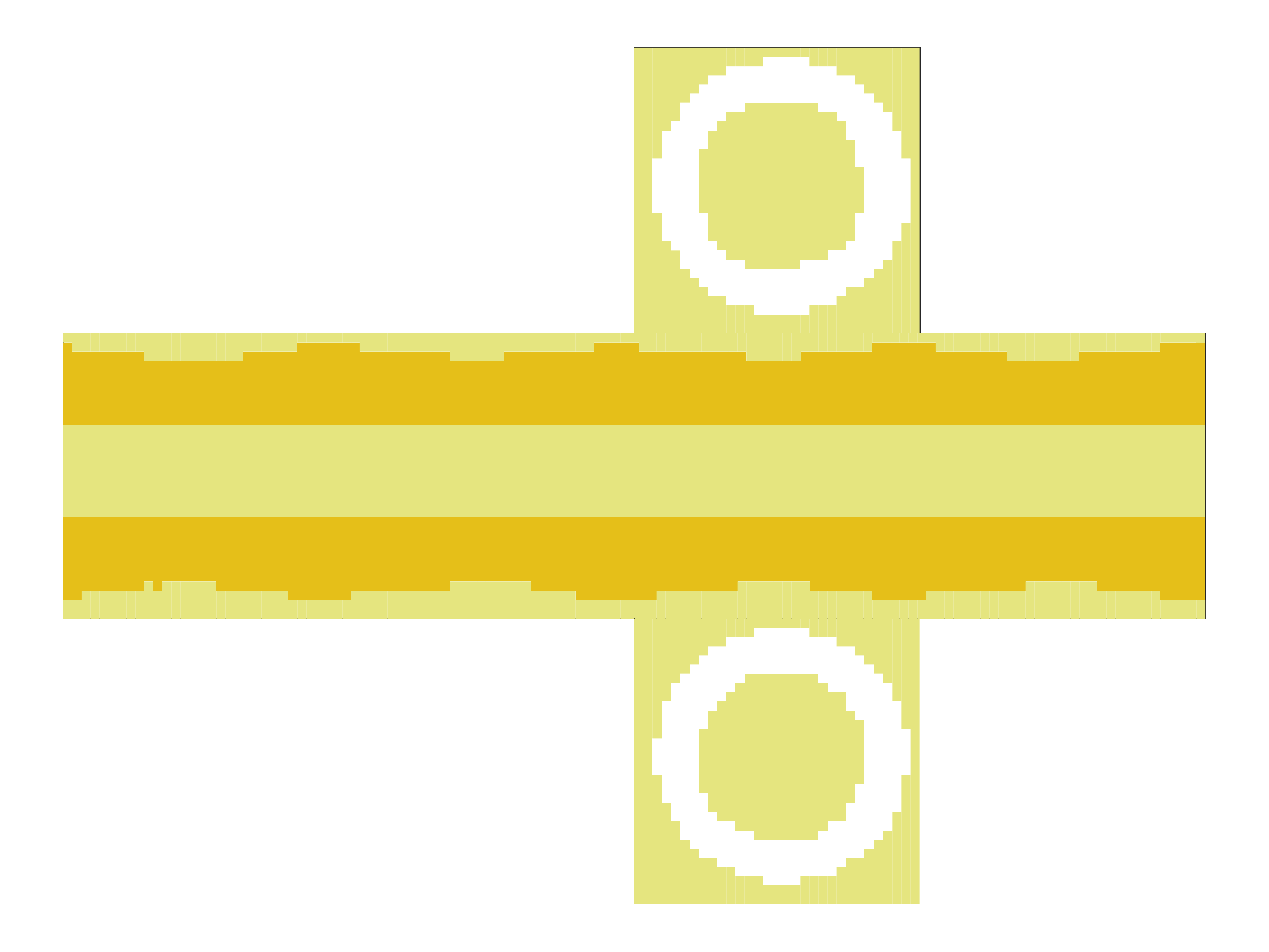}
\includegraphics[width=\linewidth]{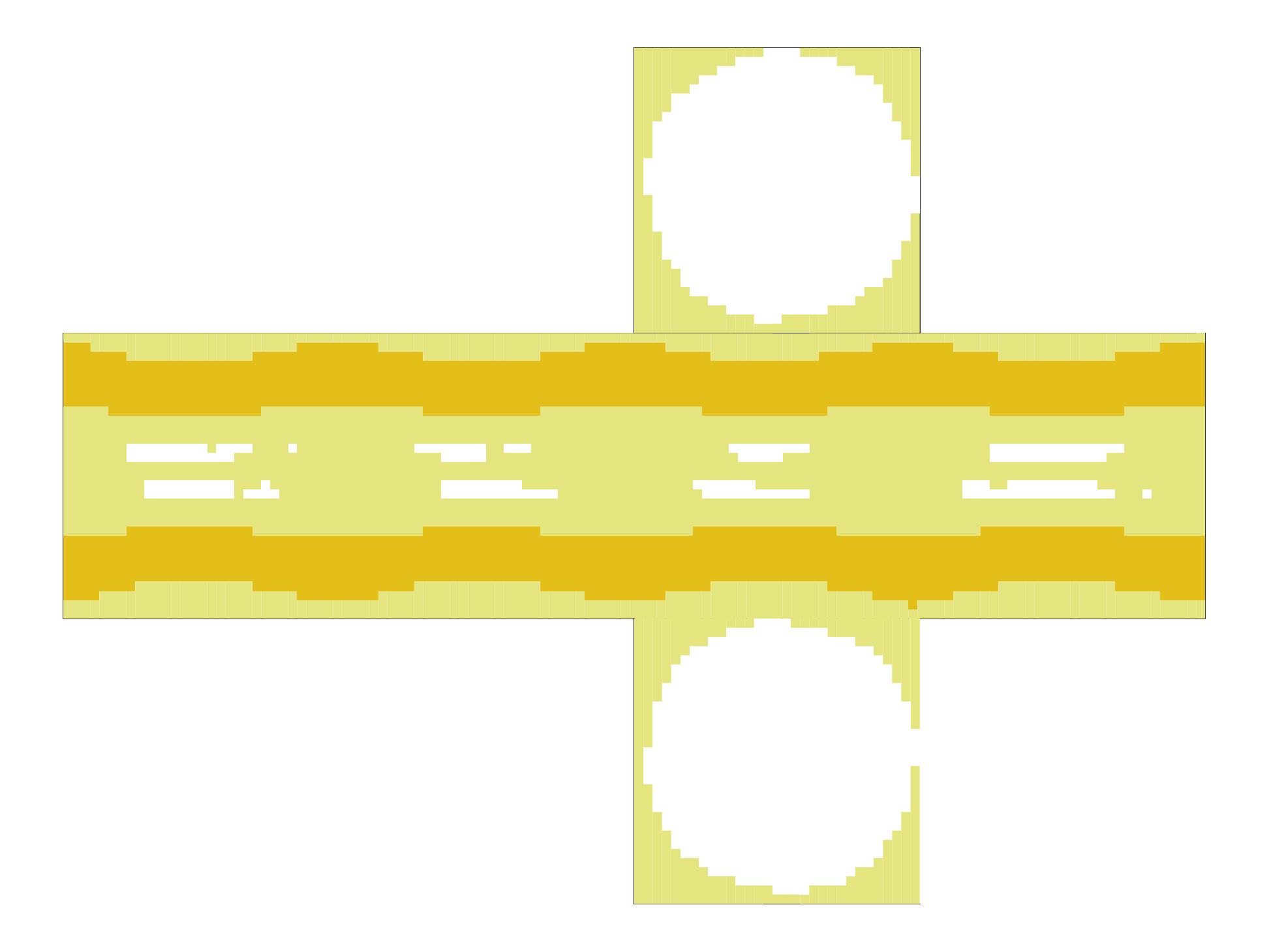}
\end{minipage}\hfill
\begin{minipage}{0.24\textwidth}
\includegraphics[width=\linewidth]{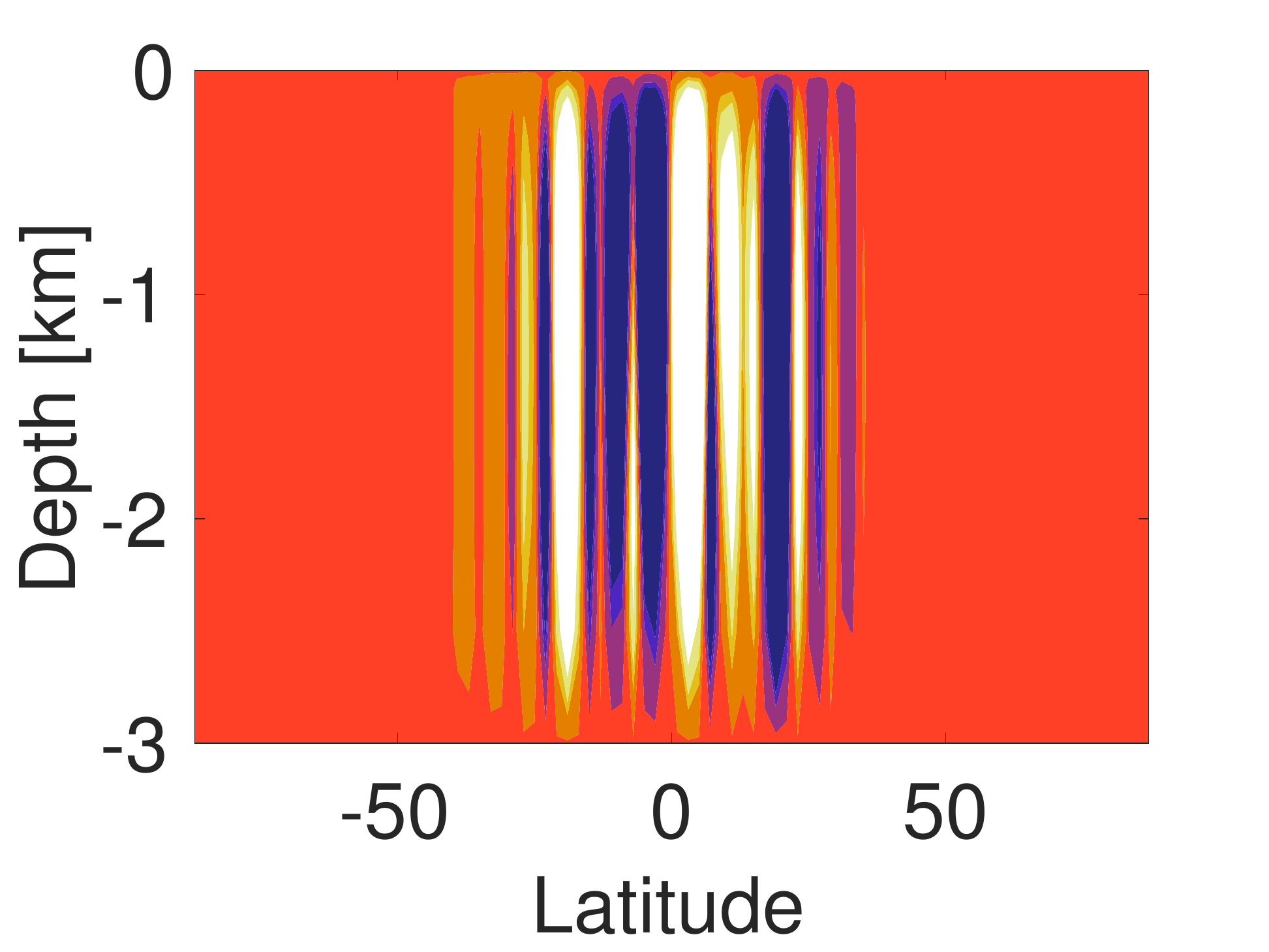}
\includegraphics[width=\linewidth]{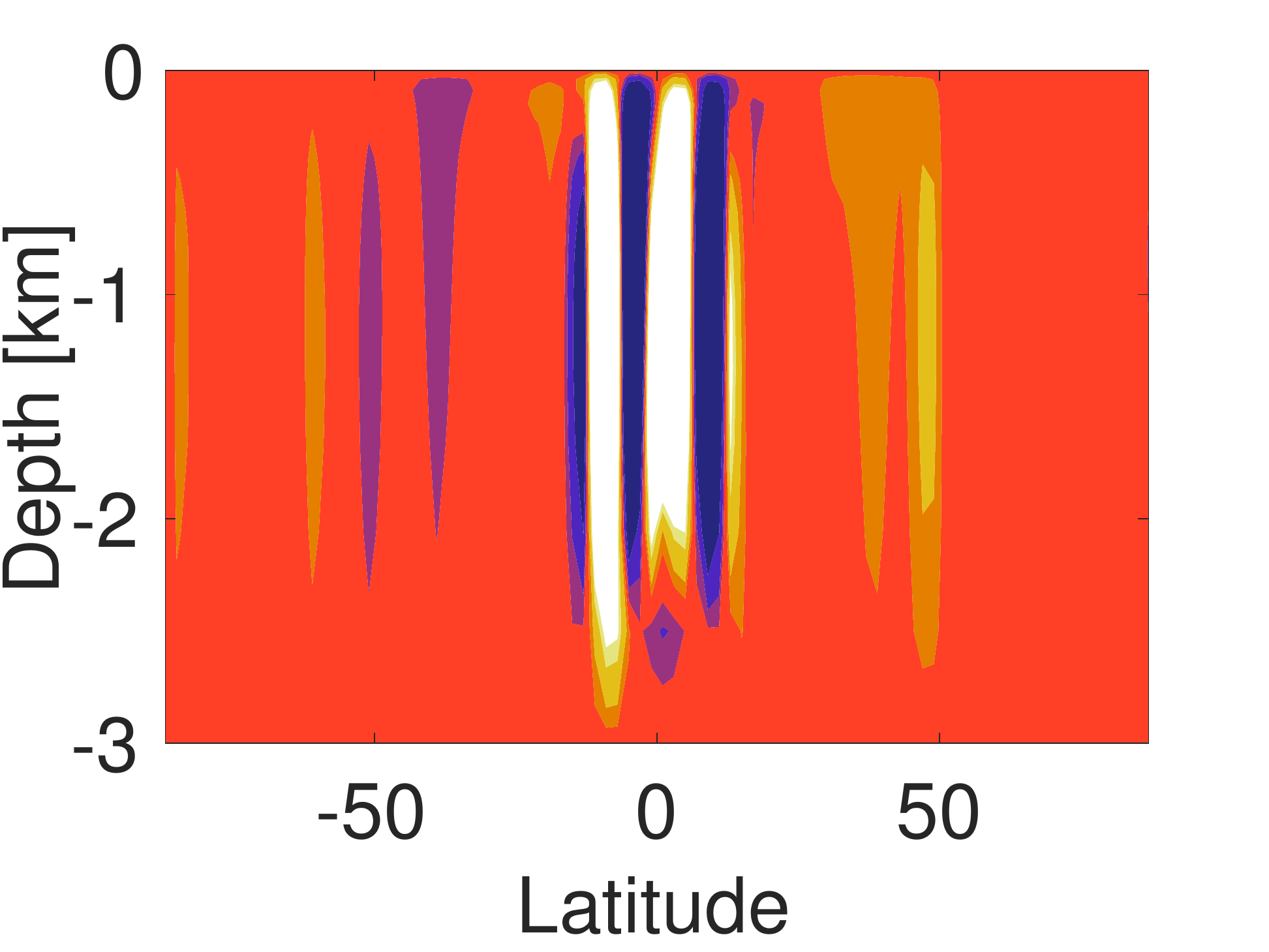} 
\includegraphics[width=\linewidth]{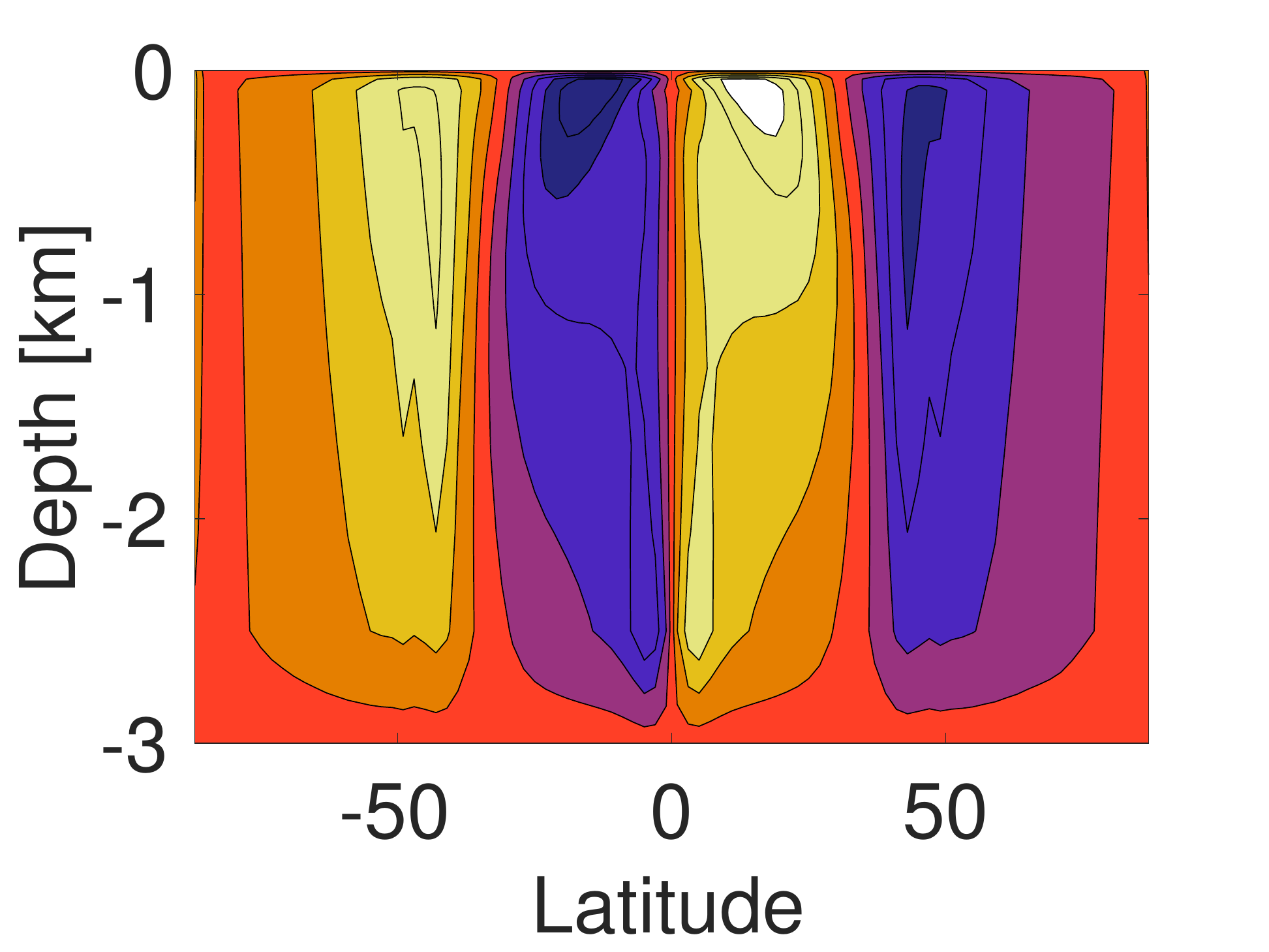}
\includegraphics[width=\linewidth]{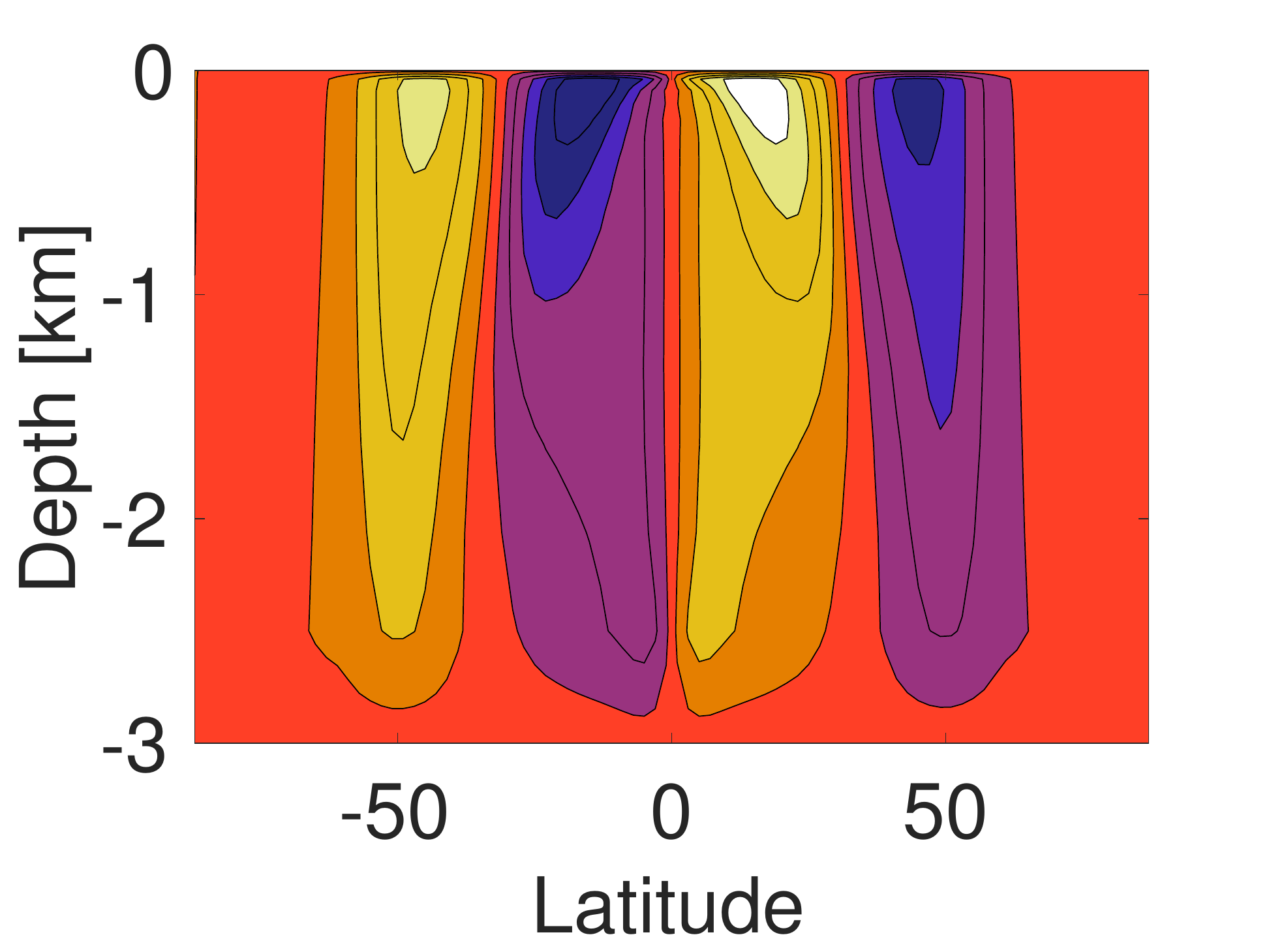}
\includegraphics[width=\linewidth]{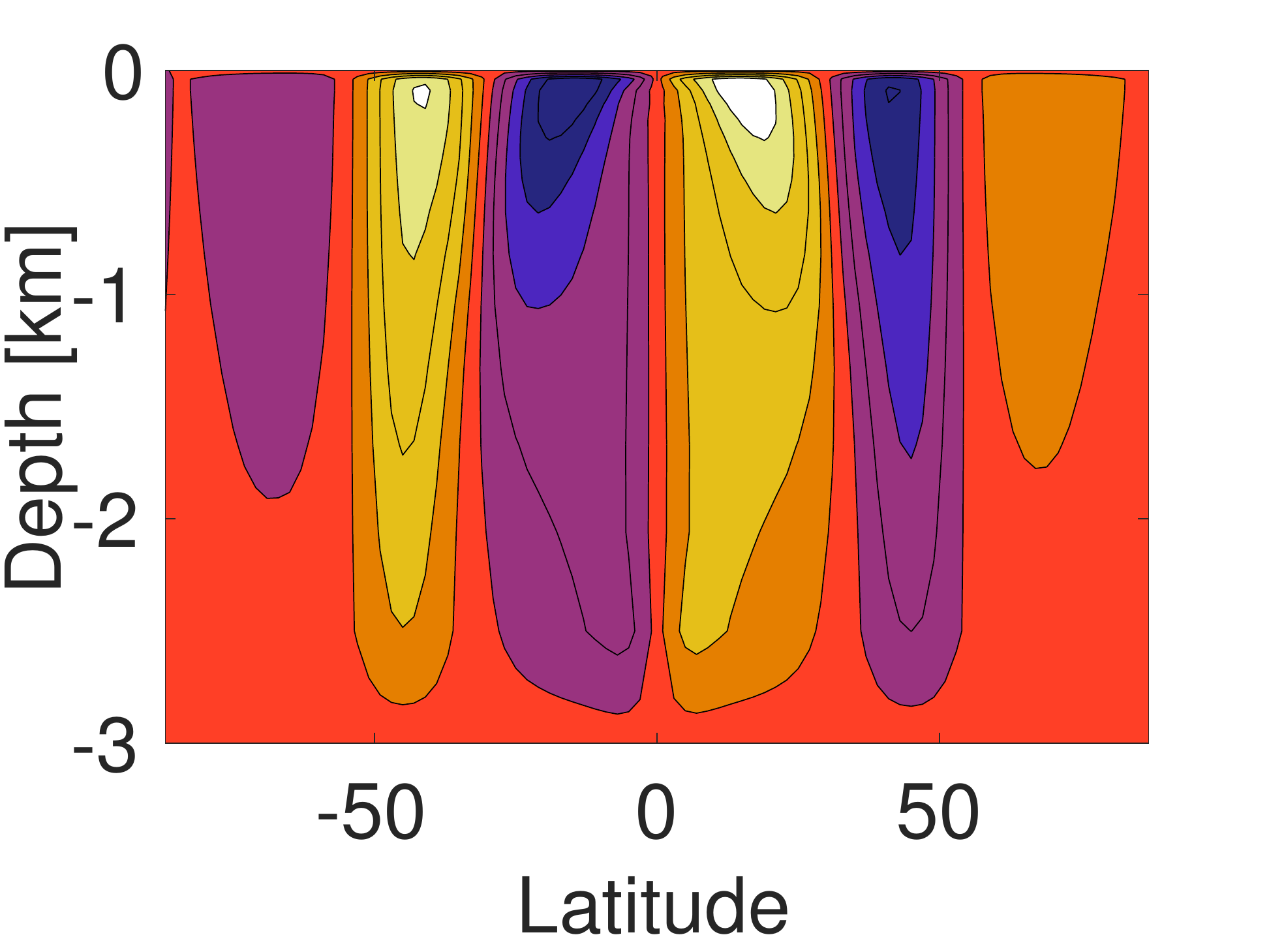}
\end{minipage}\hfill
\begin{minipage}{0.24\textwidth}
\includegraphics[width=\linewidth]{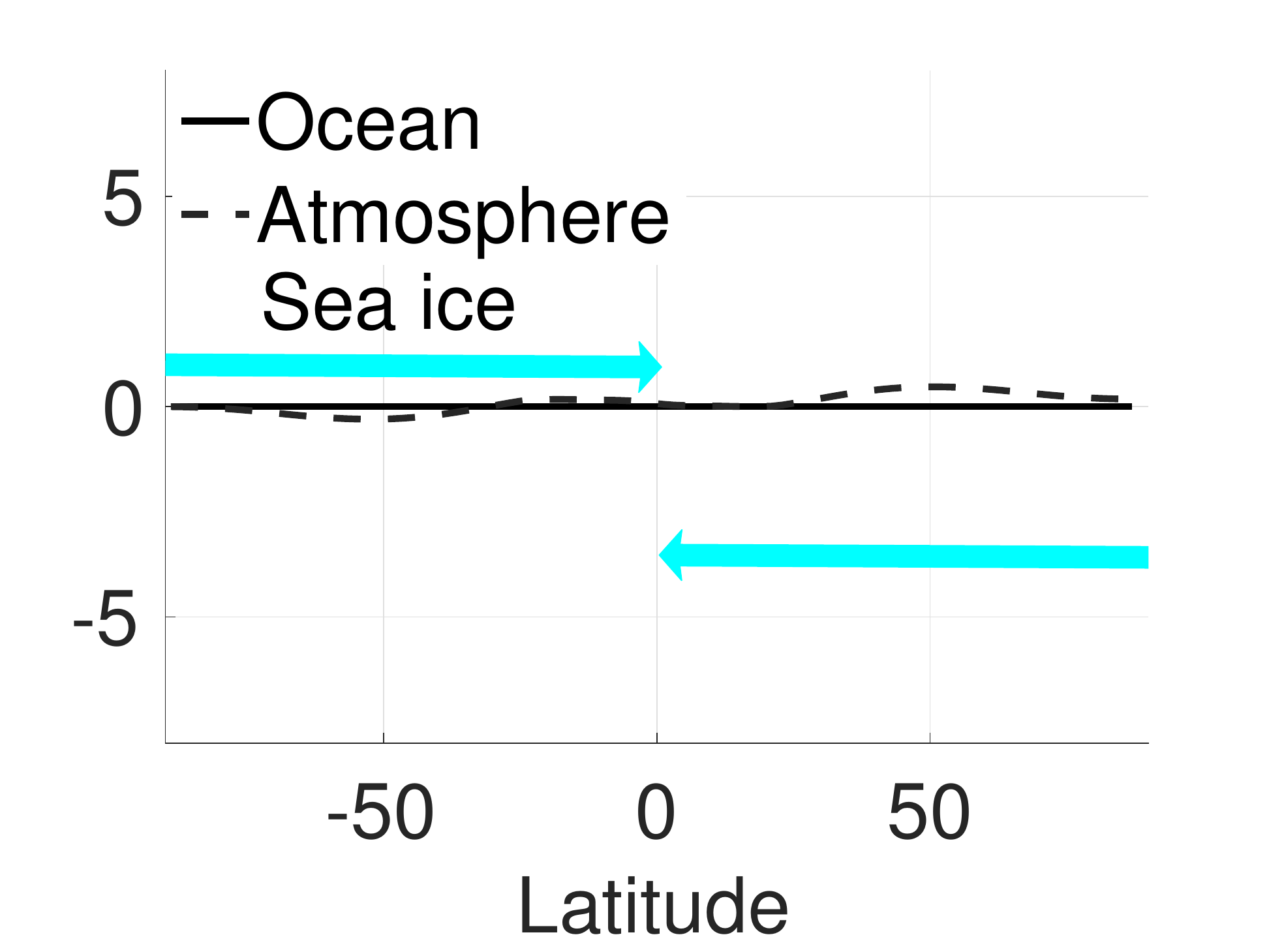}
\includegraphics[width=\linewidth]{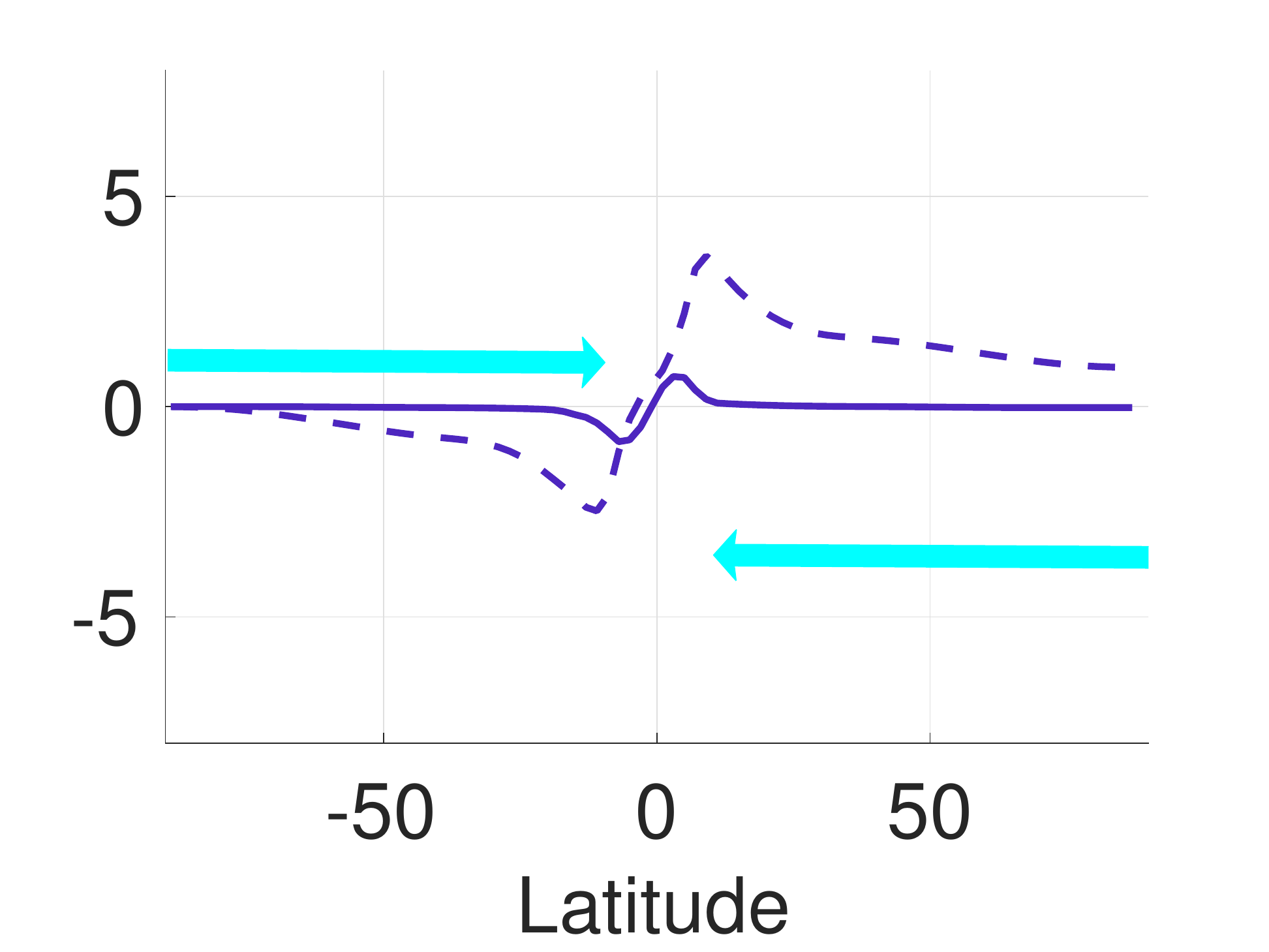}
\includegraphics[width=\linewidth]{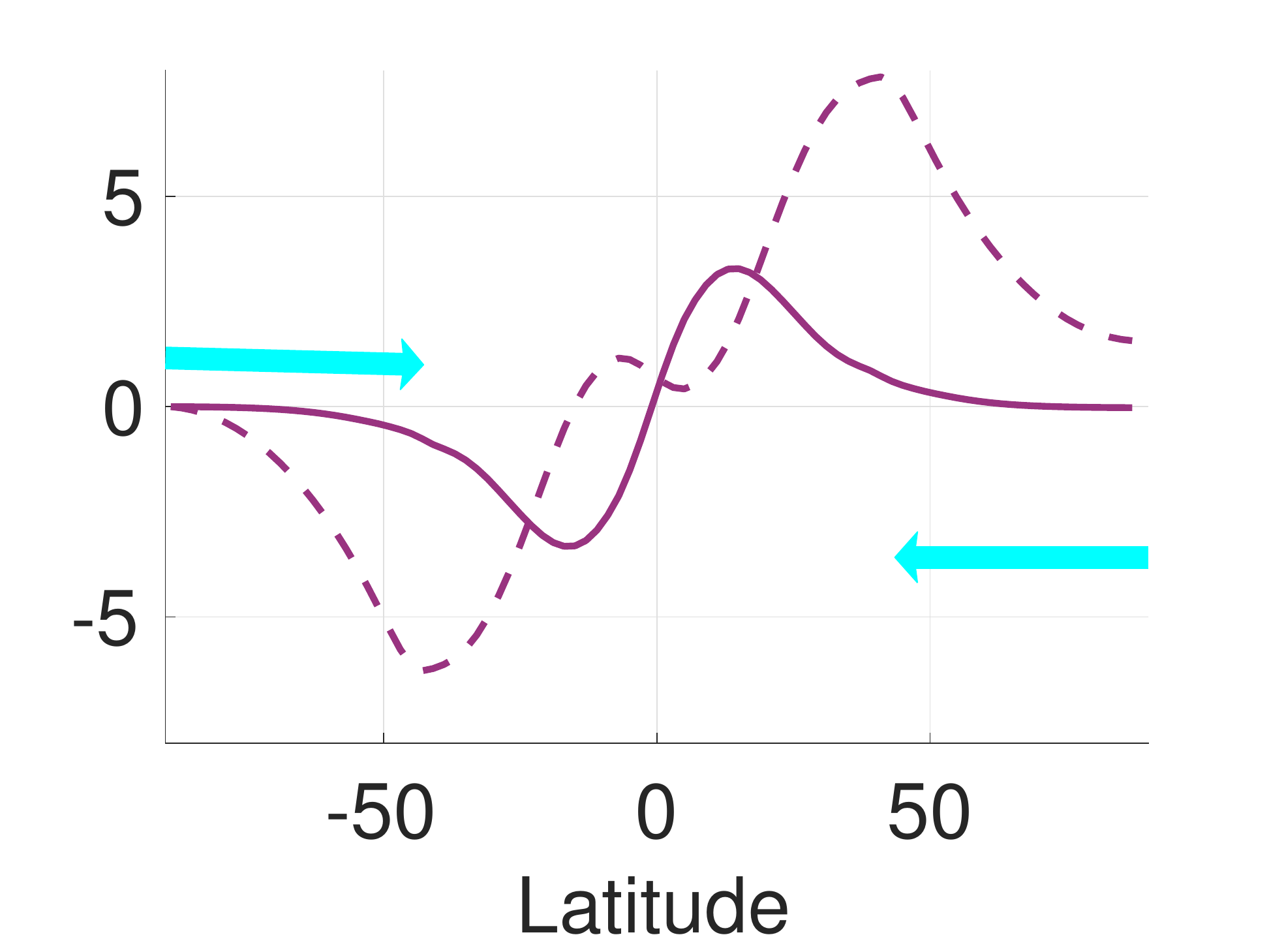}
\includegraphics[width=\linewidth]{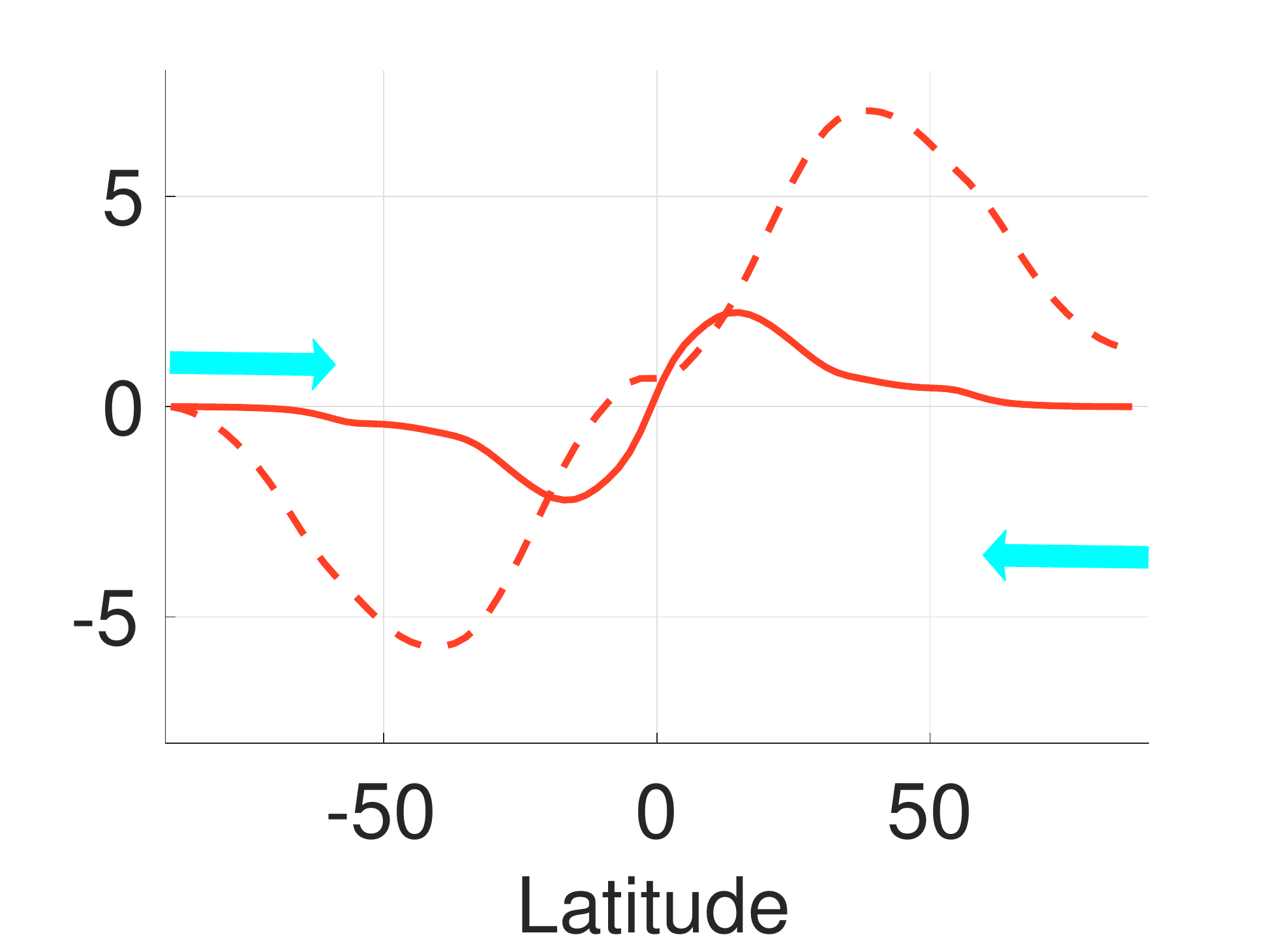}
\includegraphics[width=\linewidth]{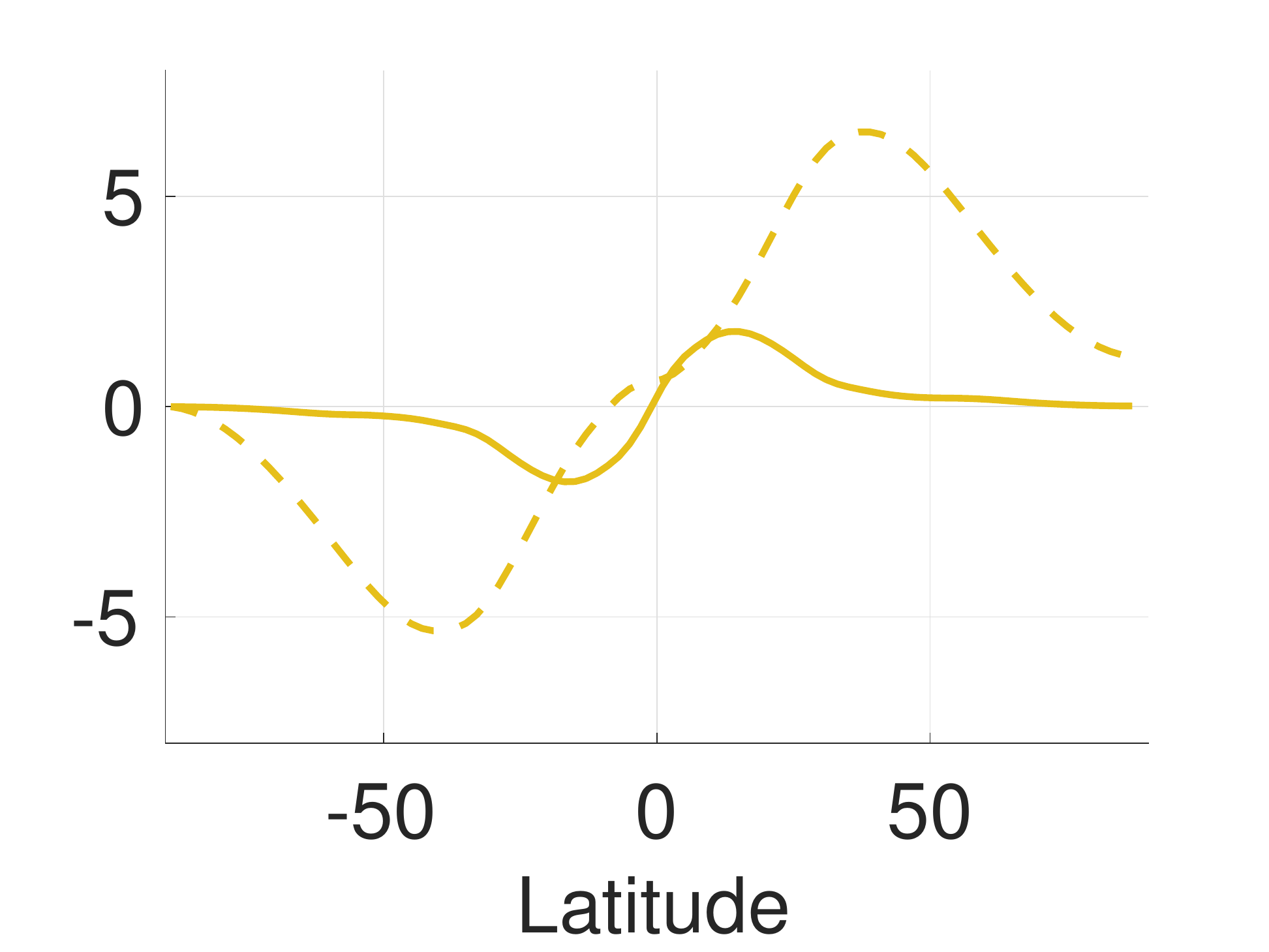}
\end{minipage}\hfill
\caption{Same as Fig.~\ref{fig:two},  for {\tt setUp2}.} 
\label{fig:three}
\end{figure}

We compare the distribution of SAT (Figs.~\ref{fig:two}-\ref{fig:three}, first column), cloud fraction (second column), ocean overturning circulation (third column), heat transport and sea ice extent (fourth column) for each attractor in {\tt setUp1} (Fig.~\ref{fig:two}) and in {\tt setUp2} (Fig.~\ref{fig:three}). The components of the radiation budget at TOA, in the atmosphere and at the surface are listed in Table~\ref{tab:3}. In the snowball, where ice completely covers the ocean surface, the ice-albedo feedback dominates, since clouds are practically absent at all latitudes, the effective transmissivity of the atmosphere in the long-wave spectrum 
being very close to one, $\tau = 0.92$ (see Table~\ref{tab:3})\footnote{Note that the transmissivity is defined as the ratio between outgoing thermal radiation at TOA and upward thermal radiation at surface, see Table~3, and it is governed by the cloud feedback as well as clear-sky processes.}. 
In the waterbelt, the sea ice boundary is found at latitudes of 16$^\circ$ in {\tt setUp1} (9$^\circ$ in {\tt setUp2}), the free ocean surface absorbs more energy due to its lower albedo, and clouds form above the warmer equatorial region. In agreement with \citet{Rose2015}, the waterbelt implies that the ocean circulation is sufficiently organised so that heat can be emitted over a restricted range of latitudes where the ice is absent. Thus the waterbelt can be viewed as the result of the competition between the destabilising ice-albedo feedback over the ice-covered surface and the stabilising effects of the ocean circulation over the ice-free surface. 
In the cold state, where ice boundary reaches latitudes of $49^\circ$ in {\tt setUp1} 
(43$^\circ$ in {\tt setUp2}), clouds start to play a more relevant role, also in the polar regions, the trasmissivity reaching an average value of $\tau = 0.63-0.68$ (see Table~\ref{tab:3}). 
In the warm state, clouds are almost homogeneously distributed over all latitudes, and the competition is essentially between the 
Boltzmann and the cloud feedbacks, the transmissivity being $\tau = 0.57$, lower than the present-day transmissivity that is of the order of 0.6~\citep{2013ClDy...40.3107W}. Finally, in the hot state cloud cover is high everywhere and the transmissivity reaches a very low value of $\tau = 0.53$, meaning that within this hot climate a large fraction of long-wave radiation remains trapped at the surface in particular because of the presence of clouds.

\begin{table}
\begin{minipage}{1\linewidth}
\caption{Magnitudes of the radiation balance components (global annual mean in W/m$^2$), as defined in the schematic diagram below for the two configurations, {\tt setUp1} in black and {\tt setUp2} in gray. Averages are calculated from the last 100 yr of simulations.}
\label{tab:3}
\centering
\small
\begin{tabular}{llllll}
\hline
\\
Attractor & Snowball & Waterbelt & Cold state & Warm state & Hot state \\
{\bf Solar radiation} & & & & & \\
 Reflected at TOA   &  184.3  &  163.8  &  117.1 &   114.3 &   ---  \\
      & {\color{gray} 183.9} & {\color{gray} 171.9} & {\color{gray}  117.7} & {\color{gray}  109.8} &  {\color{gray}  105.3} \\
 Absorbed by atmosphere   &  112.0  &  110.5  &  101.5 &   97.2 &   ---  \\
      & {\color{gray} 112.3} & {\color{gray} 112.2} & {\color{gray}  104.8} & {\color{gray}  95.3} &  {\color{gray}  86.9} \\
Absorbed by surface   &  45.7  &  67.7  &  123.4 &   130.5 &   ---  \\
      & {\color{gray} 45.8} & {\color{gray} 57.9} & {\color{gray}  119.5} & {\color{gray}  136.9} &  {\color{gray}  149.8} \\
 Reflected at surface   &  70.7  &  63.9  &  36.6 &   26.3 &   ---  \\
      & {\color{gray} 71.0} & {\color{gray} 68.0} & {\color{gray}  44.4} & {\color{gray}  23.9} &  {\color{gray}  11.3} \\
 {\bf Thermal radiation}  & & & & &  \\    
 Outgoing at TOA   &  158.2  &  179.3  &  225.2 &   228.0 &   ---  \\
      & {\color{gray} 157.8} & {\color{gray} 168.5} & {\color{gray}  221.3} & {\color{gray}  229.6} &  {\color{gray}  234.4} \\
Up surface   &  171.9  &  209.2  &  356.1 &   397.5 &   ---  \\
      & {\color{gray} 171.2} & {\color{gray} 186.2} & {\color{gray}  326.2} & {\color{gray}  402.0} &  {\color{gray}  438.6} \\
Down surface   &  87.2  &  134.4  &  307.6 &   352.3 &   ---  \\
      & {\color{gray} 86.5} & {\color{gray} 106.1} & {\color{gray}  273.8} & {\color{gray}  354.6} &  {\color{gray}  392.2} \\
 $\tau = {\rm{Thermal~outgoing~TOA}}/{\rm{Thermal~up~surface}}$  &  0.92  &  0.86  &  0.63 &   0.57 &   ---  \\
      & {\color{gray} 0.92} & {\color{gray} 0.90} & {\color{gray}  0.68} & {\color{gray}  0.57} &  {\color{gray}  0.53} \\
      \\
{\bf Sensible heat}   &  17.7  &  17.5  &  15.3 &   11.1 &   ---  \\
      & {\color{gray} 18.1} & {\color{gray} 18.0} & {\color{gray}  18.6} & {\color{gray}  12.3} &  {\color{gray}  9.3} \\
{\bf Latent heat}  &  14.0  &  39.3  &  96.2 &   100.5 &   ---  \\
      & {\color{gray} 14.0} & {\color{gray} 27.8} & {\color{gray}  92.9} & {\color{gray}  101.1} &  {\color{gray}  105.4} \\
 \hline
\end{tabular}
\end{minipage}\hfill
	\begin{minipage}{\linewidth}
\includegraphics[width=0.95\linewidth]{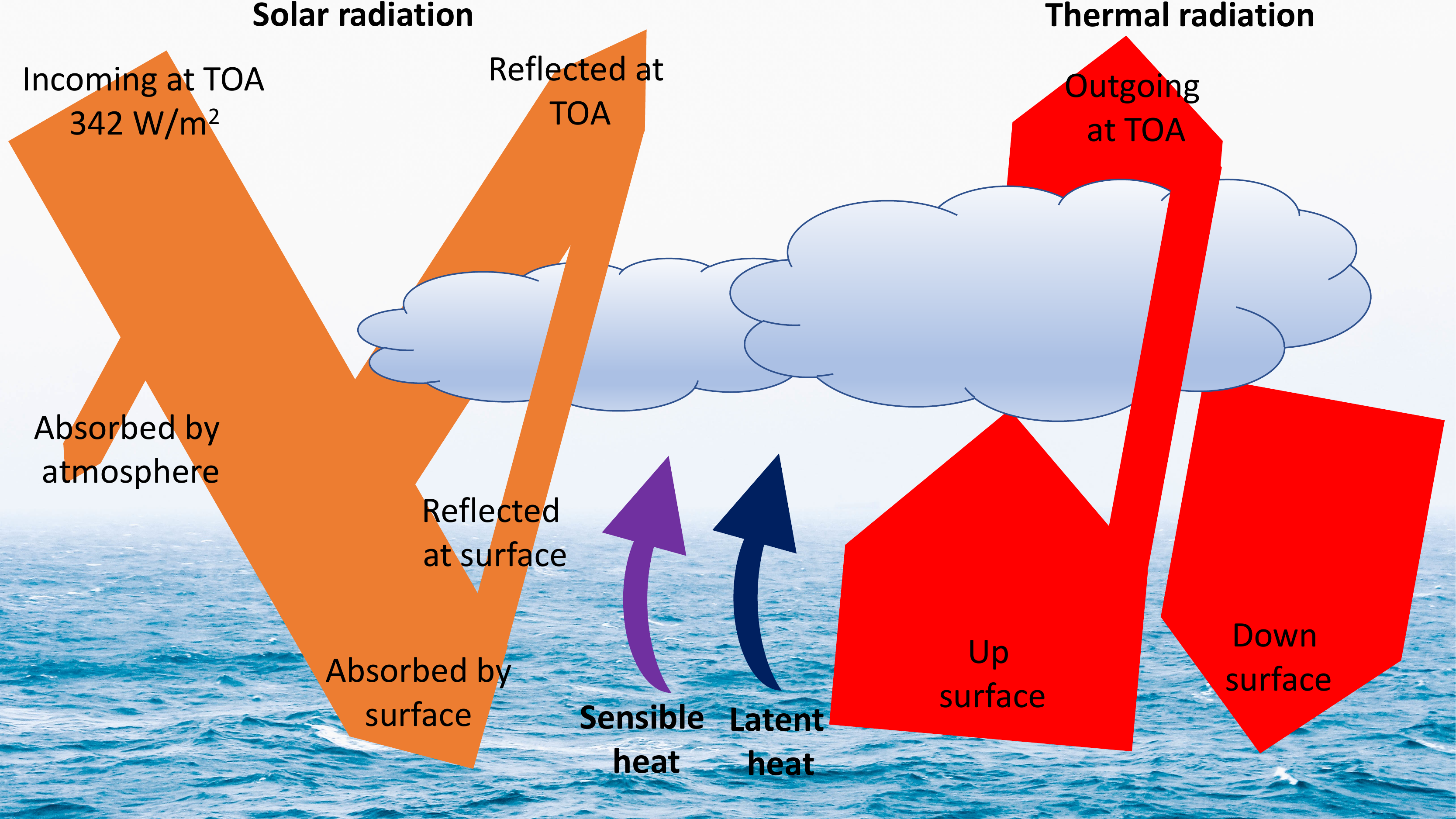}  
\end{minipage}
\end{table}

The ocean overturning circulation, shown in the third column of 
Figs.~\ref{fig:two}-\ref{fig:three}, 
is overall very weak in the snowball, except near the Equator, where a series of small and extremely intense cells develop. Since these cells are disconnected in latitude, the associated meridional heat transport turns out to be very weak (see first panel along the fourth column in Figs.~\ref{fig:two}-\ref{fig:three}). 
As the global SAT increases, the circulation within the ocean is organised in larger overturning cells, anti-symmetrical with respect to the Equator. 
Positive values correspond to clockwise circulation in Sv 
(1~Sv $=10^6$~m$^3$~s$^{-1}$). Even if these overturning cells are still quite irregular in the waterbelt (much less in {\tt setUp1} than in {\tt setUp2}), 
the ocean circulation turns out to be responsable of heat transport over a restricted
range of latitudes where the ice is absent (see last column in Figs.~\ref{fig:two}-\ref{fig:three}), in qualitative agreement with~\citet{Rose2015}.
The overturning cells reach high intensity and extent in cold and warm states. In particular, the polar cells become slightly deeper and stronger in the cold state than in the warm one. In the hot state that occurs in {\tt setUp2}, additional overturning cells 
can develop in the polar regions as they become ice-free (see last panel in the third column of Fig.~\ref{fig:three}).  

The poleward heat transport is proportional to the strength of the circulation multiplied by the temperature gradient at the concerned latitudes, and it is dominated by the surface circulation~\citep{2005GeoRL..3210603B}. Thus we see from the panels in the fourth column of 
Figs.~\ref{fig:two}-\ref{fig:three} that the ocean heat transport is negligible in the snowball, 
where the overturning cells have small meridional extent and temperatures are  
quite homogeneous, while it reaches several petawatts (PW) when the surface circulation is well organised through the appearance of overturning cells that connect latitudes with large meridional temperature gradient. In particular, we observe that the peak of the ocean heat transport is of 3~PW  in the cold state, larger than the corresponding value in the warm state. This can be explained by the larger meridional temperature gradient in the cold state, 
the intensity of the overturning cells in the tropical regions being very similar in the two attractors.  The robust feature observed in~\citet{2011JCli...24..992F} for the existence of the cold state, namely a meridional structure of the ocean heat transport with a shoulder at mid-latitudes, is also present in our cold attractor, as can be seen from the behaviour of the ocean heat transport near the sea ice 
boundary (see Figs.~\ref{fig:two}-\ref{fig:three}, fourth column, arrow and solid line in the panel corresponding to the cold state).  In the warm state, such shoulder appears at higher latitudes, while in the hot state in {\tt setUp2} it disappears and heat can reach the North/South pole with a complete melting of sea ice. 

\begin{figure}[ht!]
\includegraphics[width=0.5\linewidth]{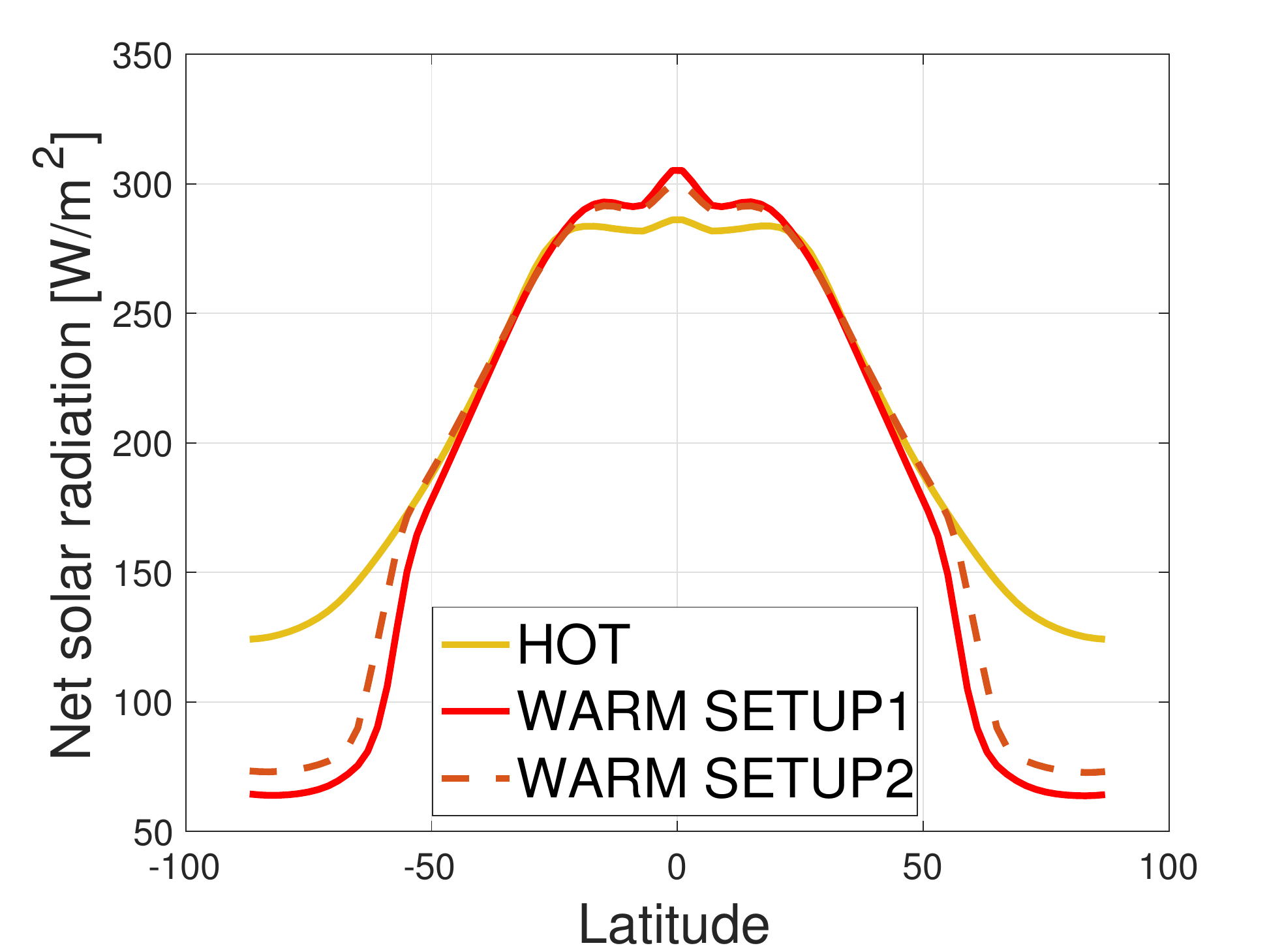} 
\includegraphics[width=0.5\linewidth]{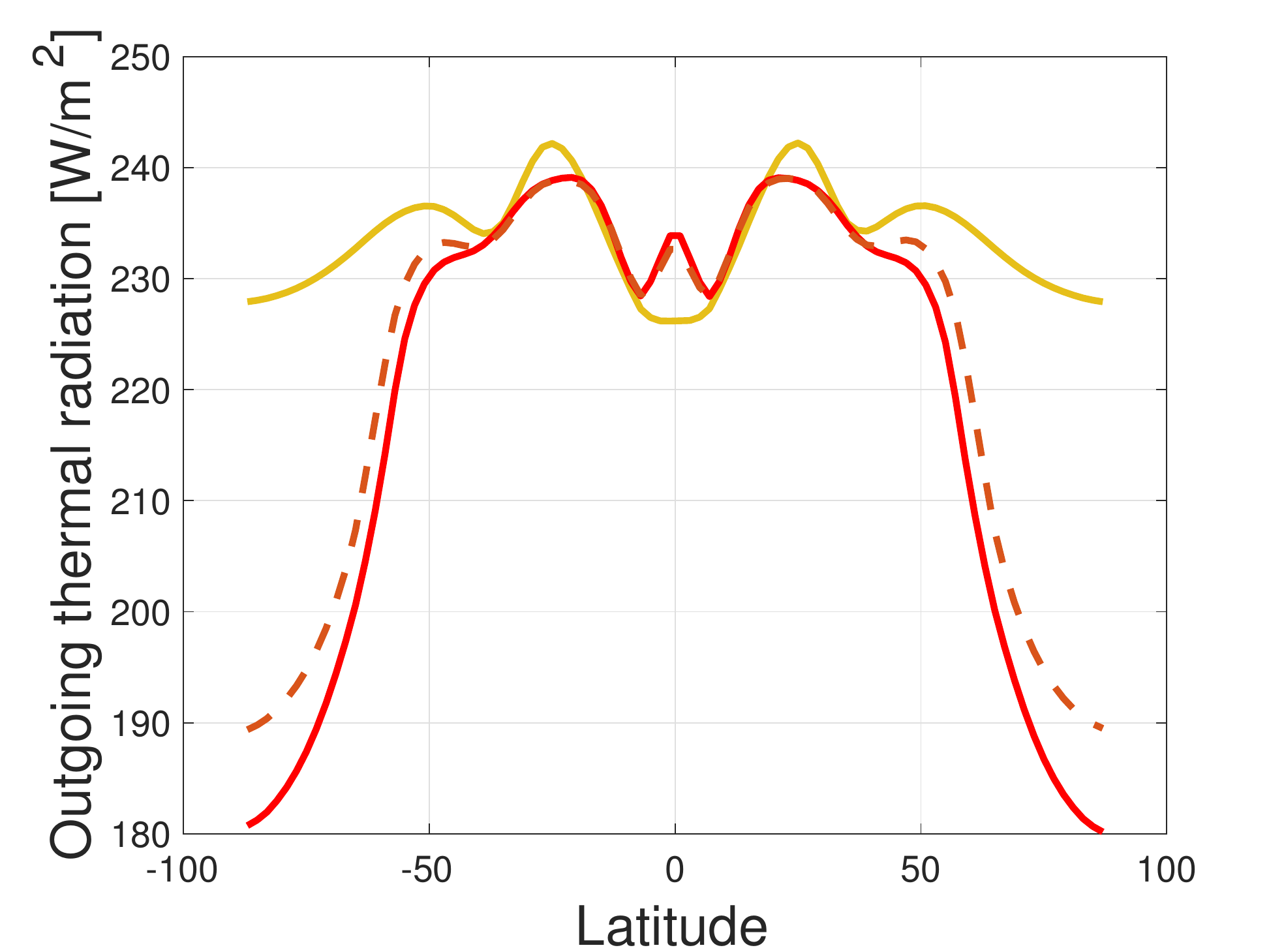} 
\caption{Zonal averages of net solar radiation ({\it left}) and outgoing thermal radiation ({\it right}) at TOA for warm and hot states in both 
set-ups. }
\label{fig:4}
\end{figure}

We have verified that the hot state is still present if we turn on the re-injection of dissipated kinetic energy, while keeping the cloud albedo constant in latitude. This shows that the appearance of the hot state  depends on the cloud albedo parameterisation, in particular on cloud reflection properties~\citep{2012JCli...25.3832D} and on the amount of solar radiation that is allowed to enter at high latitudes. In Fig.~\ref{fig:4}  the net solar radiation and the outgoing long-wave radiation at TOA are shown as a function of latitude for warm and hot states in both set-ups. 
We can see that the radiation at TOA for the hot state agrees with the one shown in \citet{2011JCli...24..992F}. 
The zonal averages of both TOA thermal and solar radiation for the warm state differ at high latitudes in the two set-ups, showing that the main difference between the two considered parameterisations of cloud albedo occurs in the polar regions. We cannot exclude however that changing the solar constant or CO$_2$ content would allow for the presence of the hot state also in {\tt setUp1}.  
     
We note that the atmospheric heat transport is very similar in cold, warm and hot states. Therefore, it 
does not play a relevant role in distinguishing 
between these attractors (see Figs.~\ref{fig:two}-\ref{fig:three}, fourth column, dashed line), in agreement 
with~\citet{2005GeoRL..32.2705C,2013Icar..226.1724B} where it is shown that for sufficiently large mean global temperature and small meridional gradient the atmospheric heat transport stays constant. 
Values of atmospheric heat transport not returning to zero in the north polar region are the signature of a remaining energy imbalance at TOA, which we have not corrected to display the limitations within the considered set-up of the numerical model~\citep{2018ClDy...50.4425B}. 
The TOA imbalance is larger in {\tt setUp2} than in {\tt setUp1} (compare also values in Table~\ref{tab:2}), confirming that the former is indeed a less accurate configuration for MITgcm.

\subsection{Complexity assessment of time series}

\begin{figure}[ht!]
\includegraphics[width=0.5\linewidth]{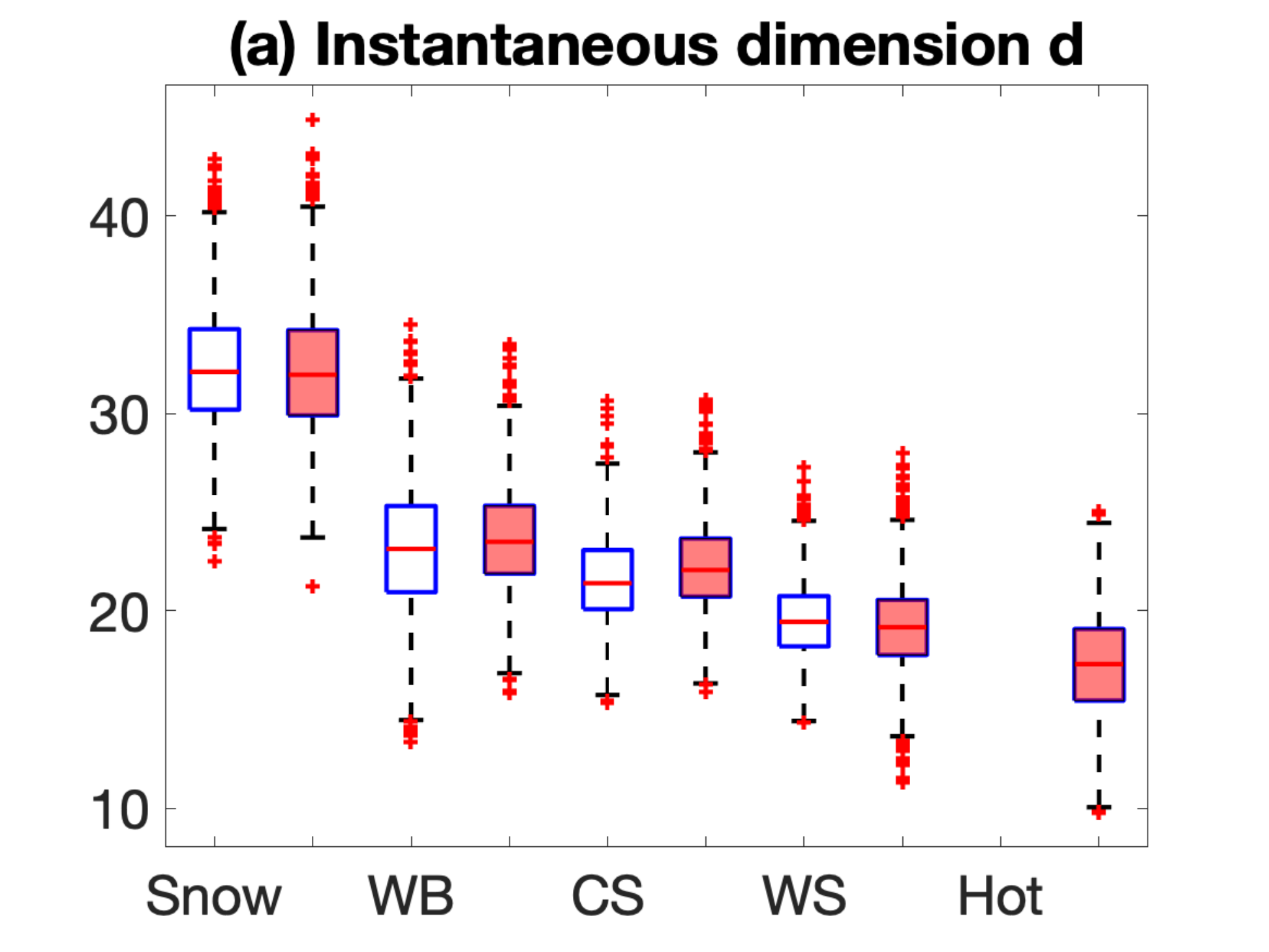} 
\includegraphics[width=0.5\linewidth]{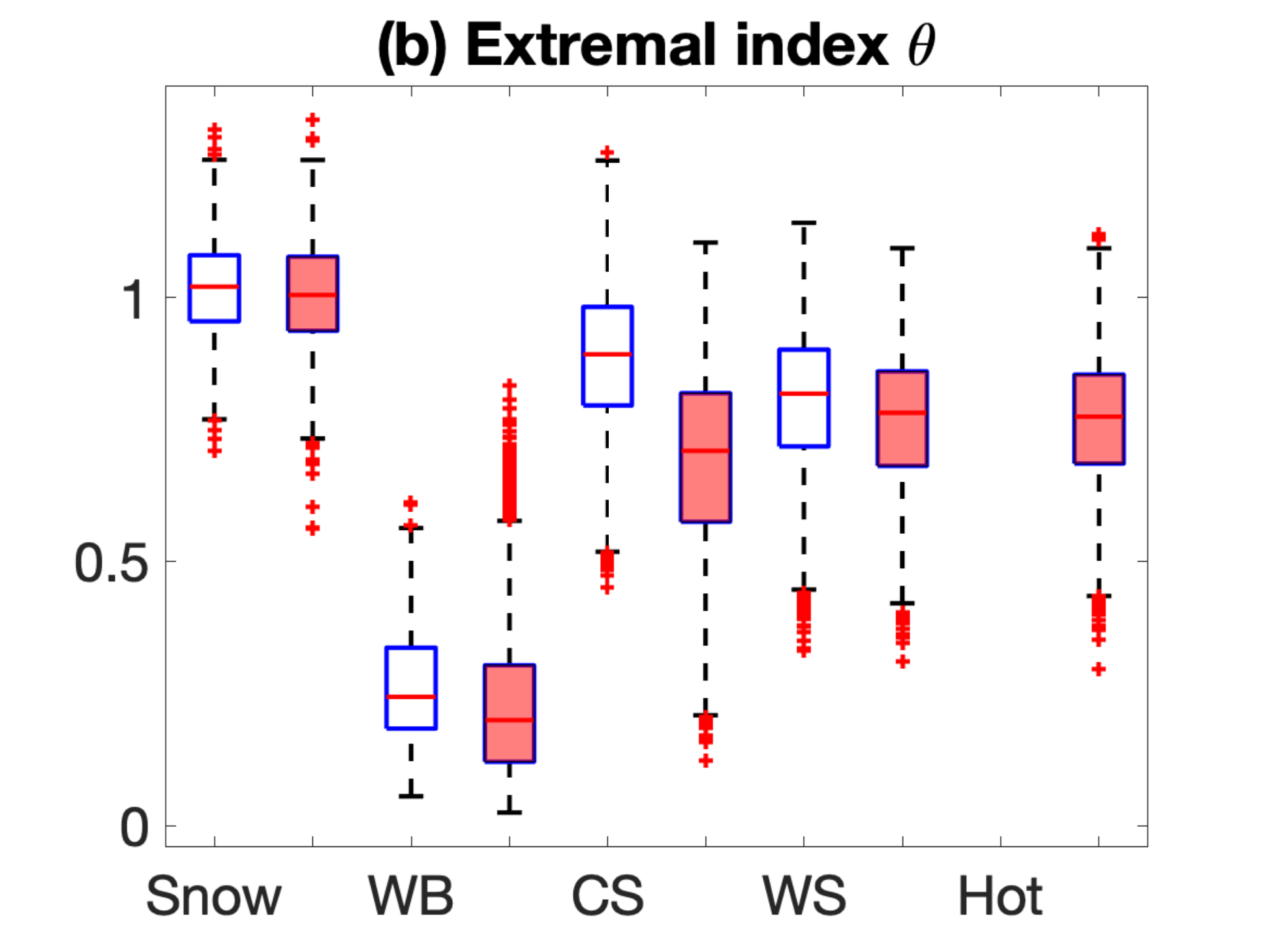}
\caption{Dynamical systems analysis of the attractors. Box plots of the distribution of {\it (a)} instantaneous dimension $d$ and {\it (b)} extremal index $\theta$, for the  attractors corresponding to 
{\tt setUp1} (blue boxes) and {\tt setUp2} (filled red boxes). On each box, the central mark is the median,
the edges of the box indicate the 25th and 75th percentiles, the whiskers extend to 1.5 times the interquartile range, and outliers are plotted individually using the `$+$' symbol.  The attractors are denoted as Snow, WB, WS, CS, Hot, respectively for snowball, waterbelt, warm state, cold state, hot state.
}
\label{fig:five}
\end{figure}

The attractors can be characterised by quantities widely used in the study of dynamical systems, namely the instantaneous dimension and the persistence~\citep{lucarini2016extremes,2017NatSR...741278F,2017NPGeo..24..713F}. 
These quantities have been calculated from $32\times 32\times 6 =6144$ yearly averaged SAT time series, {\it i.~e.}, for each point on the horizontal cubed-sphere grid (see Fig.~\ref{fig:zero}). 
The time series length is at least 1500 yr for each attractor. 
In Fig.~\ref{fig:five}$(a)$-$(b)$,  we show the resulting box plots for all the attractors 
and both set-ups. The instantaneous dimension $D = \overline{d(\zeta)}$, that represents the minimal number of degrees of freedom able to describe the system, decreases from a value around 32 in both set-ups for the snowball, to a value around 18 for the 
hot state (Fig.~\ref{fig:five}$(a)$). The extremal index $\theta =[0,1]$ is related to the inverse of trajectories persistence: 
low (respectively high) values of $\theta$ correspond to long (short) persistence of trajectories near a given point on the attractor.  $\theta$ is in general near 0.8 in all the attractors, except in waterbelt where it is of the order of 0.2, and in snowball where it is close to 1 (see  
Fig.~\ref{fig:five}$(b)$).

In order to understand the meaning of these dynamical measures, it is helpful to remember what happens in the Lorenz system~\citep{1963JAtS...20..130L}. 
In such prototypical system the maxima of the instantaneous dimensions are found between the two wings, where the trajectories diverge the most and the instability is stronger (see~\citet{2017NatSR...741278F} and Fig.~A.1 in the Supplementary Information of that paper). The fact that stronger instabilities are associated to higher instantaneous dimensions is also supported by~\citet{2016QJRMS.142.2143S,2017NPGeo..24..713F,2017NatSR...741278F} where it is shown that higher meridional temperature gradients, associated to blocking events, give rise to higher values of Liapunov exponents and of instantaneous dimension, respectively.  

The waterbelt state is characterised by strong meridional gradients localised at the ice edges, as can be seen in Figs.~\ref{fig:two}-\ref{fig:three}.
This can explain the higher instantaneous dimension of this attractor with respect to the cold, warm and hot ones. 
On the contrary, since the snowball  SAT is very uniform, as can be seen from Figs.~\ref{fig:two}-\ref{fig:three}, one would expect a low value of the instantaneous dimension. However, the fact that the extremal index $\theta$ is close to 1 in snowball means that clustering of exceedances is not occurring in snowball time series and that we are measuring irrelevant dynamical features that correspond to independently and 
identically distributed random fluctuations~\citep{2019Chaos..29b2101M} giving a high value of the instantaneous dimension. 
It is important to recall that the value of the dynamical estimator does not correspond to the true dimensionality of the system, since the convergence is very low, as discussed for example in~\citet{Galfi2017}. Rather, such estimators give useful informations when compared between each other in different attractors and provide an efficient way to obtain information on time series behavior.    

The fact that the instantaneous dimension decreases from the waterbelt to the hot state, that is from a state 
with strong meridional temperature gradient to one where temperature distribution becomes more homogeneous, is in agreement with~\citet{shao2017}, based on the sample entropy method, and~\citet{2017Nonli..30R..32L,Hammam2019} where it is shown that 
strong and localised temperature gradients are associated to low predictability. 

Finally, a principal component analysis evidenced the spatial structure of each attractor, providing hints about the main feedbacks at play in each of them.  
In particular, the principal components 
(PCs), or modes, assess the spatial regions where most of the variability takes place in each attractor and whether this variability is correlated between different regions.  

\begin{figure}[ht!]
\includegraphics[width=0.6\textwidth]{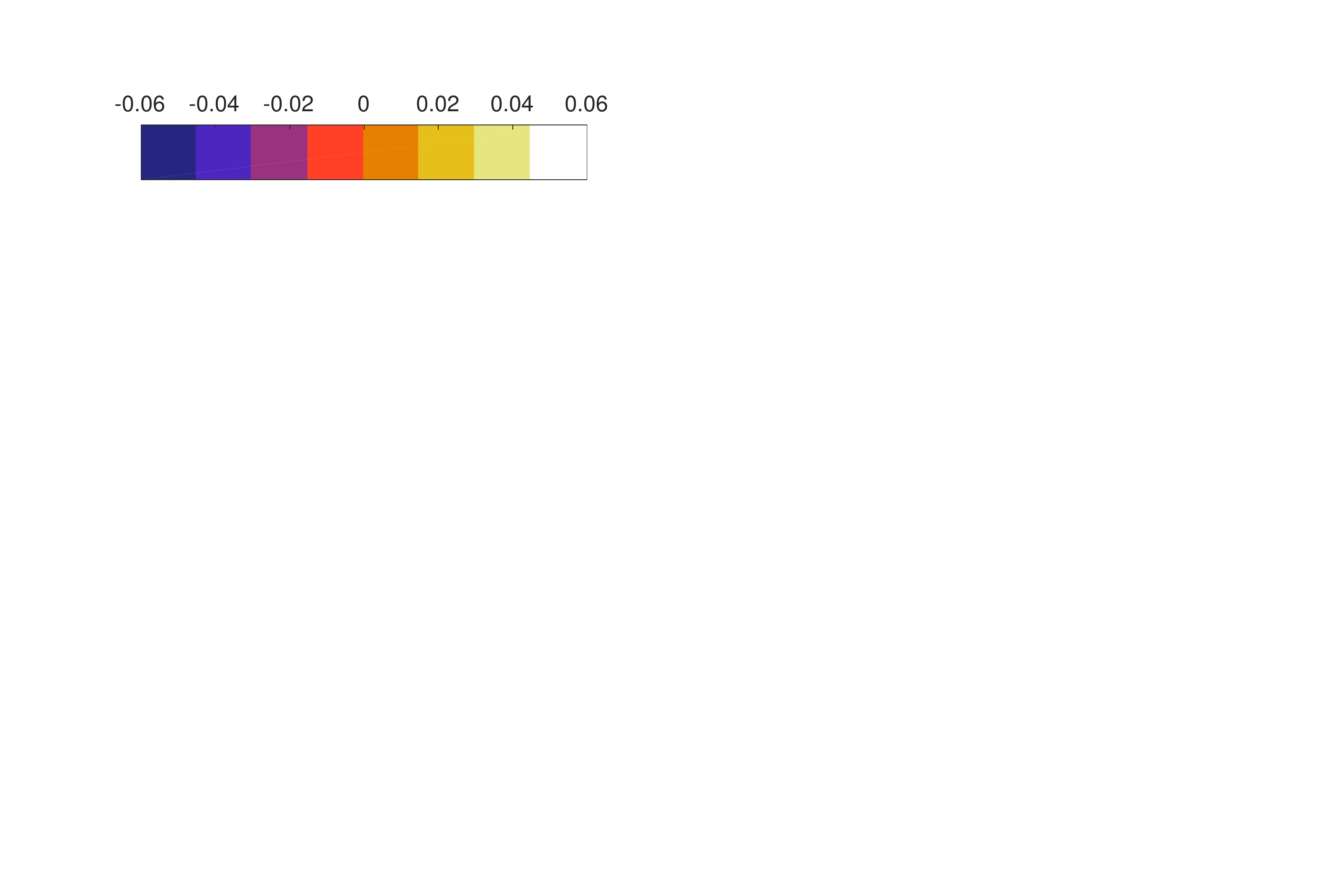}
\vspace{-4.5cm}

\begin{minipage}{0.03\textwidth}
 \vspace{0pt}\raggedright
 \vspace{0.9cm}
{\rotatebox[origin=c]{90}{Mode 1}}\vspace{1cm}
{\rotatebox[origin=c]{90}{Mode 2}}\vspace{1.cm}
{\rotatebox[origin=c]{90}{Mode 3}}\vspace{1.2cm}
{\rotatebox[origin=c]{90}{Mode 4}}\vspace{1.cm}
\end{minipage}\hfill
\begin{minipage}{0.193\textwidth}
\includegraphics[width=\linewidth]{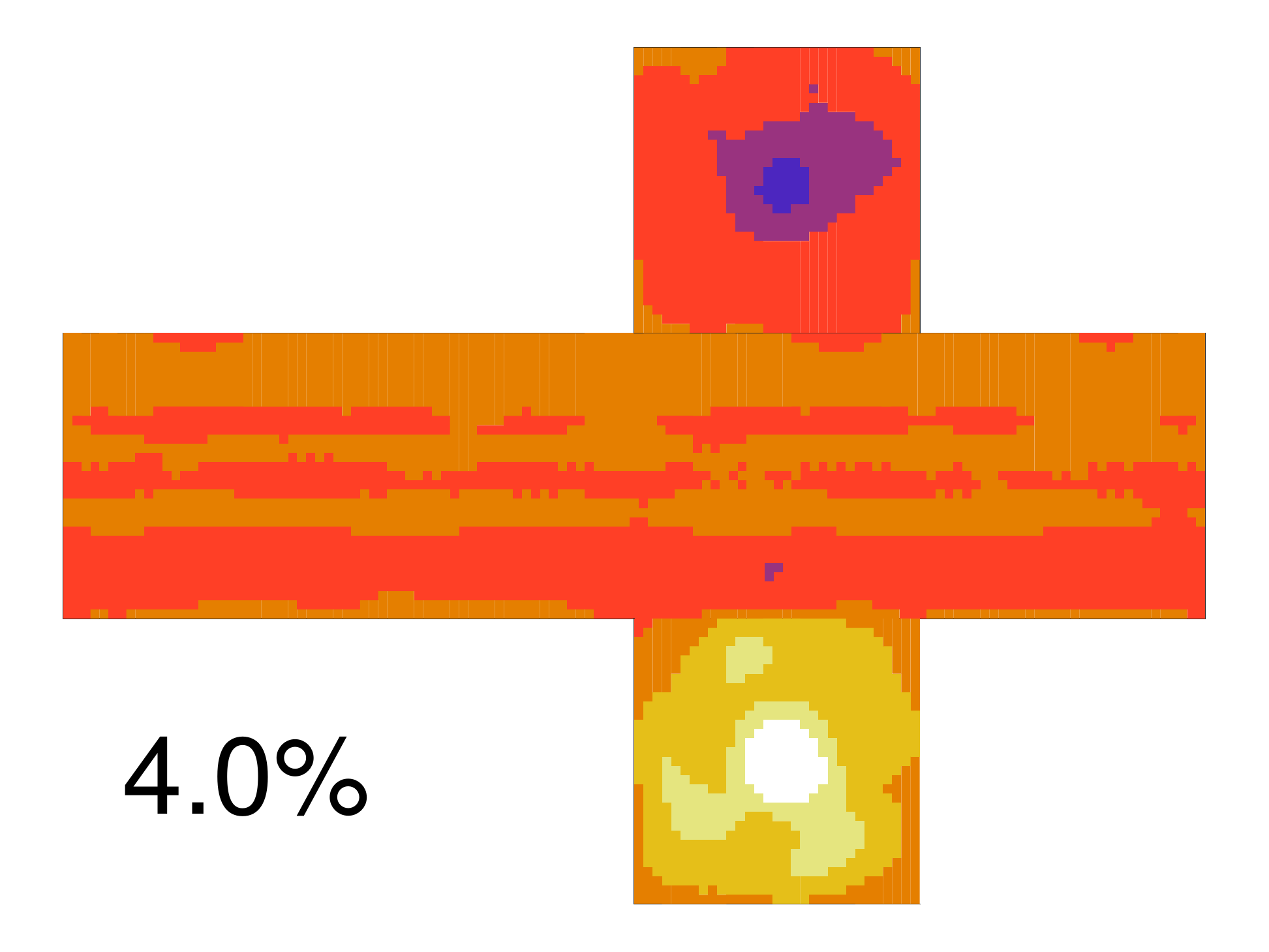} 
\includegraphics[width=\linewidth]{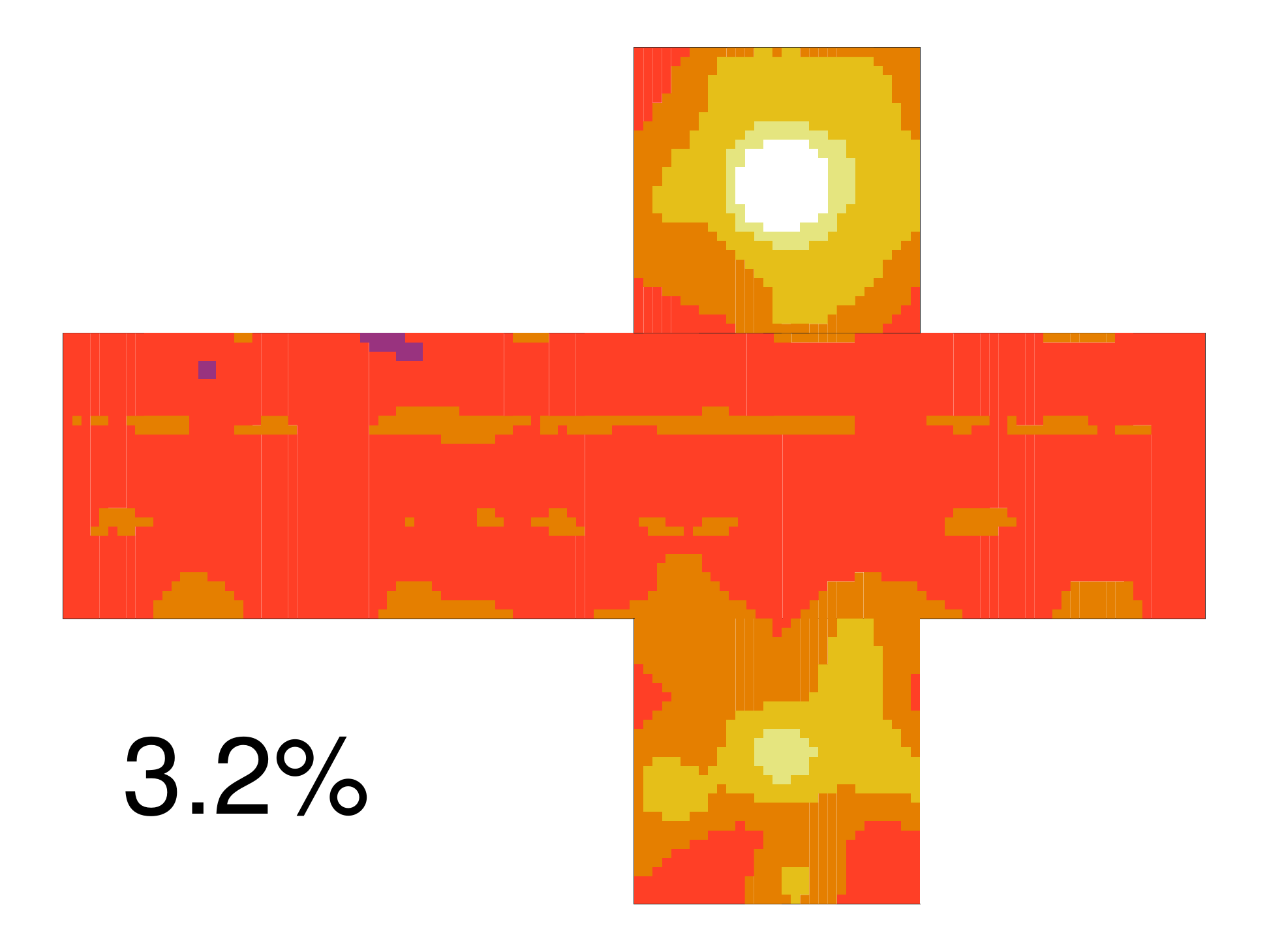}
\includegraphics[width=\linewidth]{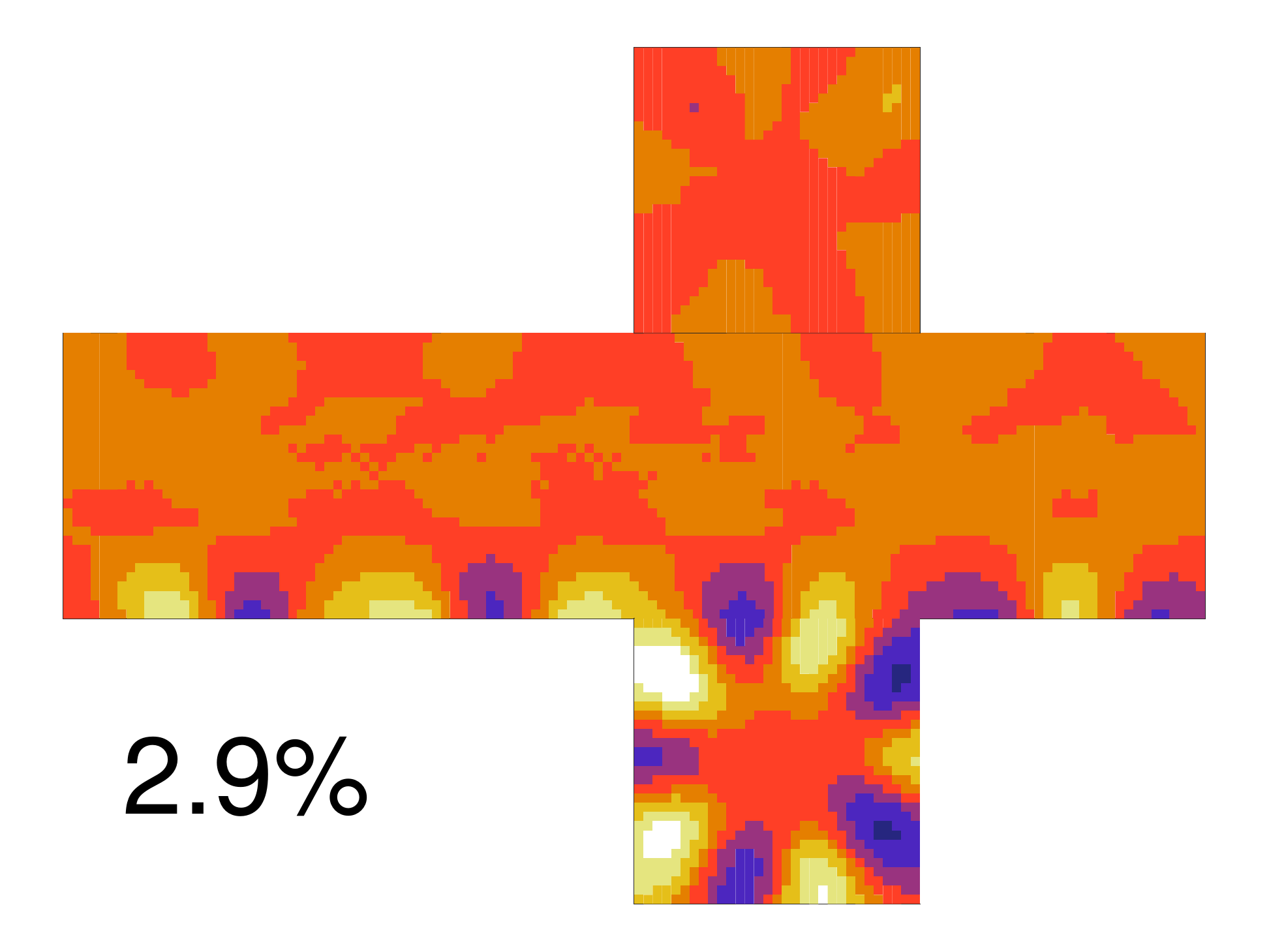}
\includegraphics[width=\linewidth]{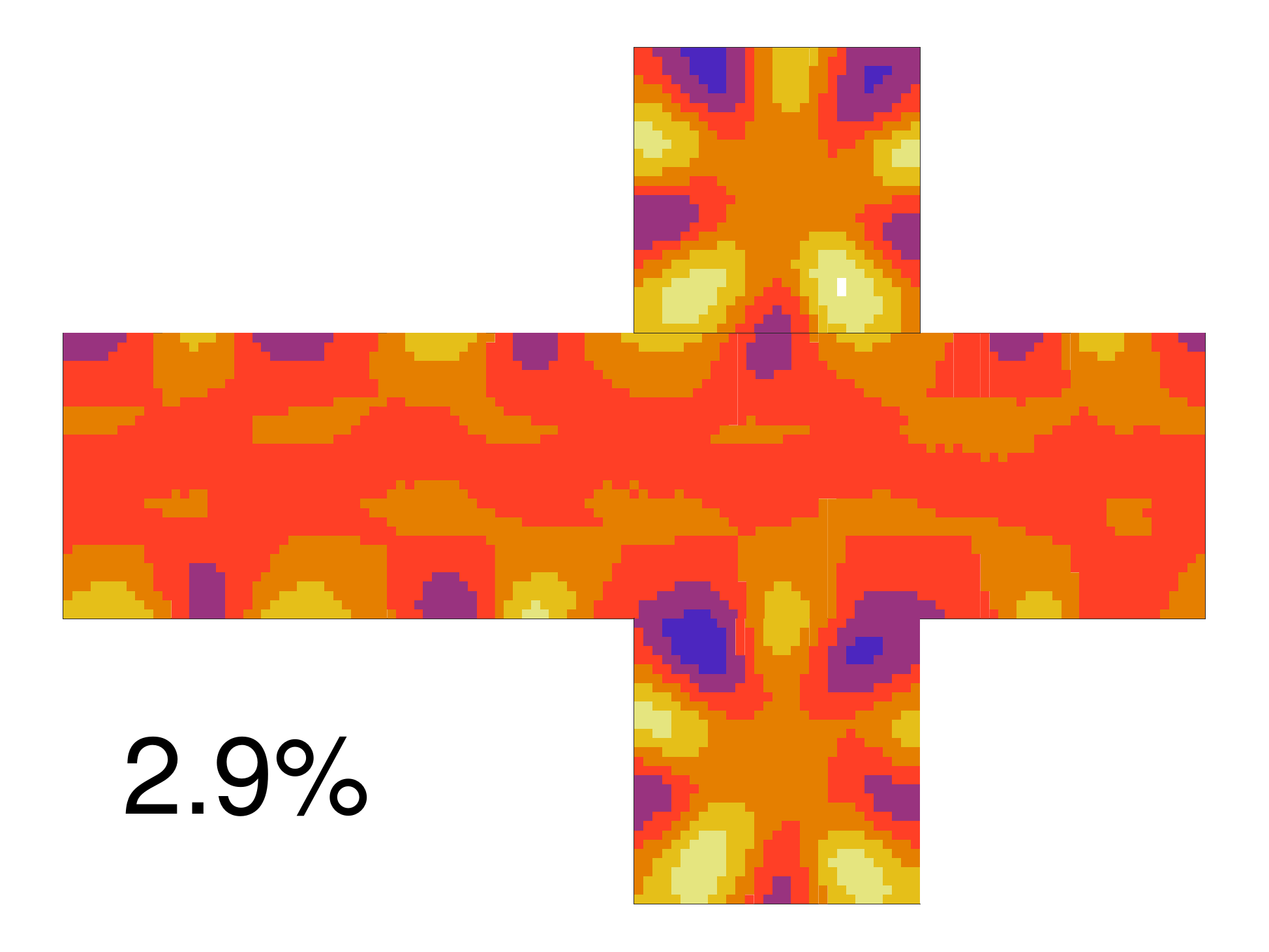}
\end{minipage}\hfill
\begin{minipage}{0.193\textwidth}
\includegraphics[width=\linewidth]{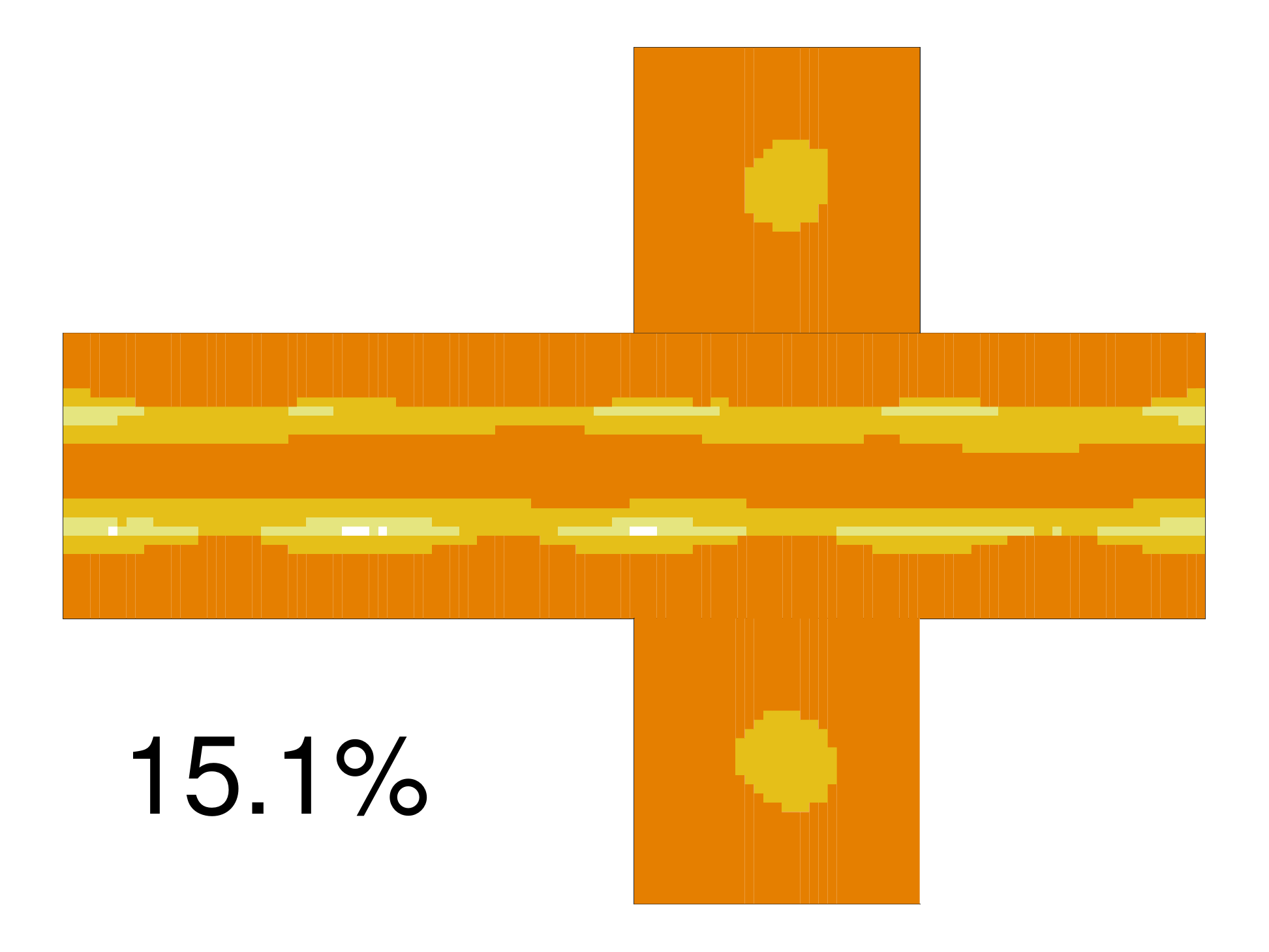} 
\includegraphics[width=\linewidth]{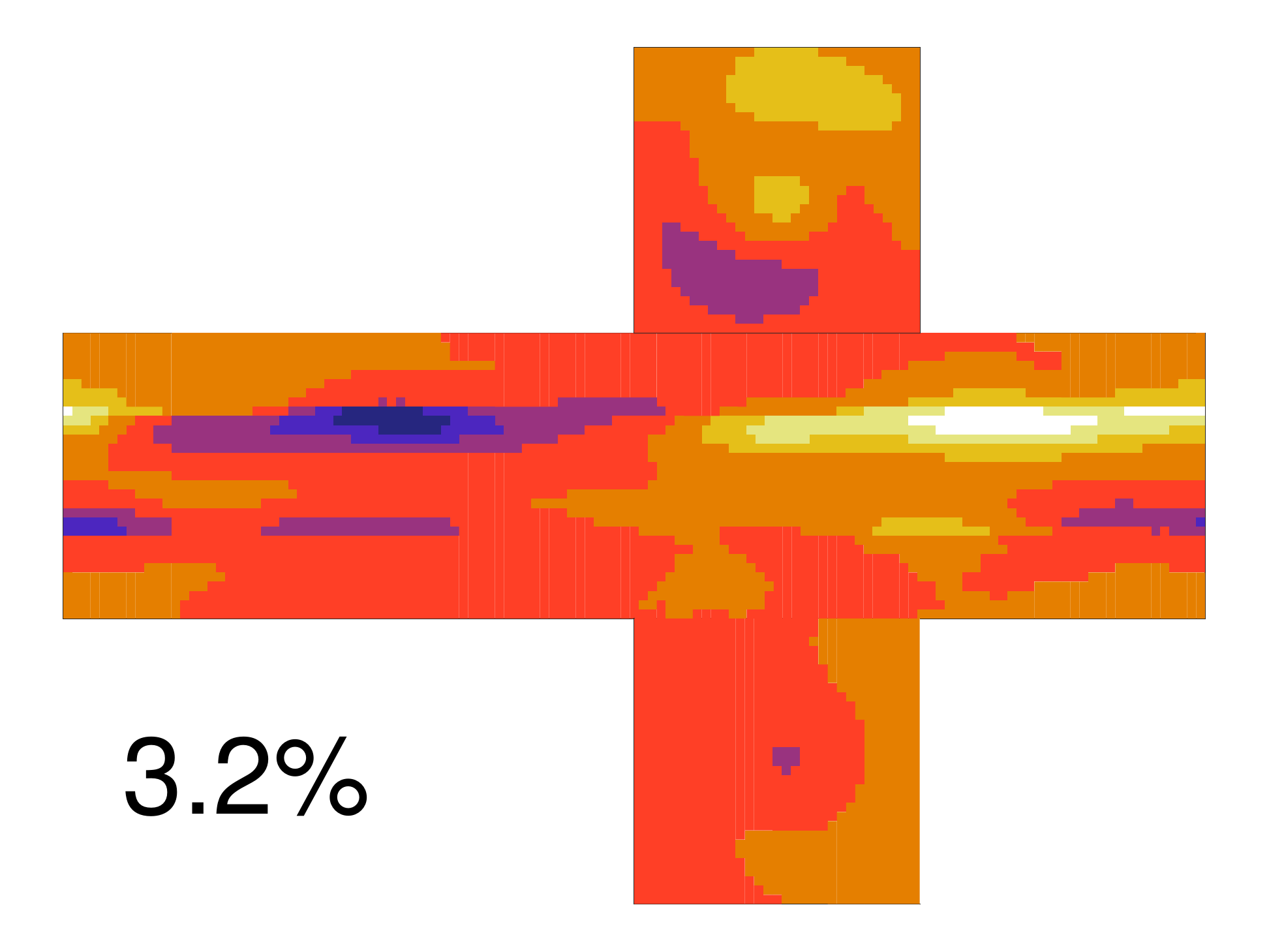}
\includegraphics[width=\linewidth]{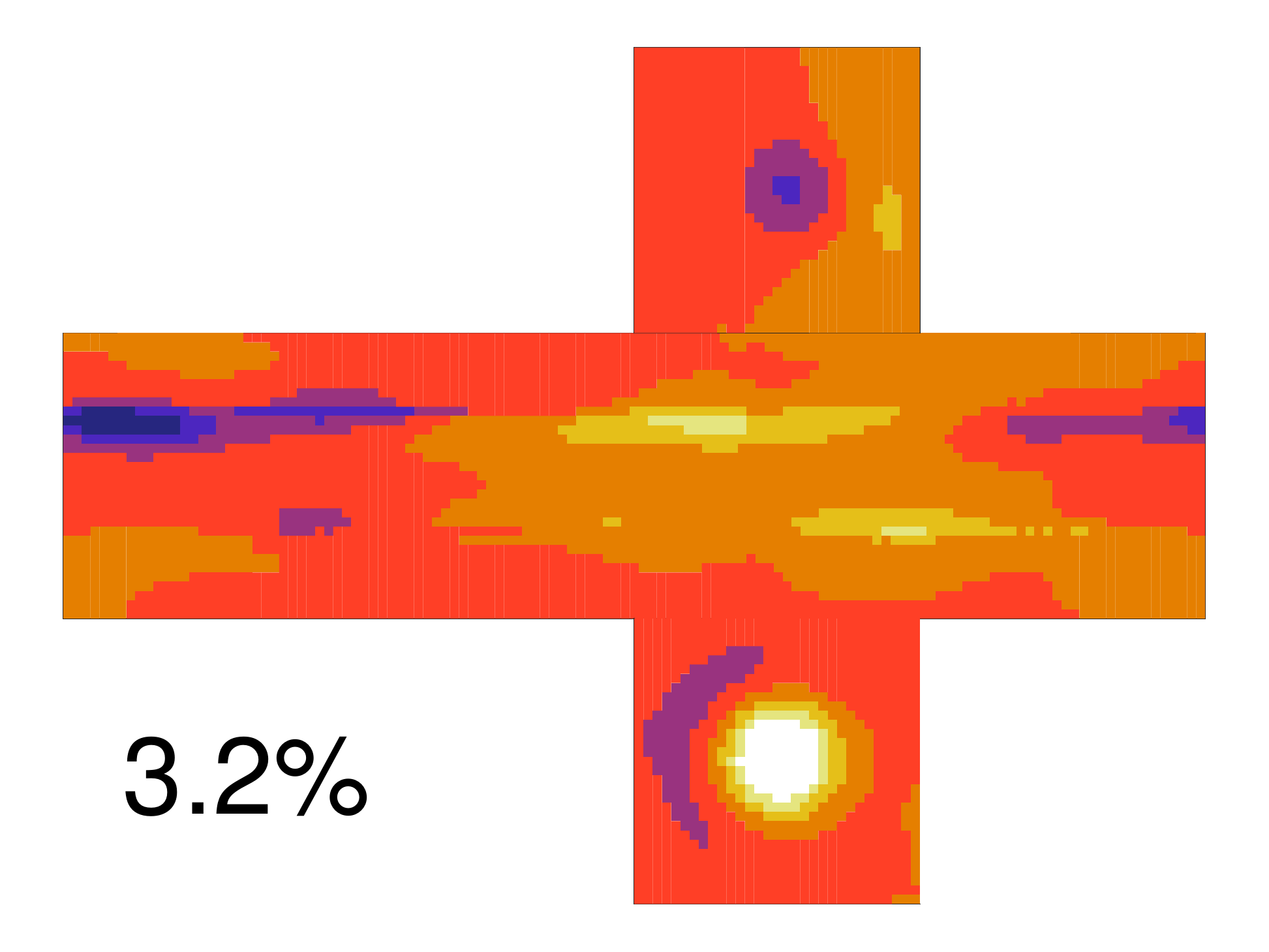}
\includegraphics[width=\linewidth]{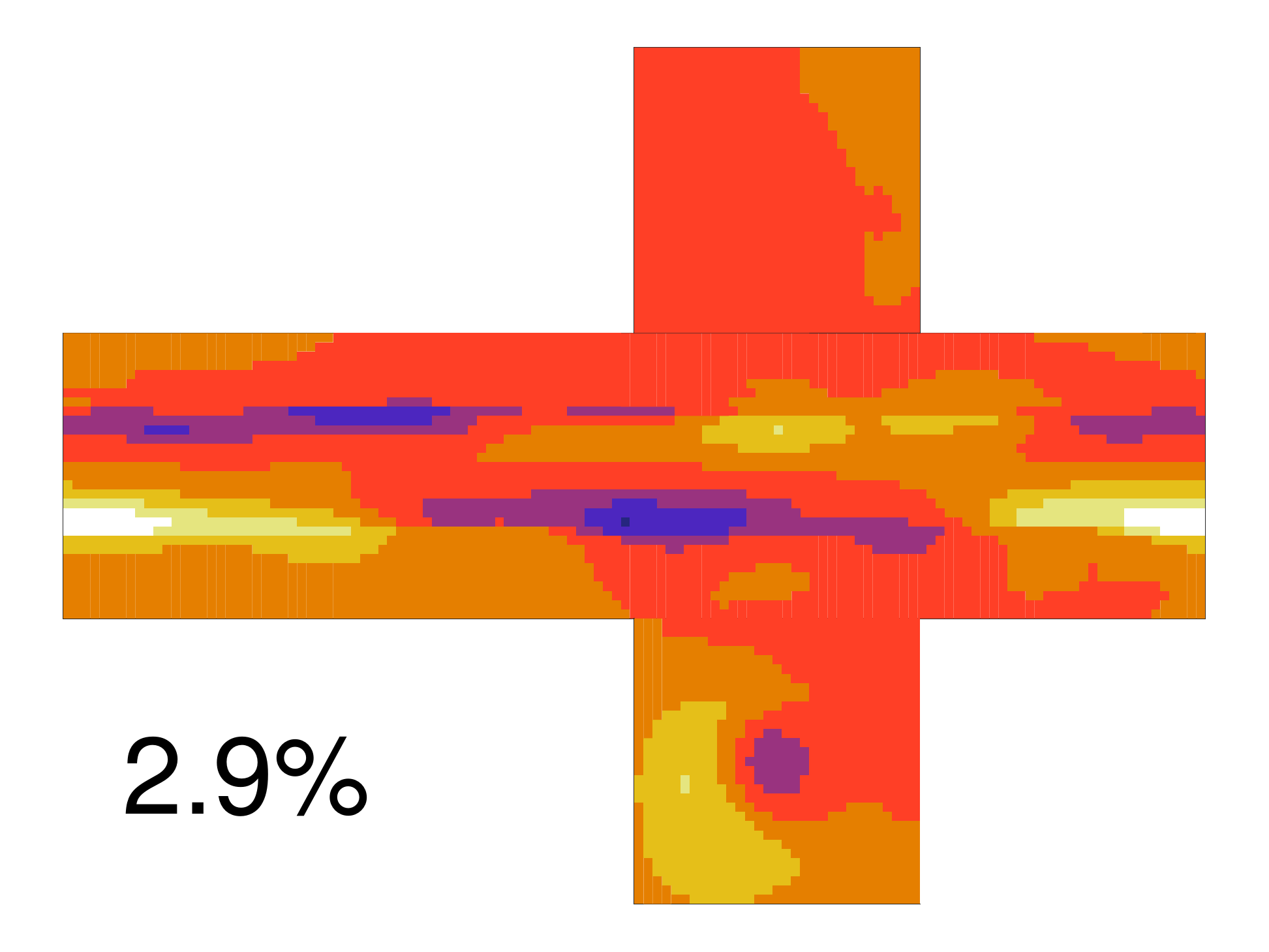}
\end{minipage}\hfill
\begin{minipage}{0.193\textwidth}
\includegraphics[width=\linewidth]{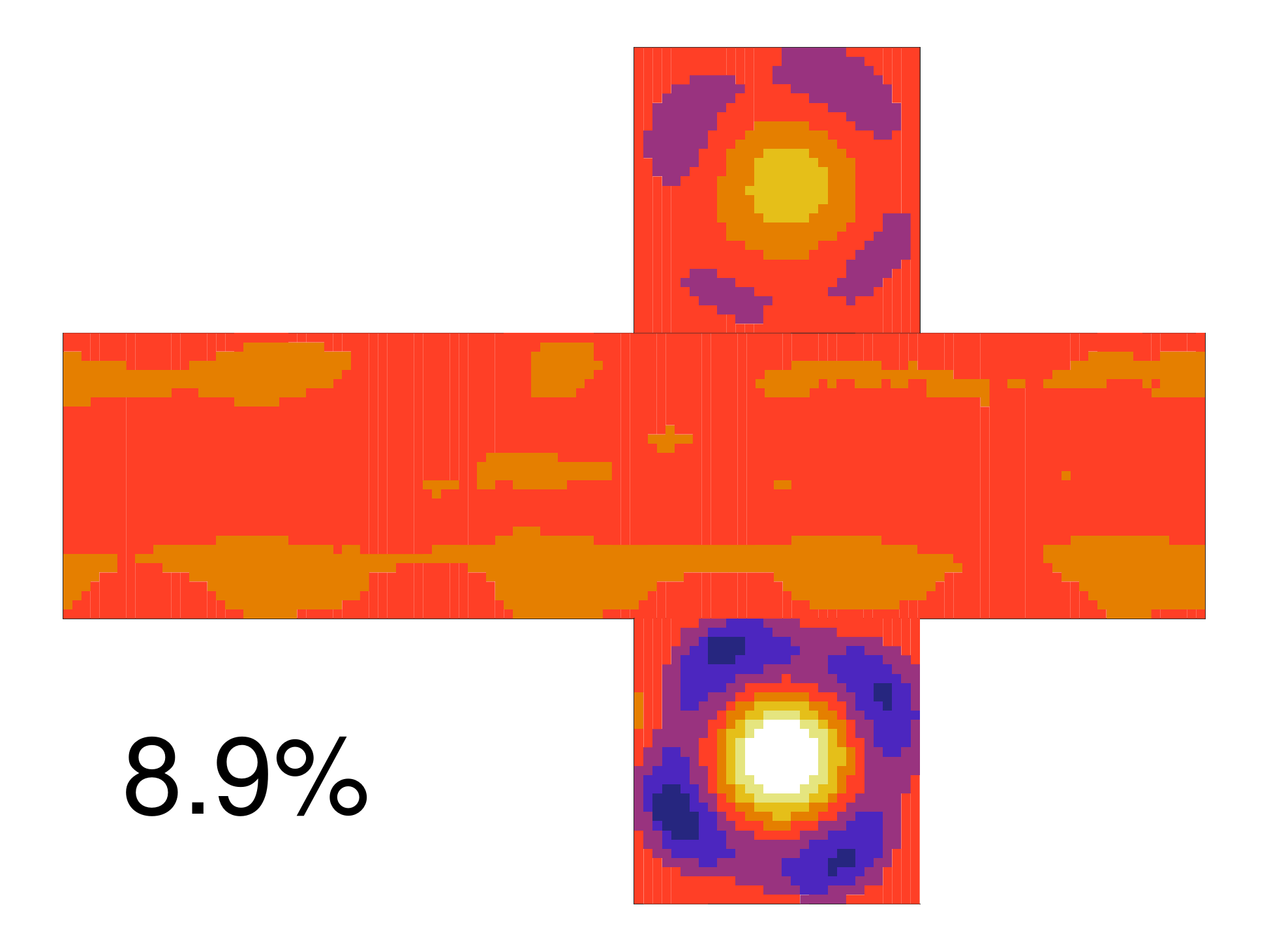} 
\includegraphics[width=\linewidth]{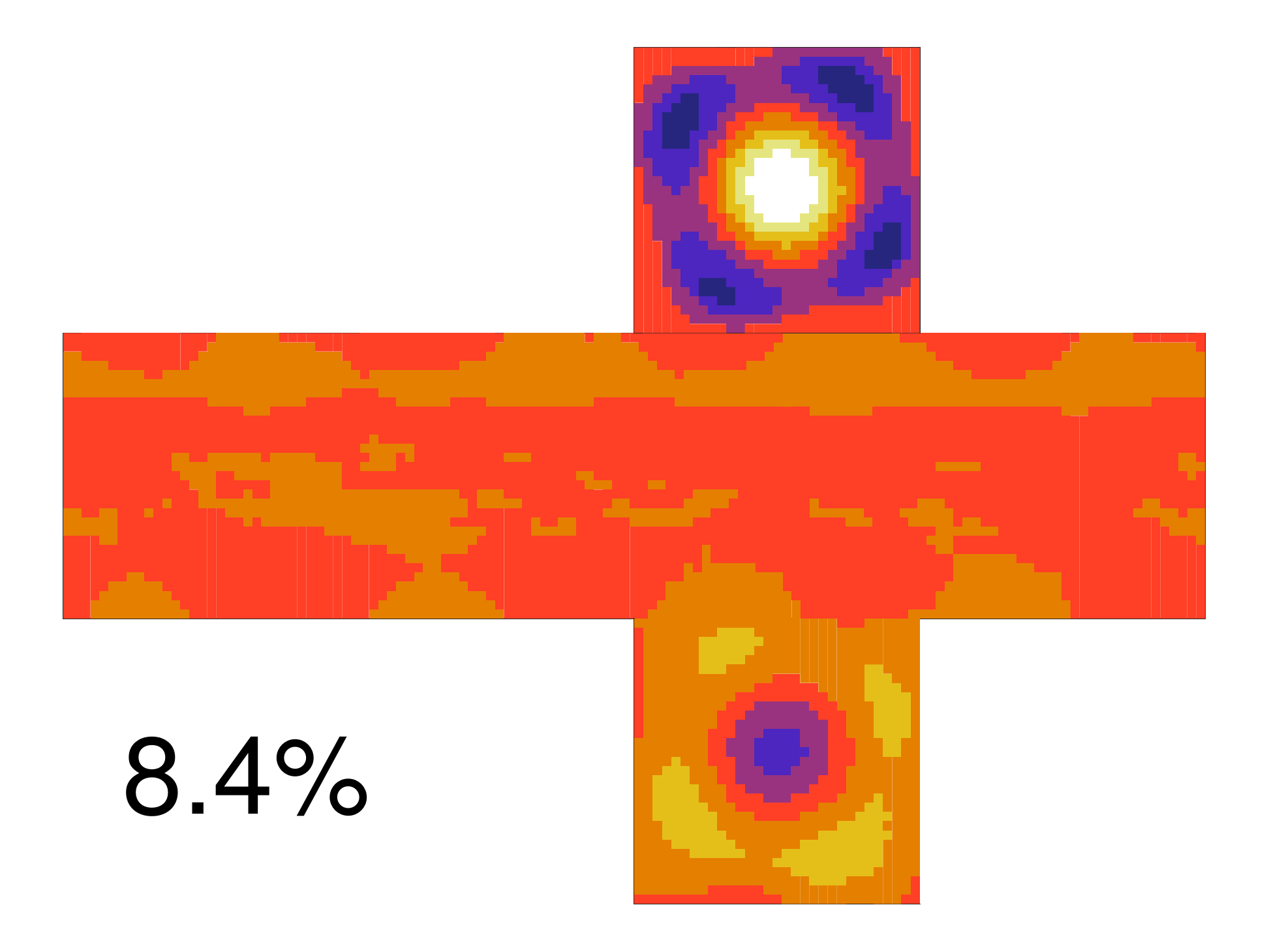}
\includegraphics[width=\linewidth]{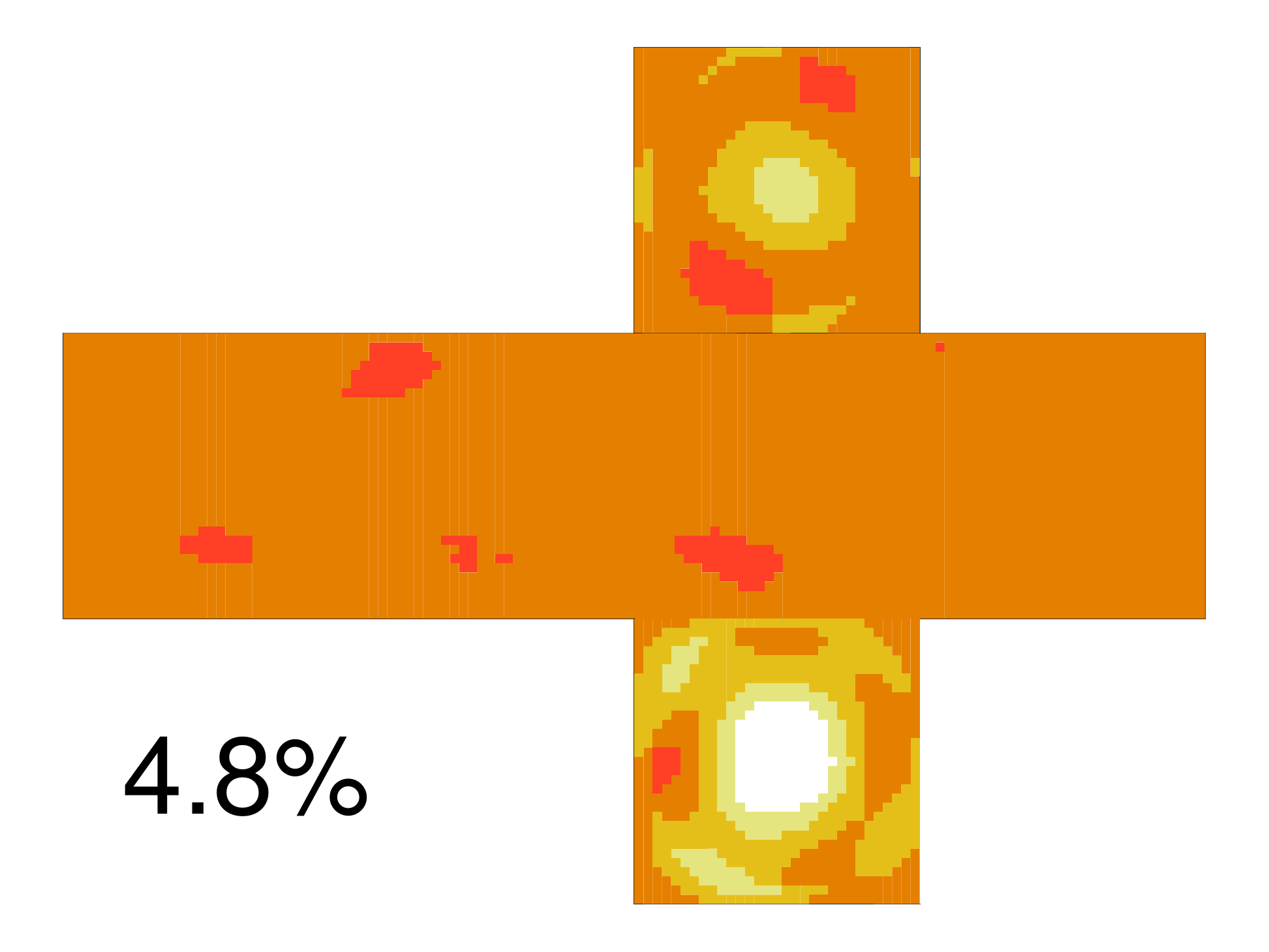}
\includegraphics[width=\linewidth]{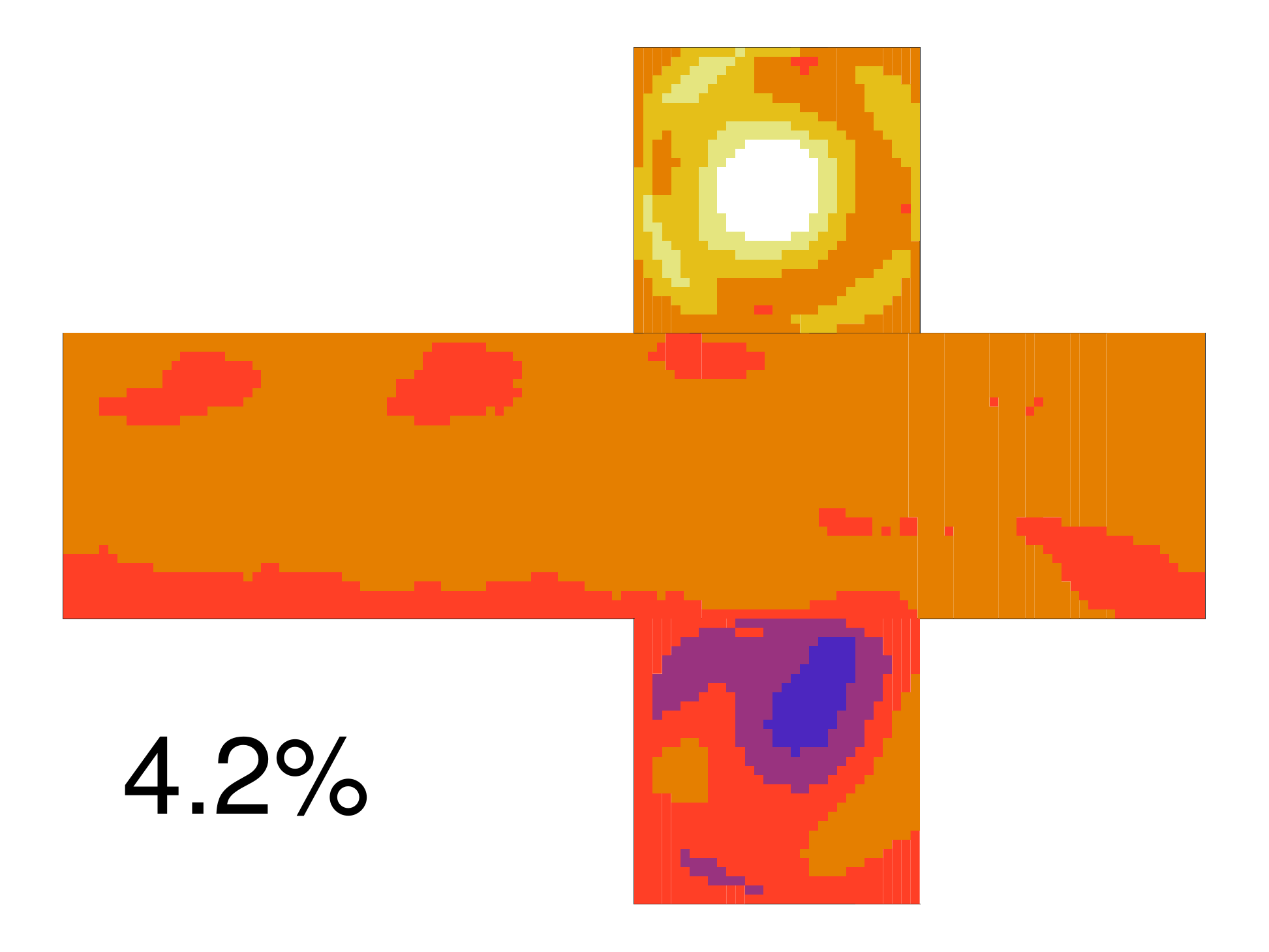}
\end{minipage}\hfill
\begin{minipage}{0.193\textwidth}
\includegraphics[width=\linewidth]{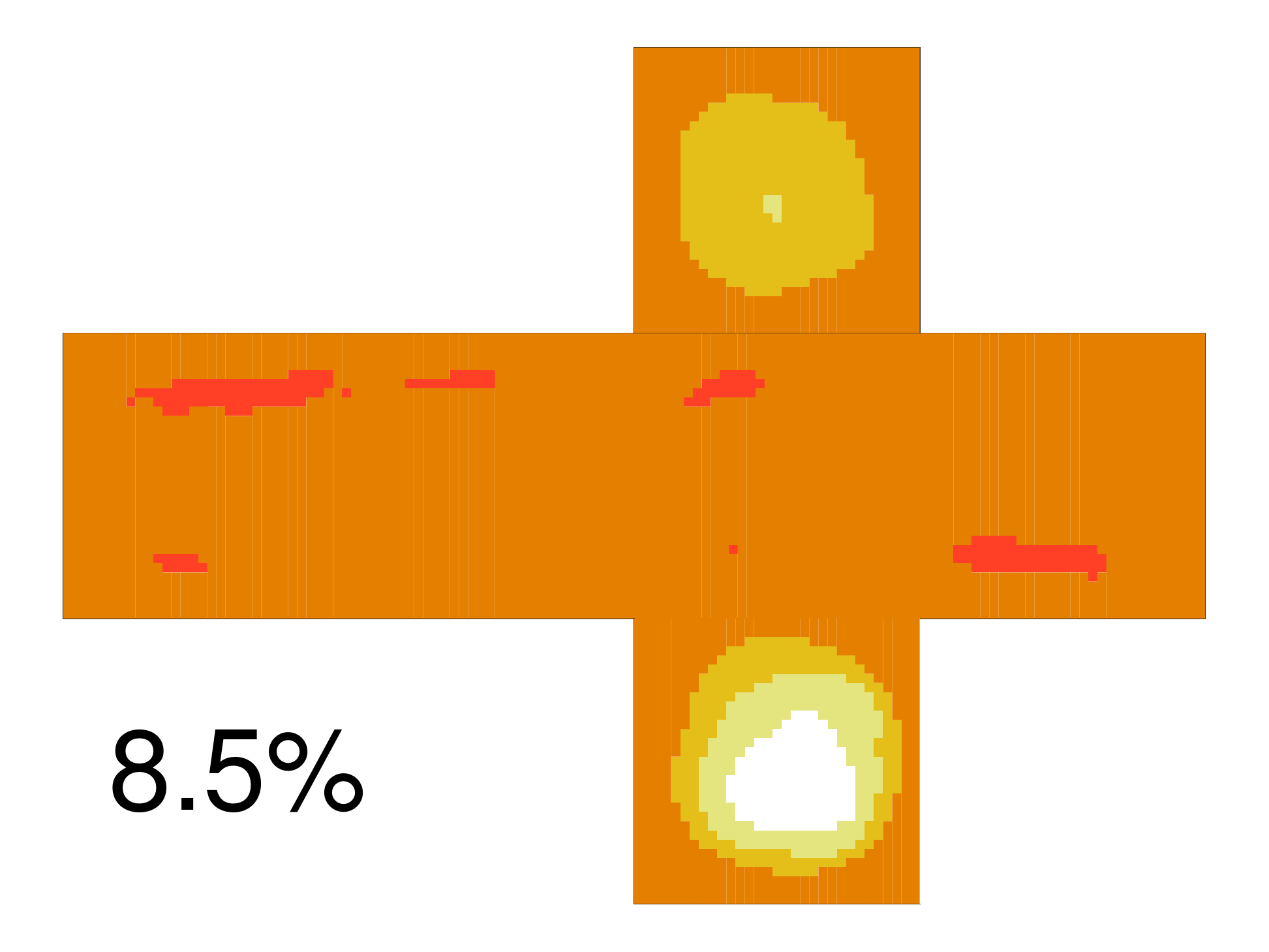} 
\includegraphics[width=\linewidth]{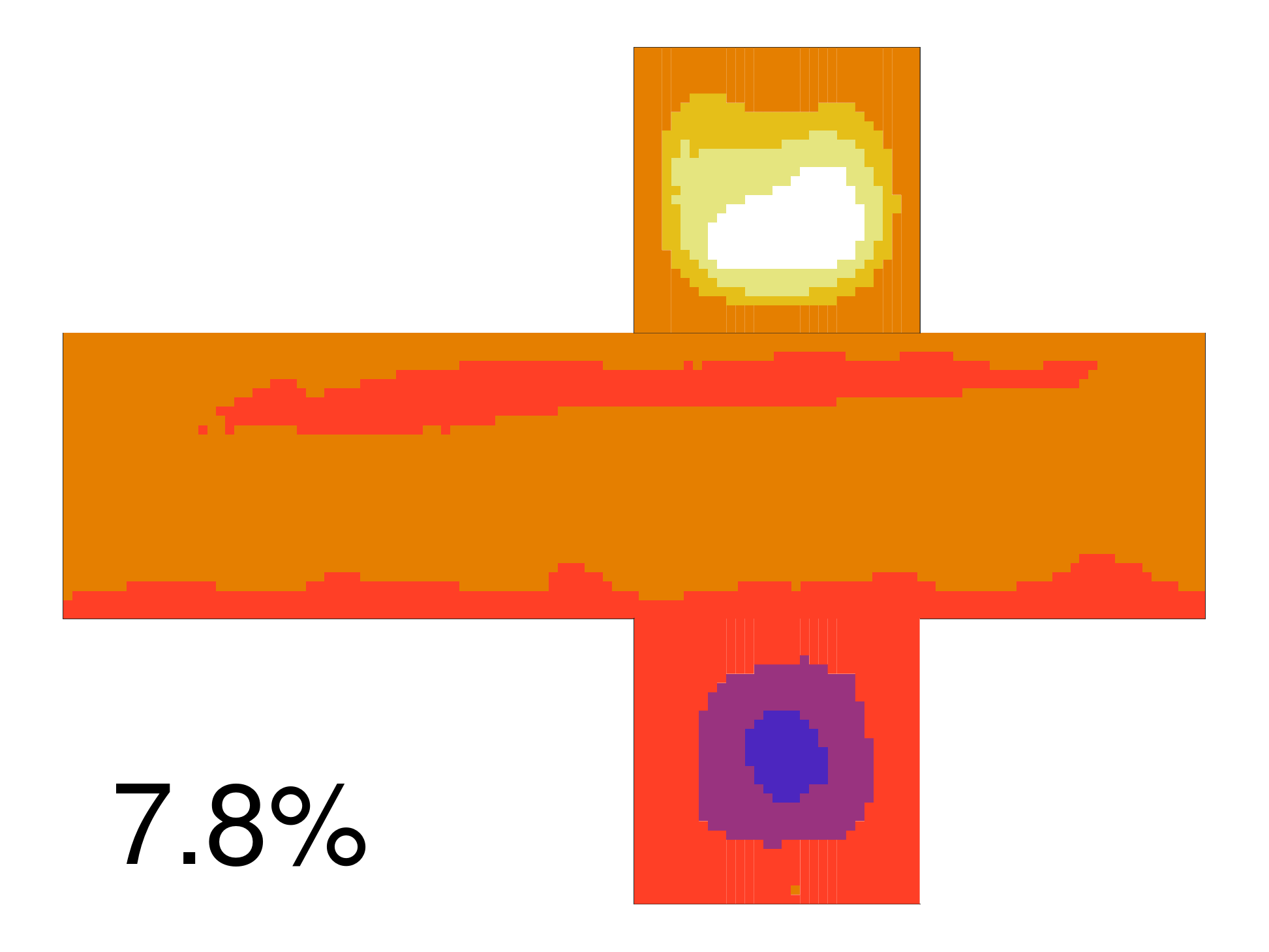}
\includegraphics[width=\linewidth]{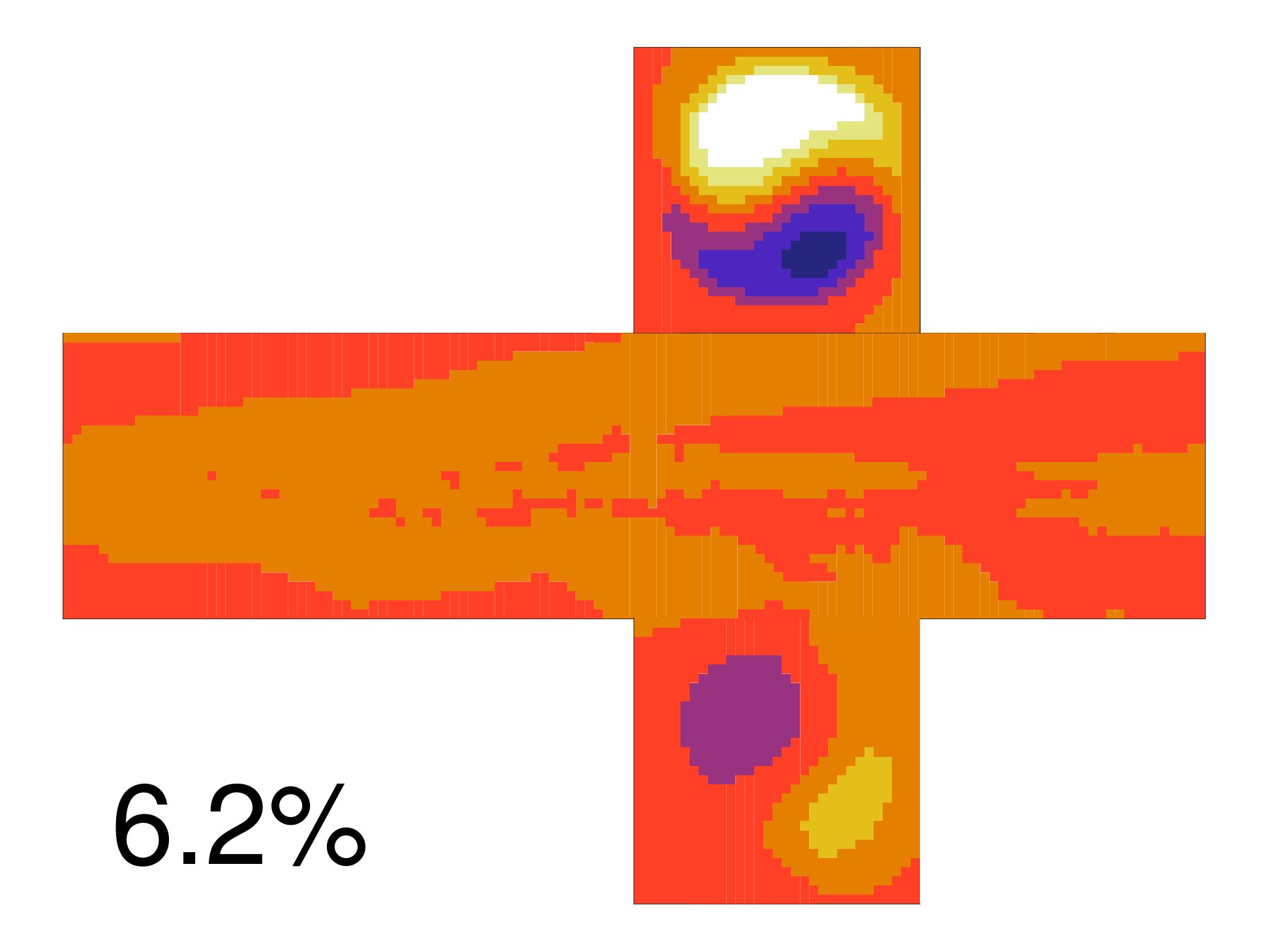}
\includegraphics[width=\linewidth]{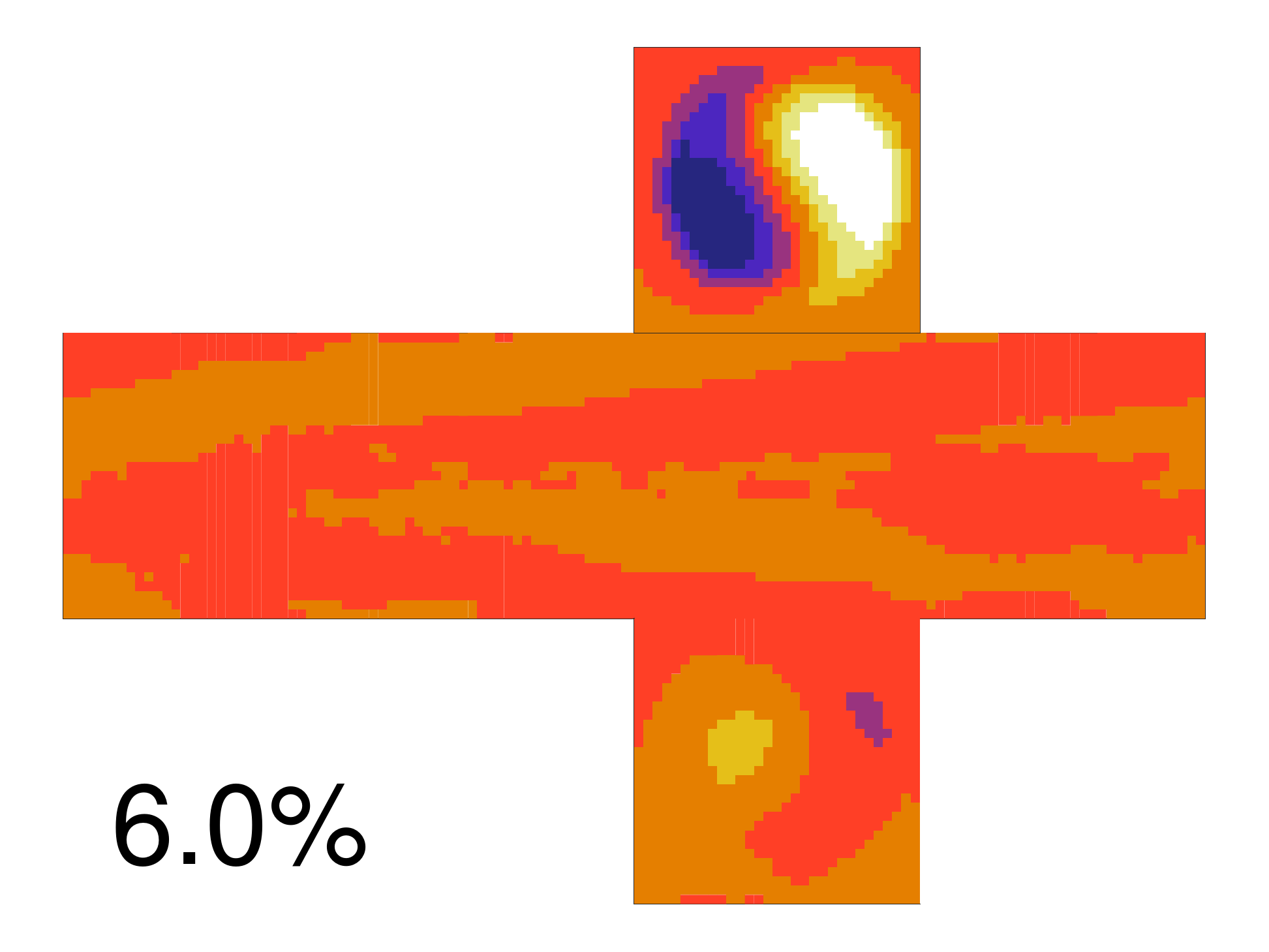}
\end{minipage}
   
\hfill\hfill\hfill Snowball \hfill\hfill Waterbelt \hfill\hfill Cold state \hfill\hfill Warm state \hfill{} 

\caption{Principal components and explained variance (in percentage) for {\tt setUp1}.}
\label{fig:seven}
\end{figure}

Fig.~\ref{fig:seven} displays the map of the four most important PCs for each attractor observed in {\tt setUp1}.
\begin{itemize}
\item
In the snowball, there are no modes describing more than 4\% of the total variance, showing the lack of marked spatial dynamics in this regime and supporting the fact that the measure given by the instantaneous dimension is spurious, as discussed before and in agreement 
with~\citet{2010QJRMS.136....2L,2017Nonli..30R..32L}. 
This is consistent with the very limited spatial heat transport, both in the ocean and in the atmosphere (Fig.~\ref{fig:two}, fourth column). However, the main feature (first two PCs) appears to be an opposition between the poles.  
The subsequent modes display rotating features in both polar regions.

\item
In the waterbelt, the four main modes display strong features around the tropics, evidencing that the dynamics occurs more at low latitudes, as opposed to the high-latitude dynamics of the cold and warm states. The first mode largely dominates the dynamics (15.1\% of the variance, as compared to 3.2\% for the second and third ones). Its main feature is a homogeneous positive sign indicating that the temperature variations are positively correlated across the whole Earth.  The two rings around 16$^\circ$ latitude (close to the ice boundary, see Table~2) are more positive than the other regions, indicating wider temperature fluctuations consistent with the locally strong temperature gradient  
(see Fig.~\ref{fig:two}-\ref{fig:three}, first column), that also determines the high value of the instantaneous dimension. 
Therefore, the main driver of the inter-decadal variability in the waterbelt appears to be fluctuations of the sea ice edge~\citep{Rose2015}. 

\item
In the cold state, the dynamics displays more or less independent poles, with modes displaying pairwise symmetry: modes 1 and 2, and 3 and 4, each pair with similar explained variance. The first pair displays an opposition between a pole and its surrounding, around 65$^\circ$ latitude, as evidenced by the positive regions surrounded by a strongly negative ring. The pattern in the cloud cover shows a strong gradient at such latitudes (see Fig.~2, second column), suggesting an increasing relevance of the cloud albedo feedback in this attractor with respect to the ice albedo feedback, that is dominant in the previous attractors. This pair of modes also displays variability at midlatitudes that can be the signature for the modulation of the jet stream and the associated surface wind stress giving rise to the onset of oceanic and atmospheric annular modes, as discussed  in~\citet{2007JAtS...64.4270M}. 
The second pair of modes  are weaker, and display two rings around the poles. They further contribute to this local dynamics in the subpolar region. 

\item
For the warm state, the first and second main modes have similar explained variance of the time series and correspond  
to temperature variations around the South (resp.~North) pole, with a weak correlation in the first mode and anticorrelation in the second mode with the other pole. 
The four next modes point to independent dynamics in the polar regions, with a dominant vortical dynamics around the North pole for 
modes 3 and 4, and around the South pole for modes 5 and 6 (not shown). 
\end{itemize}

\begin{figure}[ht!]
\includegraphics[width=0.6\textwidth]{FIGS/colorbarPCA-eps-converted-to.pdf}
\vspace{-4.5cm}

\begin{minipage}{0.03\textwidth}
 \vspace{0pt}\raggedright
 \vspace{0.9cm}
{\rotatebox[origin=c]{90}{Mode 1}}\vspace{1cm}
{\rotatebox[origin=c]{90}{Mode 2}}\vspace{1.cm}
{\rotatebox[origin=c]{90}{Mode 3}}\vspace{1.2cm}
{\rotatebox[origin=c]{90}{Mode 4}}\vspace{1.cm}
\end{minipage}\hfill
\begin{minipage}{0.193\textwidth}
\includegraphics[width=\linewidth]{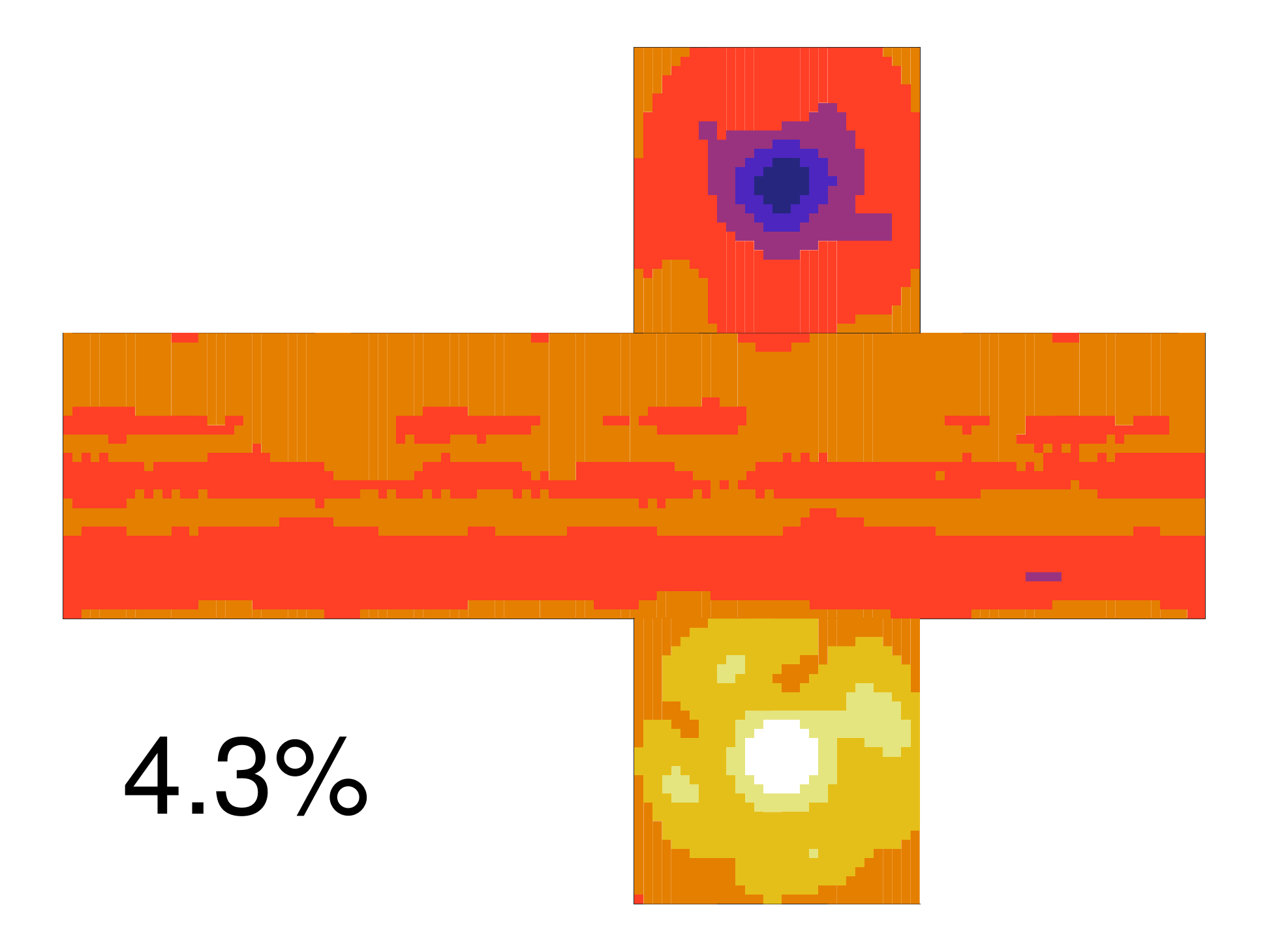} 
\includegraphics[width=\linewidth]{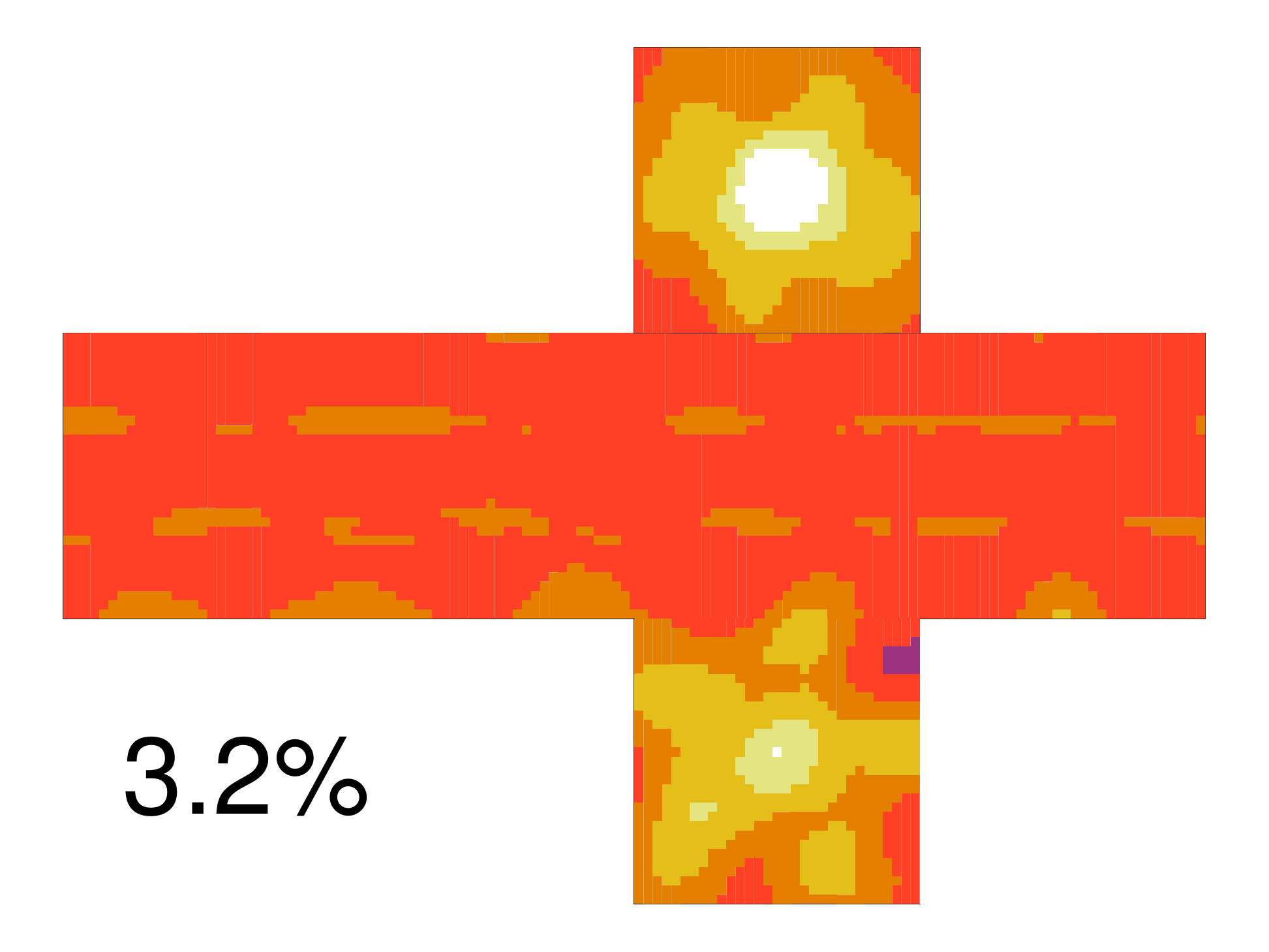}
\includegraphics[width=\linewidth]{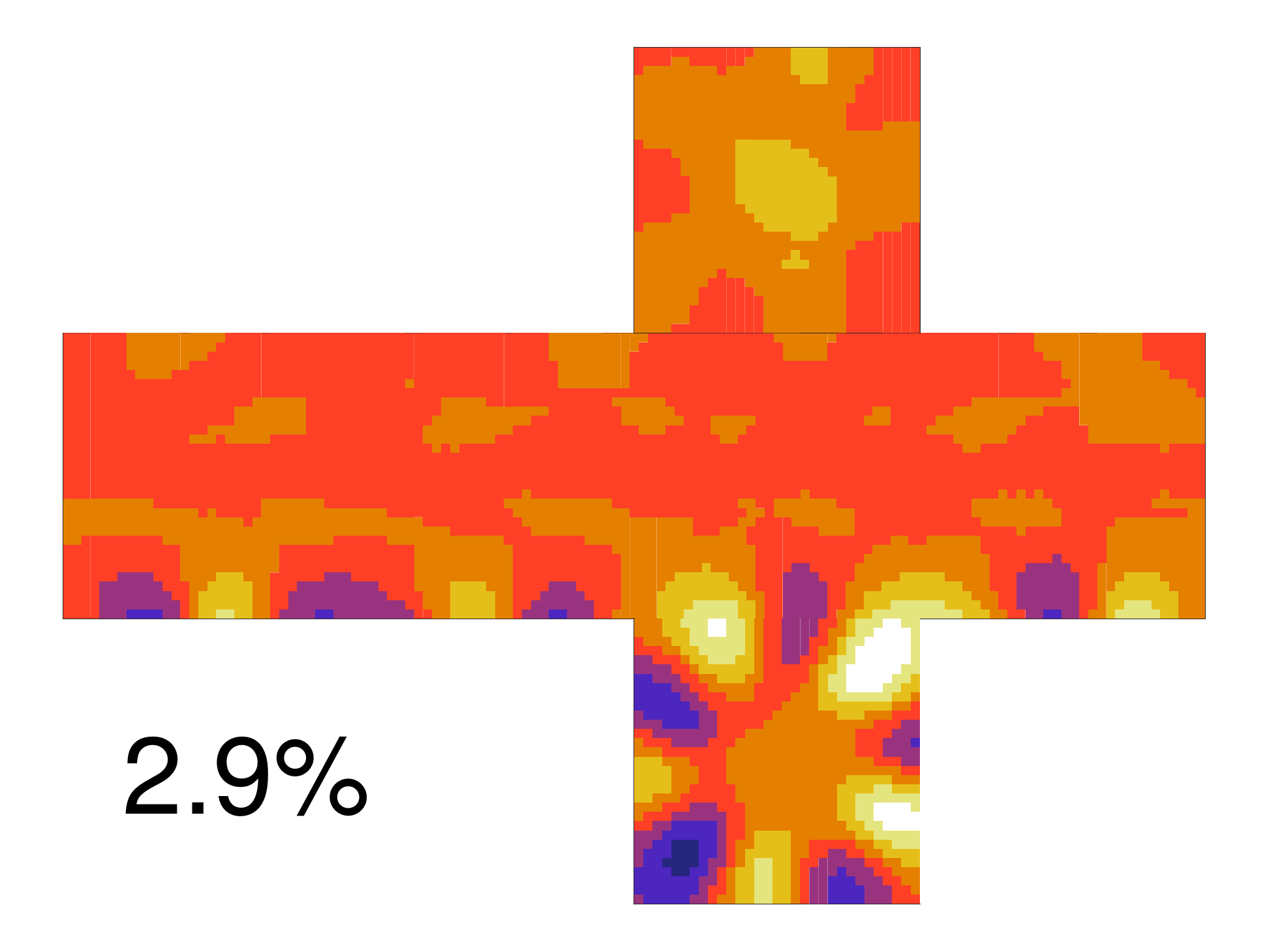}
\includegraphics[width=\linewidth]{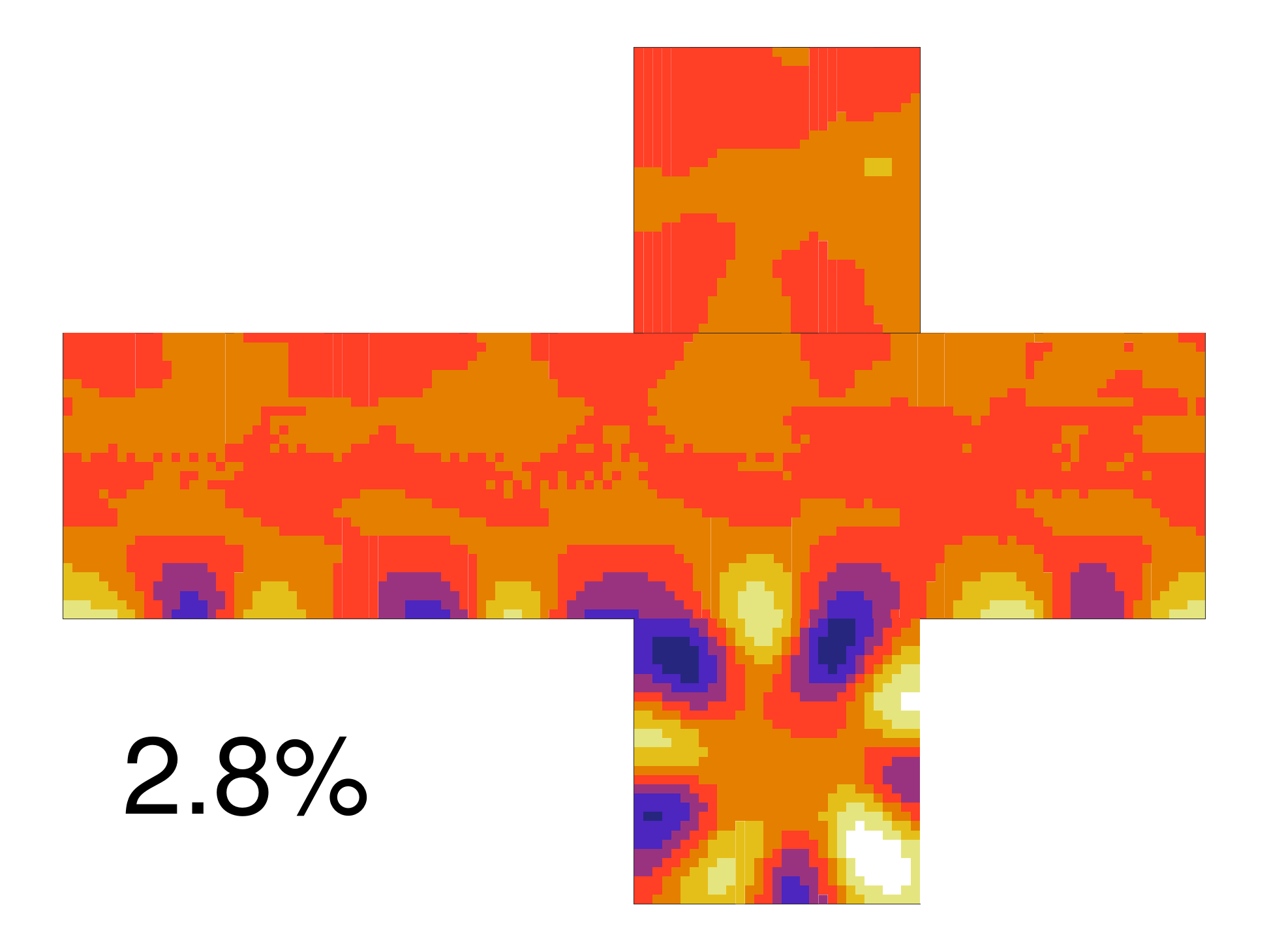}
\end{minipage}\hfill
\begin{minipage}{0.193\textwidth}
\includegraphics[width=\linewidth]{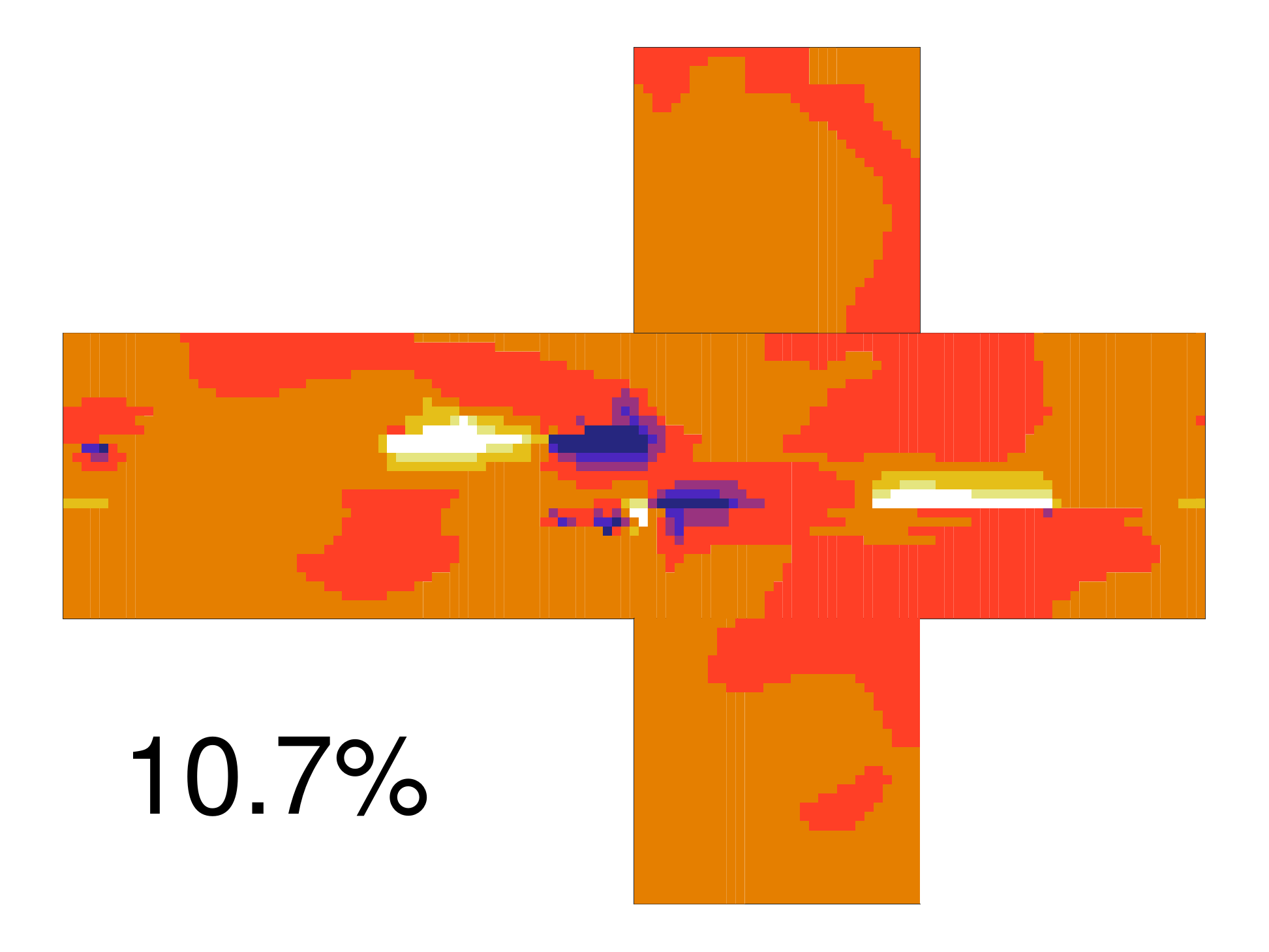} 
\includegraphics[width=\linewidth]{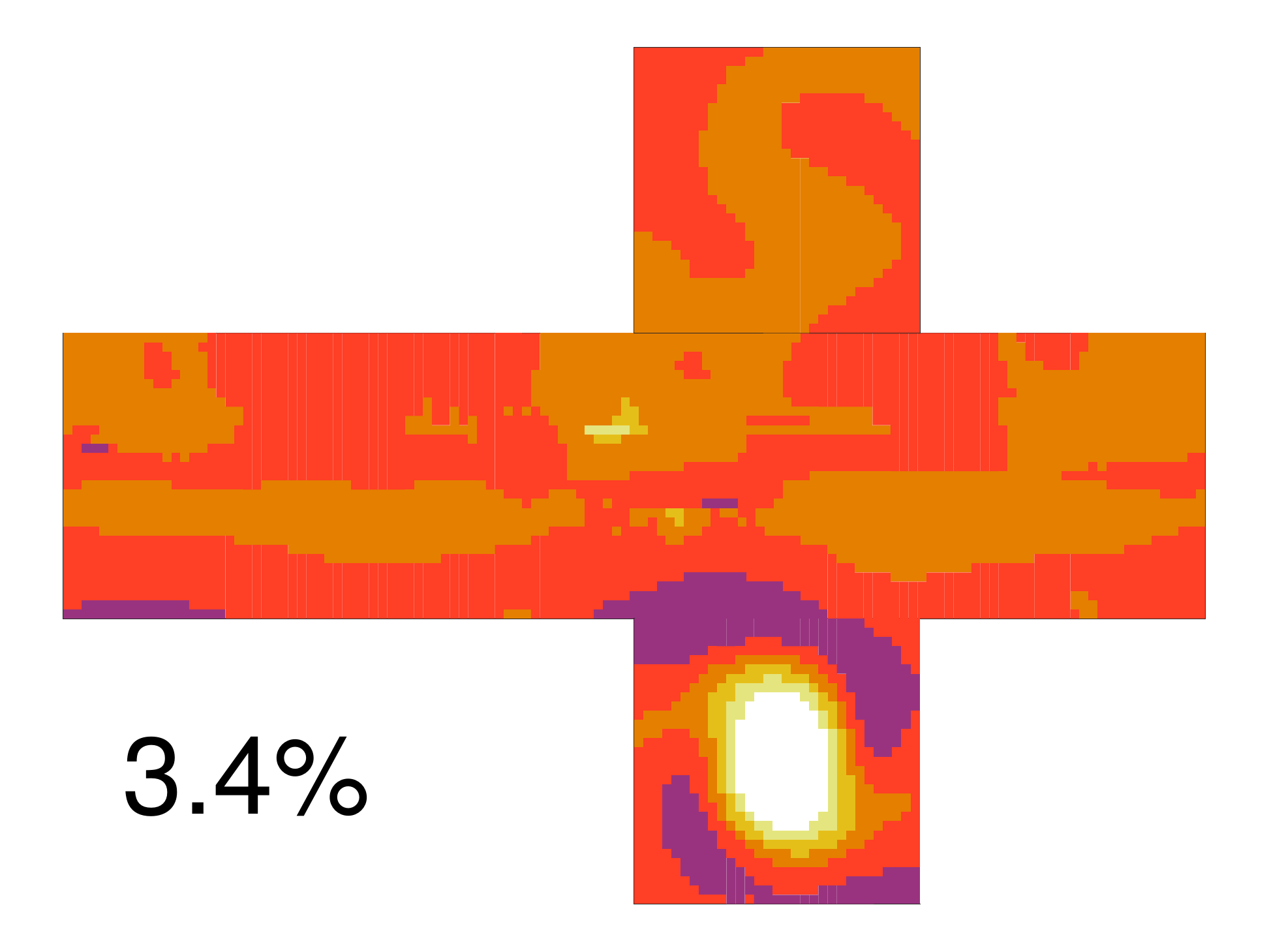}
\includegraphics[width=\linewidth]{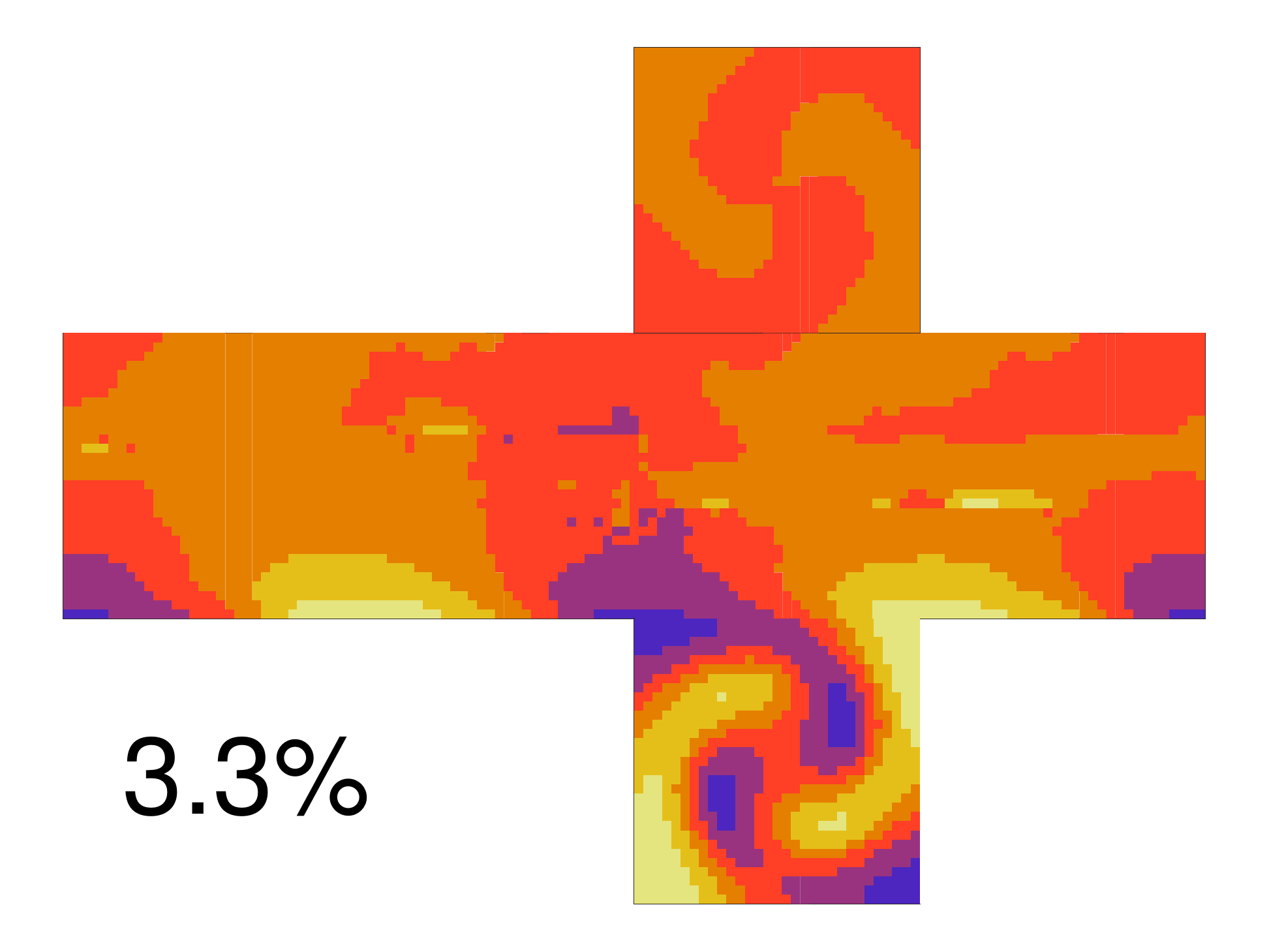}
\includegraphics[width=\linewidth]{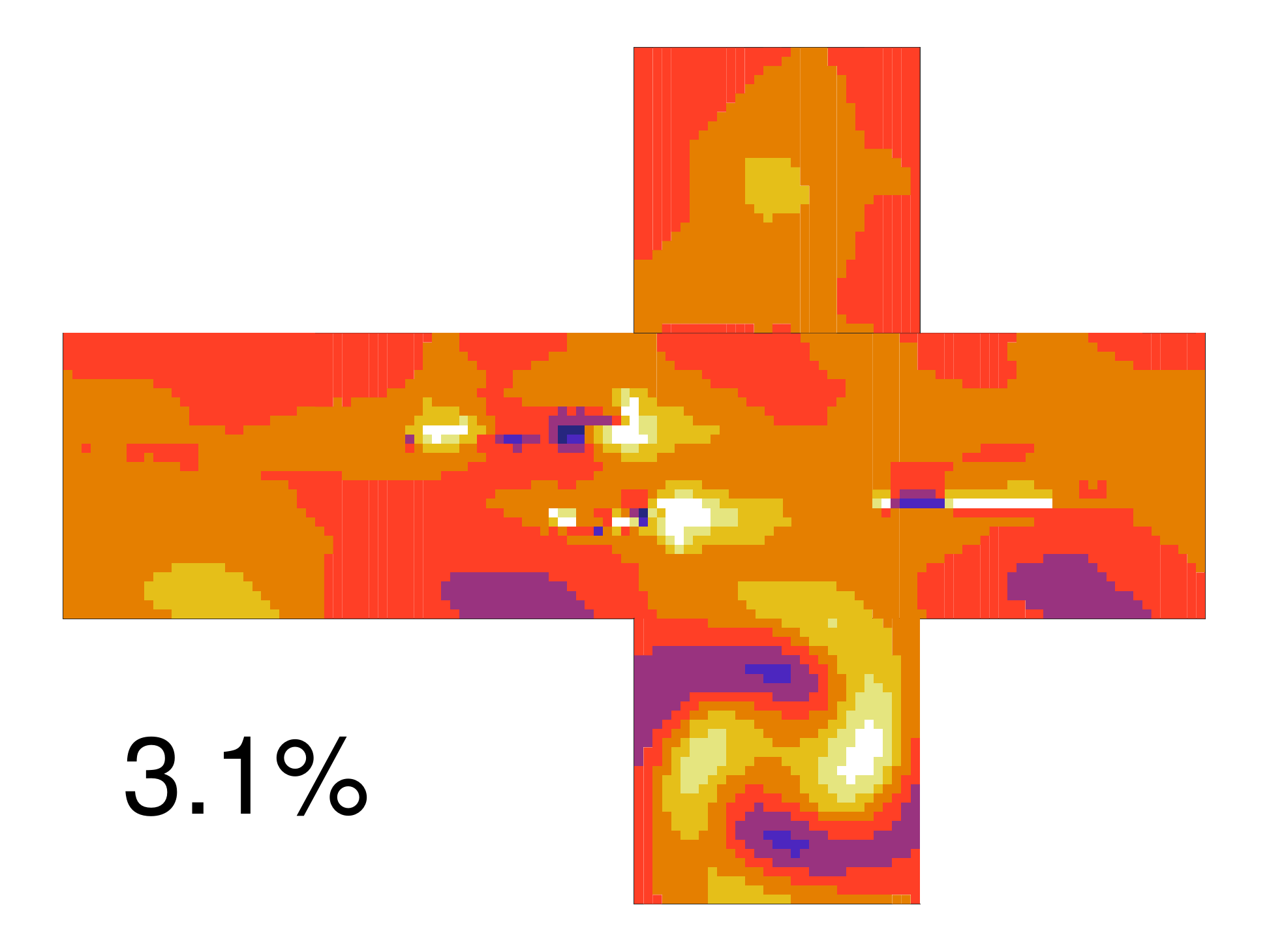}
\end{minipage}\hfill
\begin{minipage}{0.193\textwidth}
\includegraphics[width=\linewidth]{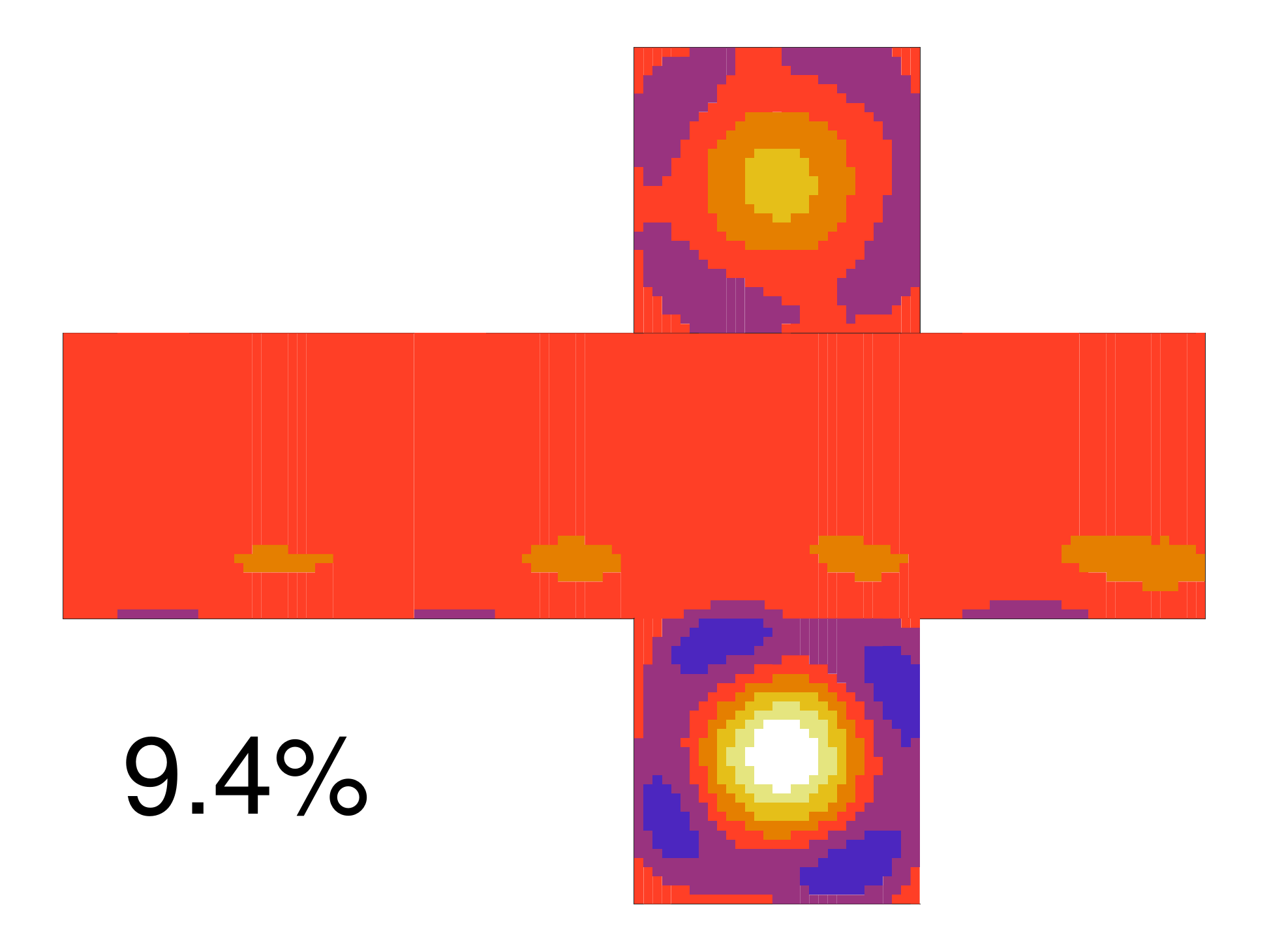} 
\includegraphics[width=\linewidth]{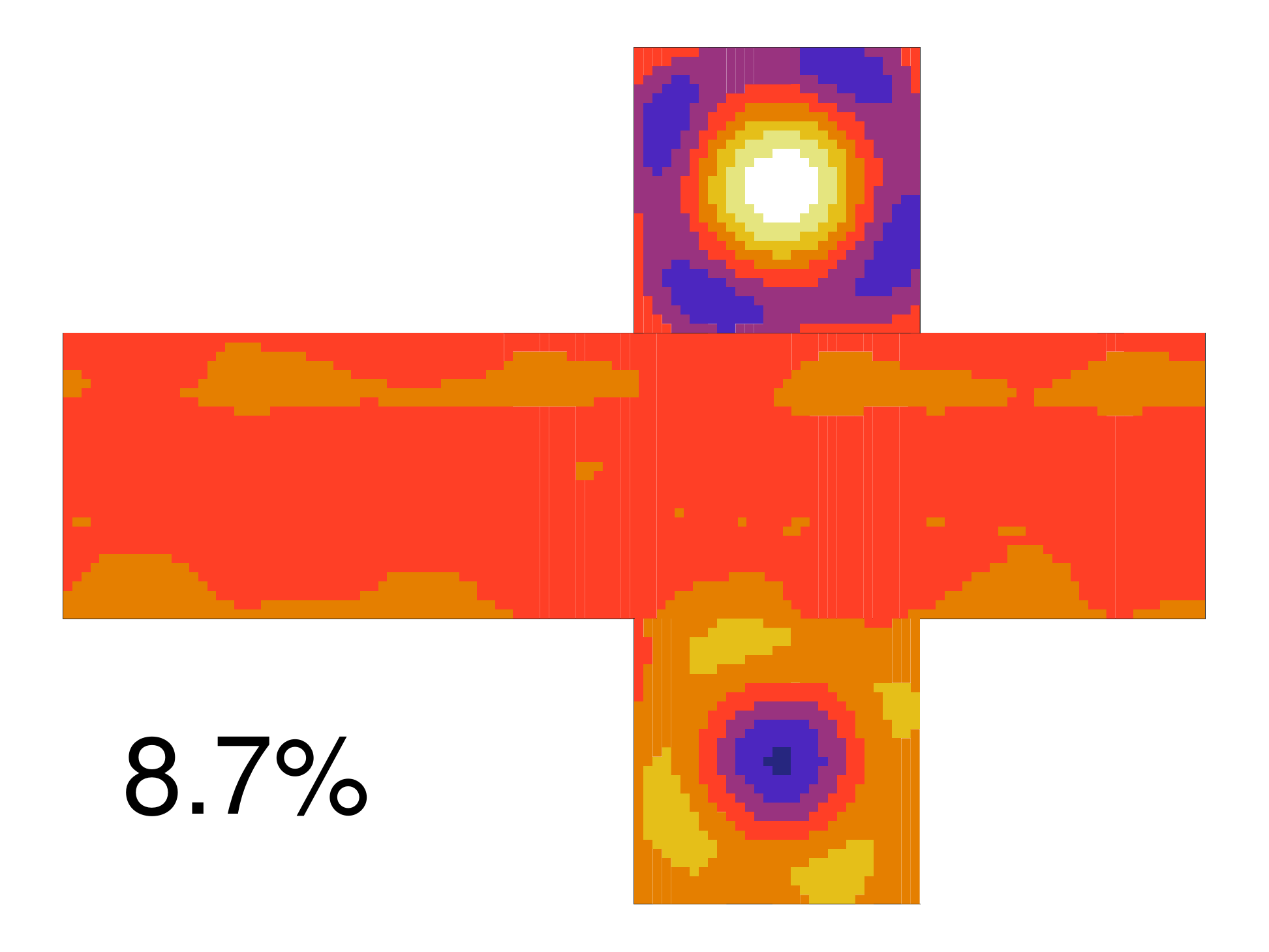}
\includegraphics[width=\linewidth]{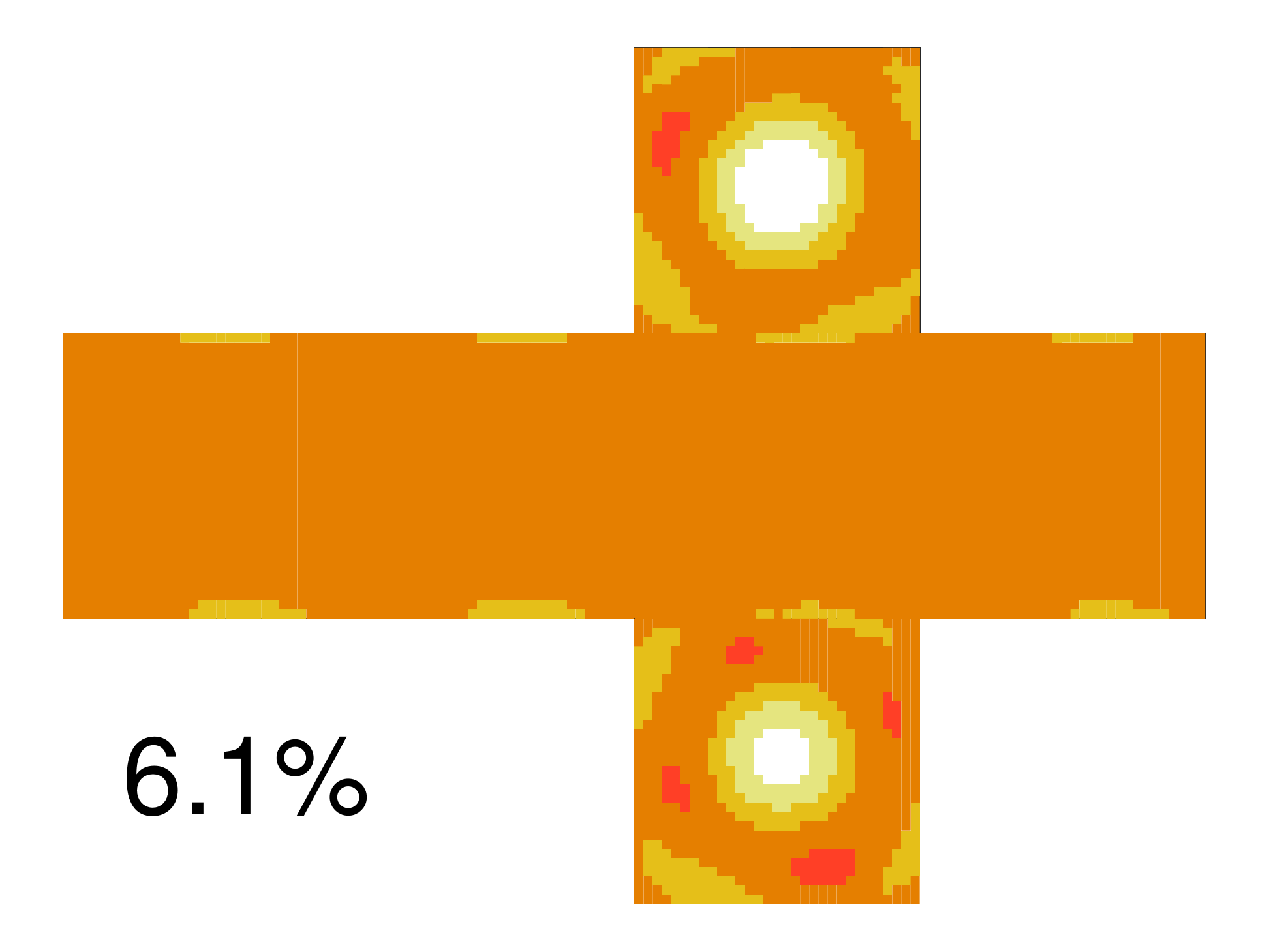}
\includegraphics[width=\linewidth]{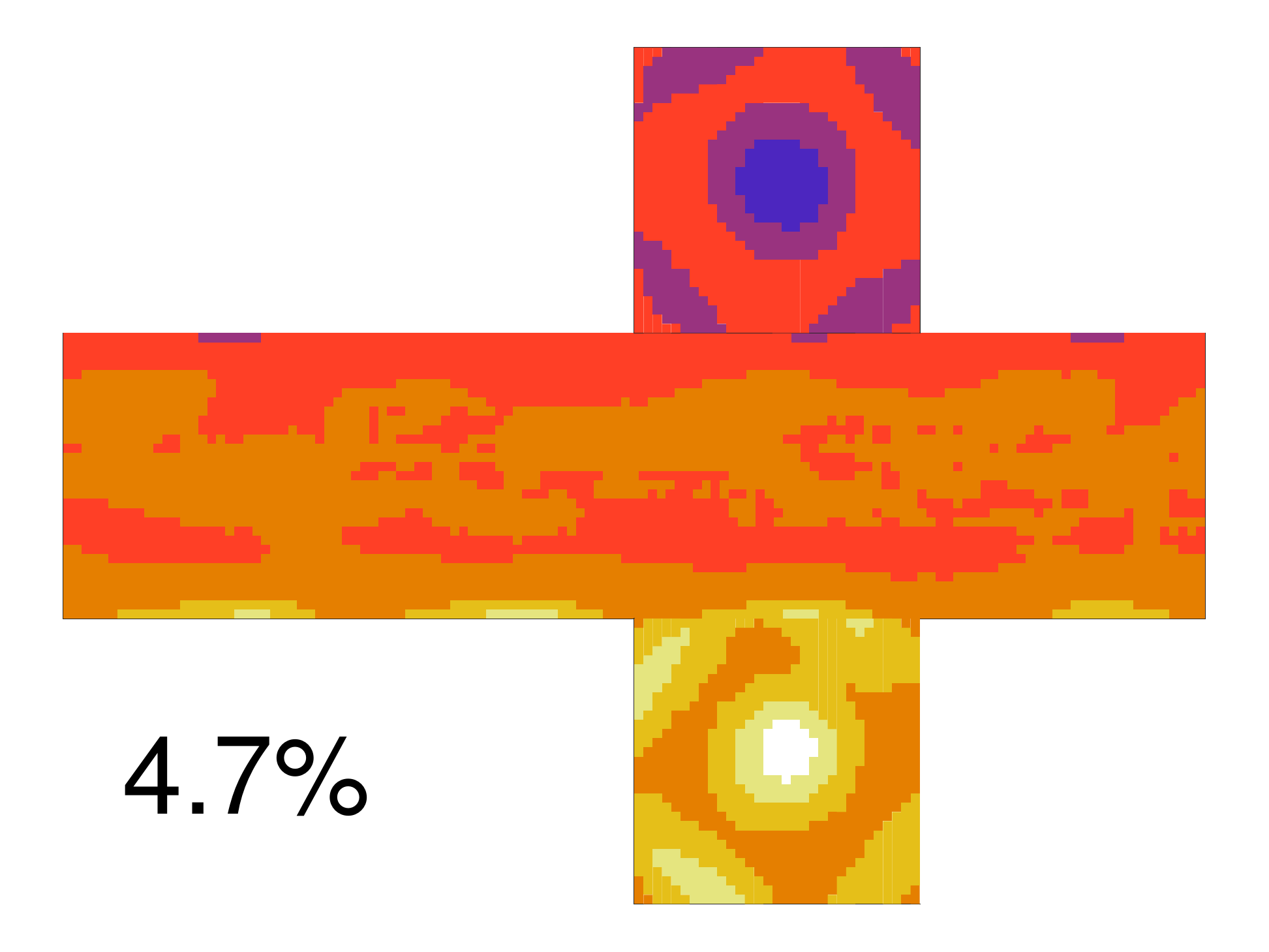}
\end{minipage}\hfill
\begin{minipage}{0.193\textwidth}
\includegraphics[width=\linewidth]{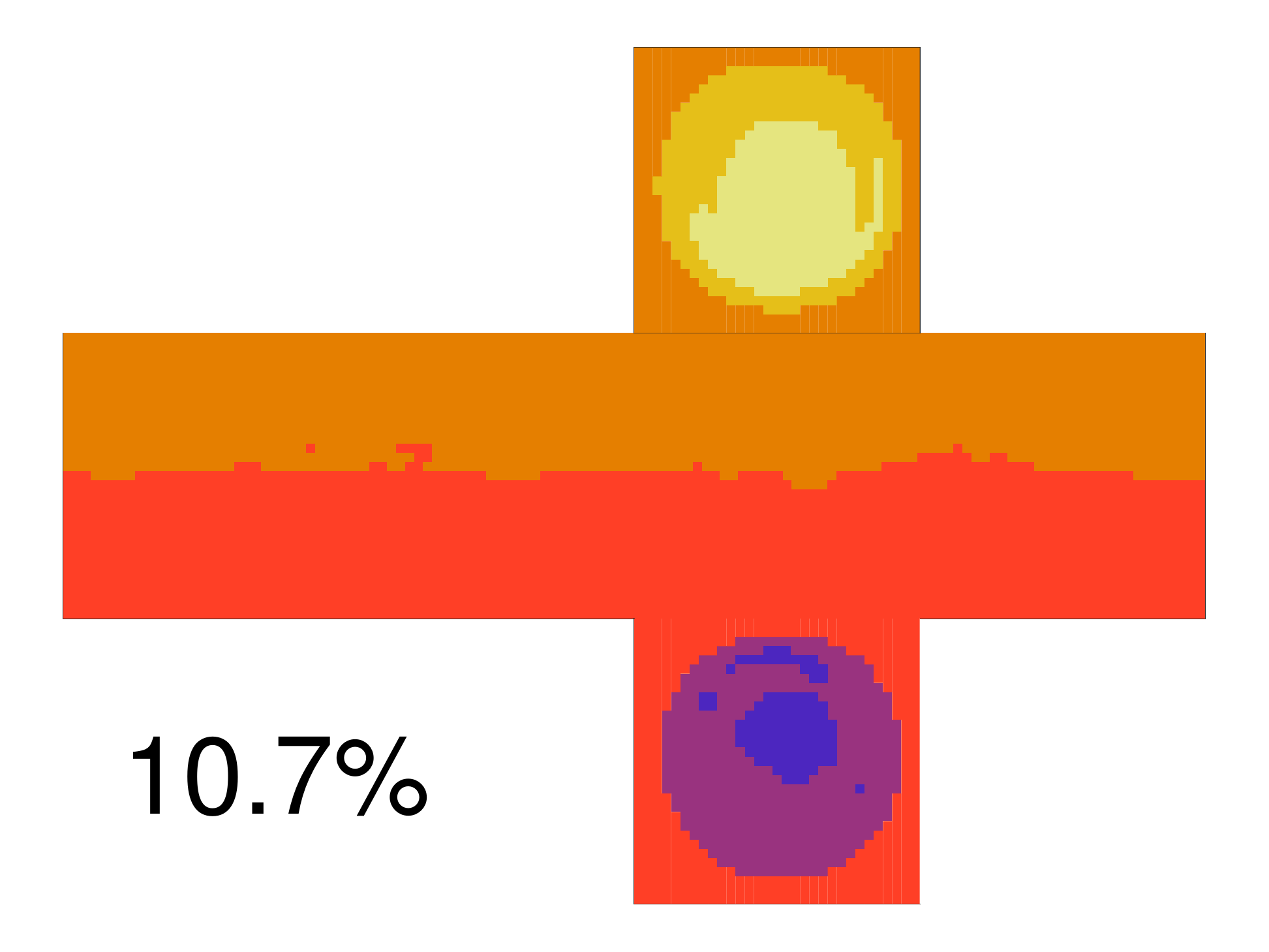} 
\includegraphics[width=\linewidth]{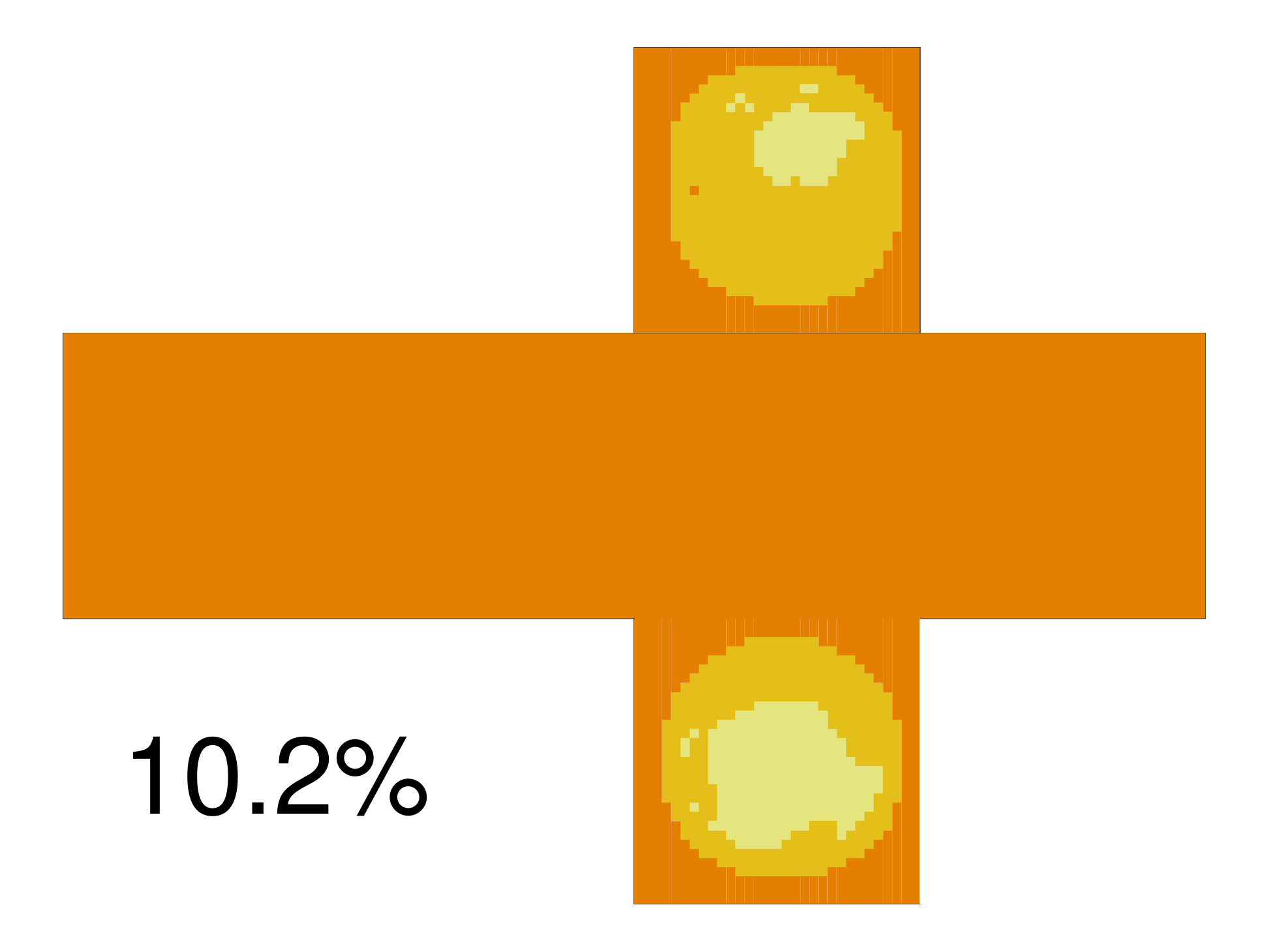}
\includegraphics[width=\linewidth]{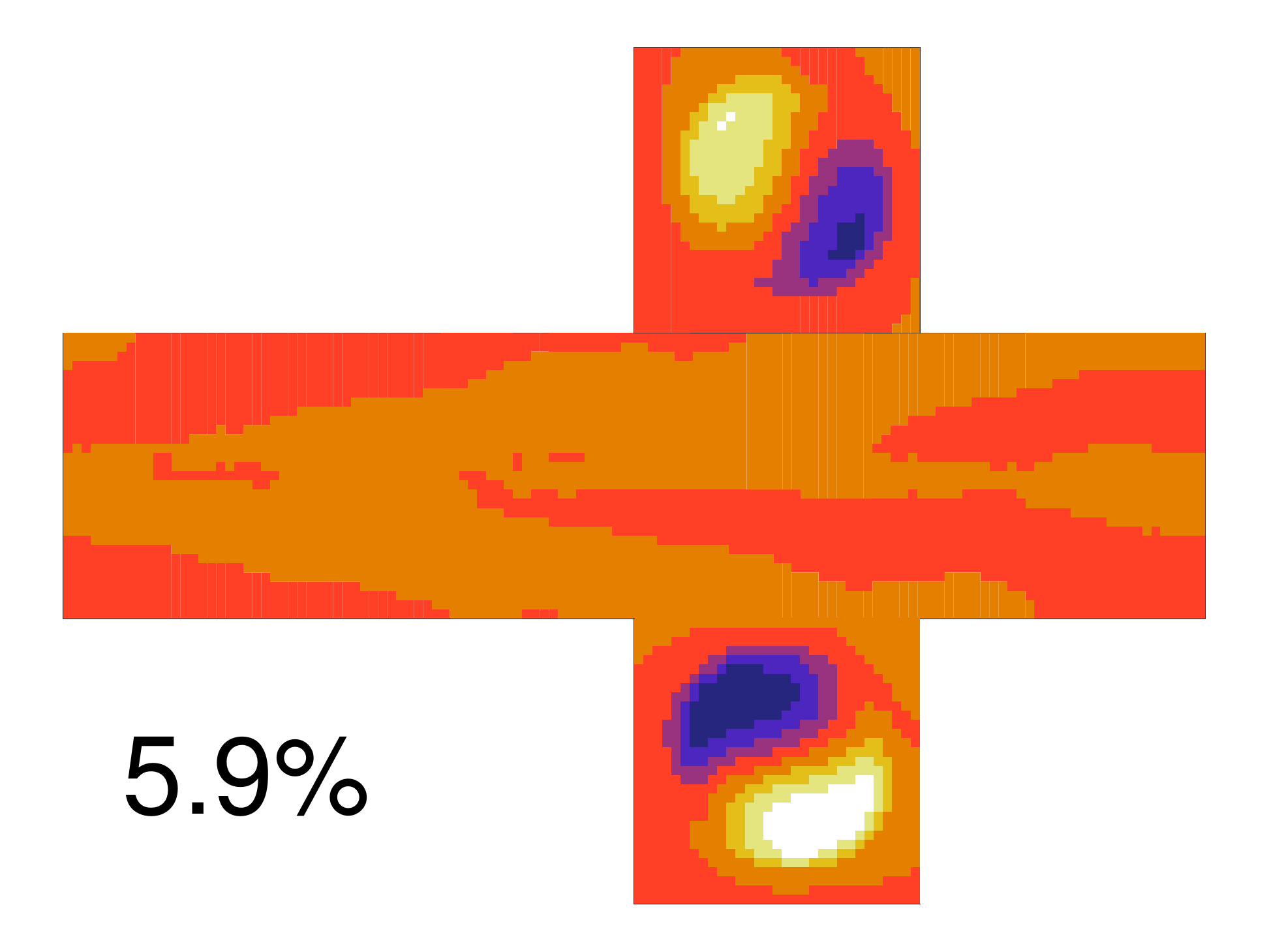}
\includegraphics[width=\linewidth]{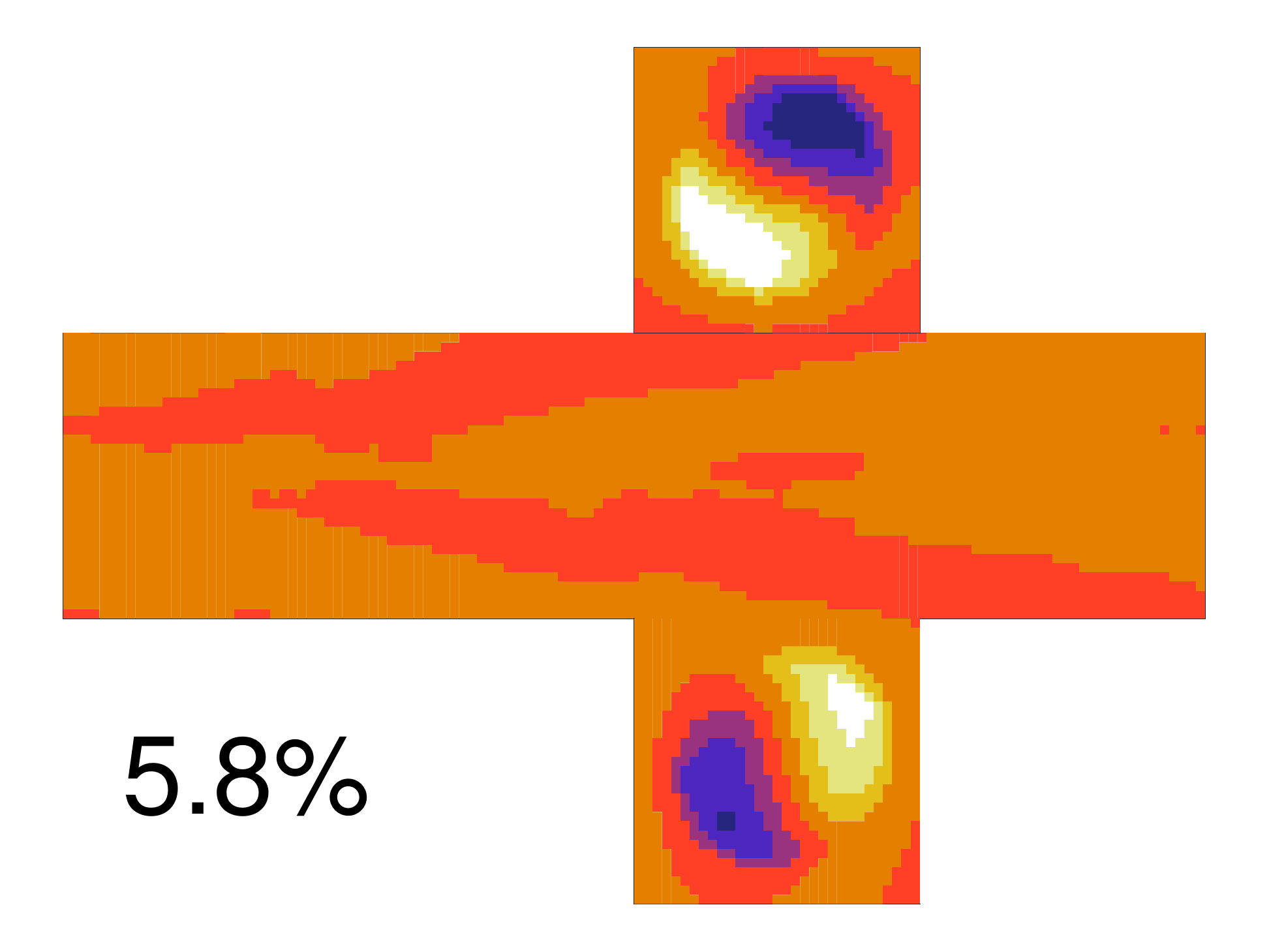}
\end{minipage}\hfill
\begin{minipage}{0.193\textwidth}
\includegraphics[width=\linewidth]{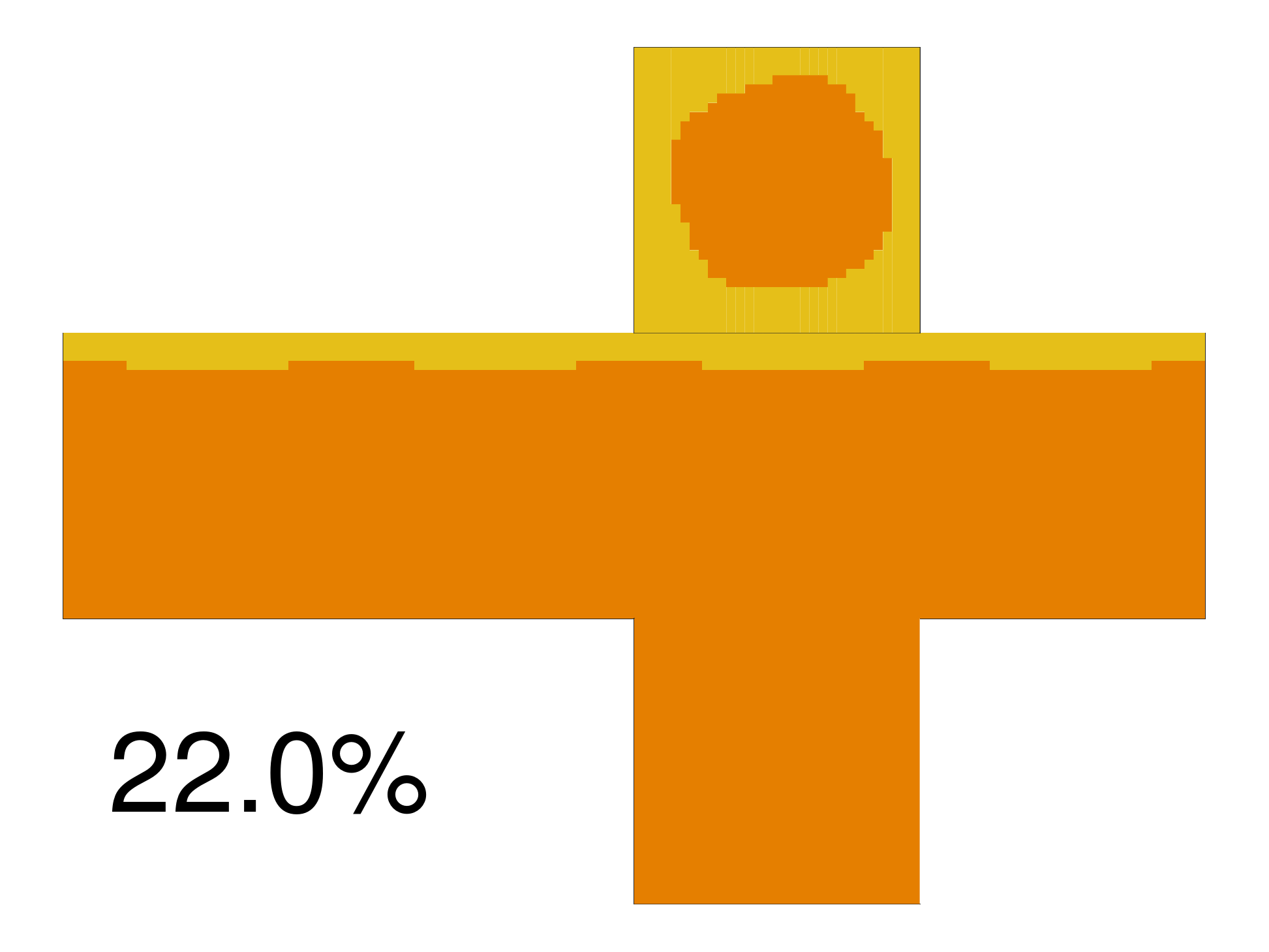} 
\includegraphics[width=\linewidth]{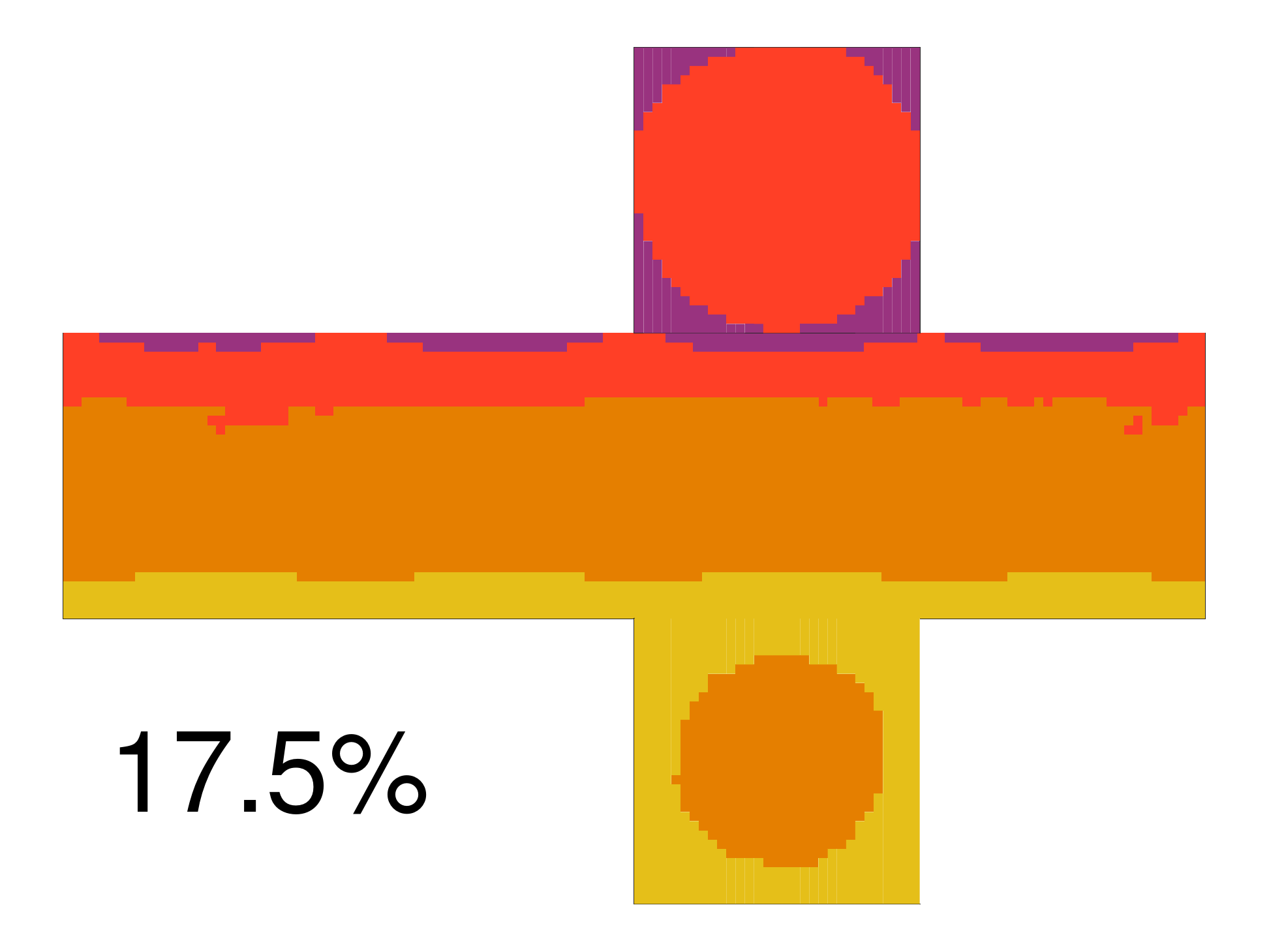}
\includegraphics[width=\linewidth]{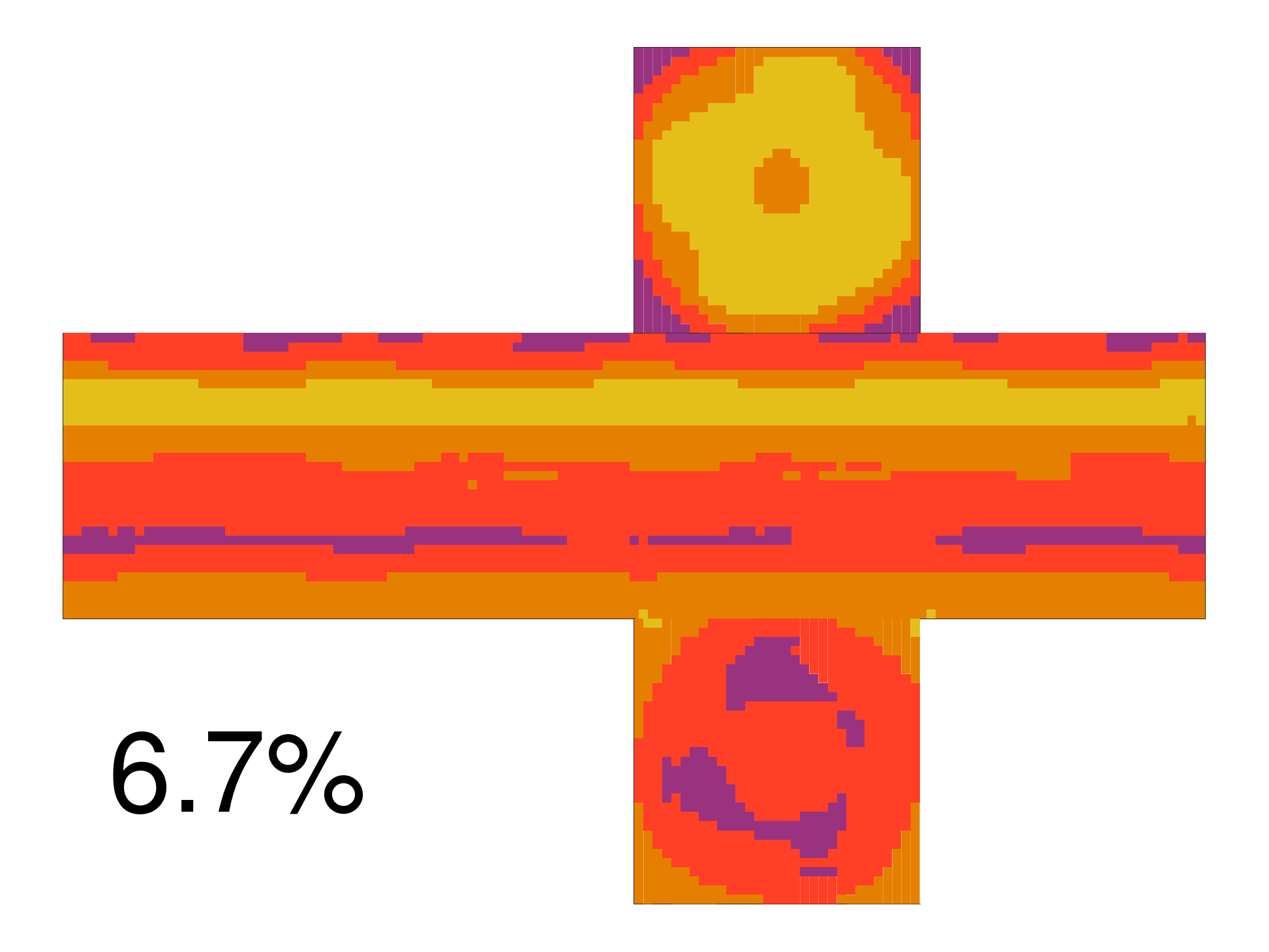}
\includegraphics[width=\linewidth]{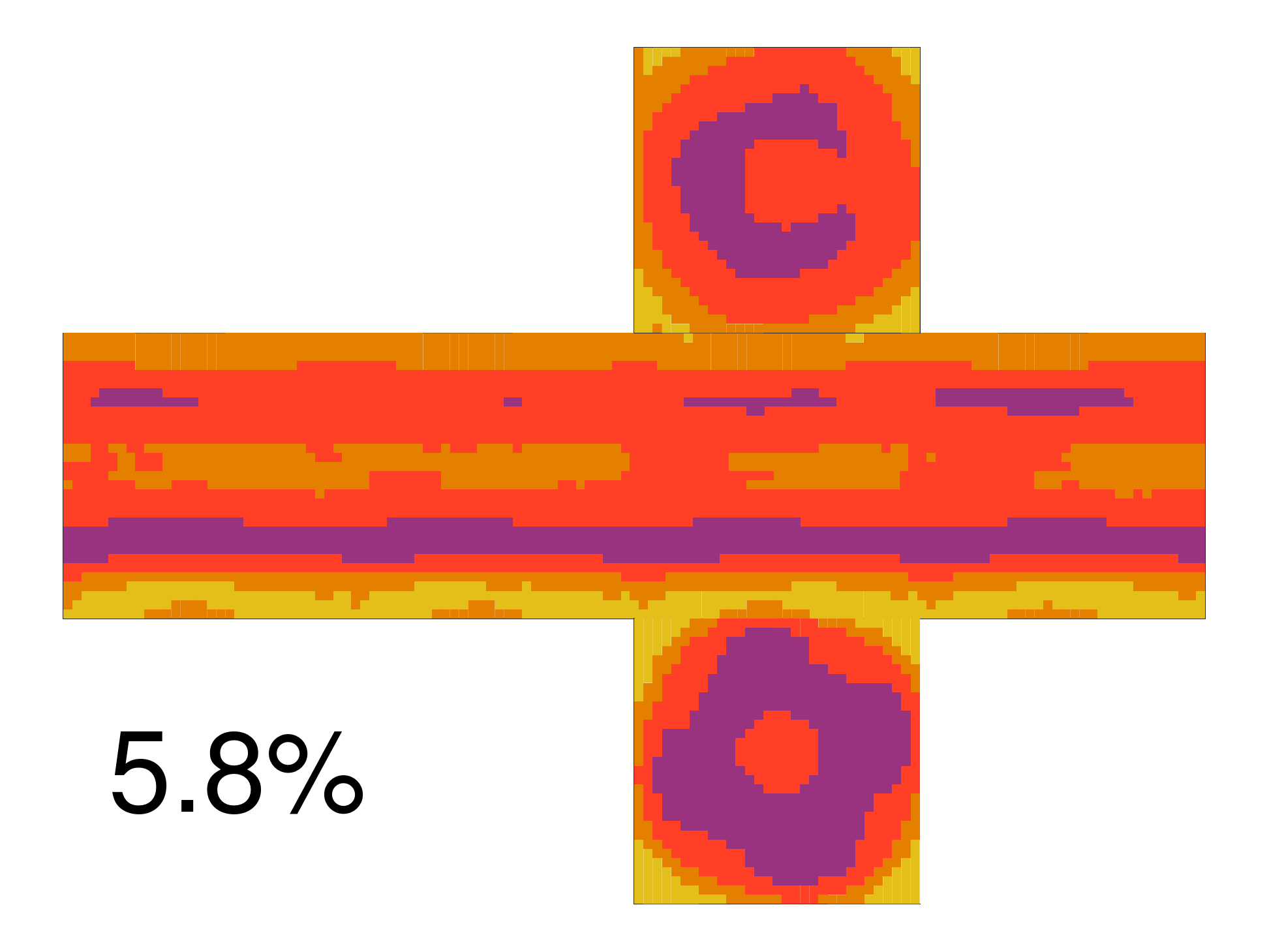}
\end{minipage}

\hfill\hfill Snowball \hfill\hfill Waterbelt \hfill\hfill Cold state \hfill\hfill Warm state \hfill\hfill Hot state \hfill{}   
\medskip

\caption{Same as Fig.~\ref{fig:seven},  for {\tt setUp2}.}
\label{fig:eight}
\end{figure}

Switching off the latitude dependence of the bulk cloud albedo and the re-injection of dissipated kinematic energy ({\tt setUp2}, Fig.~\ref{fig:eight}) has very little influence on the PCA results for the snowball, with very little apparent structure, and for the cold state, with independent poles showing strong variability at high latitudes.  
The same similarity is observed in the warm state, where the inversion of the first two PCs is not relevant as these modes have equivalent weights. 
Furthermore, the rotating modes are also present in both set-ups. 
The PCs for the waterbelt are less similar in the two set-ups. However, like for {\tt setUp1}, the first and widely dominating PC points to a strong dynamics in the tropical region at the edge of the sea ice cover, so that in both set-ups the fluctuations of the sea ice boundary turn out to be the main driver in this attractor.   
These similarities suggest that the additional processes considered in {\tt setUp1} and not in {\tt setUp2} 
have no qualitative influence on the dynamics of these attractors, although they affect their detailed quantitative properties. 

Finally, the hot state is only observed in {\tt setUp2}. It is characterised by weak fluctuations in the first mode, that represents 22\% of the total variance, and by a high variability in the equatorial region in the subsequent modes, where the cloud cover is maximal (see Fig.~3, second column). This confirms the role of the cloud feedback in the dynamics of this attractor. 

\section{Discussion and conclusions}
\label{sec:disc}

We have performed a systematic search of multiple-equilibria in  an aquaplanet where the coupling between ocean, atmosphere and sea ice dynamics is simplified by the absence of continents. Similar attempts have been done in the past using a hierarchy of models of different complexity, ranging from energy balance models~\citep{1969TellA..21..611B,1969JApMe...8..392S,1976JAtS...33....3G,Abbot2011}  and intermediate complexity models~\citep{2010QJRMS.136....2L,2013Icar..226.1724B,2017Nonli..30R..32L} to general circulation models~\citep{2011JCli...24..992F, Rose2015}. 
In these previous attempts, the model or the considered parameters range allowed for the presence of only a reduced number of attractors.  
To our knowledge, this is the first time  that five attractors are obtained under the same external forcing, demonstrating that the unperturbed 
dynamics of general circulation  models can be very complex. We have also shown in our simulations that 
by including a cloud albedo parameterisation that decreases the solar radiation in the polar regions, the number of attractors is reduced to four (the hot state disappears).   

Each attractor is the result of the competition between different nonlinear mechanisms, in particular the ice-albedo feedback, the Boltzmann radiative feedback or the cloud feedback. Within the Solar system, several examples of astronomical objects can be found where one or the other nonlinear feedback dominates the global climate, even in the absence of multi-stability. 
In the CO$_2$ atmosphere of Venus, a thick cloud layer with a maximum close to the tropical region induces a dominant cloud feedback as in the hot state, with surface temperatures as huge as 467$^\circ$C~\citep{2012Icar..217..542B}. On Europa, one of the Jupiter moons, the surface is entirely covered by ice~\citep{10.2307/27857596}, like in the snowball, the final result of a dominant ice-albedo feedback. Exo-planets within our universe will likely provide other examples of extreme conditions and possible attractors in climate systems, enriching the number of possibilities. 

The present work illustrates how complex is to construct the bifurcation diagram. Obtaining its full picture would require to systematically vary the external forcing
by changing, for example,  the inward solar radiation \citep{Rose2015}.  Once the bifurcation diagram is obtained, 
one can start to investigate the response of the system to 
perturbations, especially relevant
in the vicinity of  tipping points. This will be investigated in forthcoming studies.

In the spirit of gradually increasing the complexity of the system~\citep{2017JAMES...9.1760J}, the same analysis 
may be repeated for configurations where the position of the continents plays a role in determining the number of attractors. First attempts with simplified continent distributions 
show the robustness of the attractors~\citep{2011JCli...24..992F, Rose2015}.

Different attractors have been recently proposed to co-exist in the Earth climate dynamics. Cold/hot states may correspond to glacial/interglacial cycles over the last three millions years~\citep{Ferreira2018GRL}. 
The shift between the two states may be triggered by internal variability, changes in internal feedbacks (for example variation of vegetation or ice cover)  or in external forcing (such as Milankovitch cycles). The amplitude of the glacial/interglacial cycles is 
regular even if the incoming solar radiation is highly variable during glacial terminations~\citep{1999Natur.399..429P}, supporting the idea that such amplitude is connected to a property of the unperturbed dynamics.  Similarly,  beside the glacial/interglacial cycle, another cycle is proposed for the present-day climate that includes a hot state toward which the Earth System is approaching under the effect of global warming~\citep{Steffen8252}. The discussion in~\citet{Steffen8252} is based on a qualitative analysis; here we have started 
to quantify the existence of such possible shift from a warm to a hot state starting from a simple aquaplanet configuration and we have 
shown that the appearance of the hot state strongly depends on the cloud parameterisation and on the amount of solar radiation entering in the polar regions.   

A comprehensive understanding of attractors and their associated climates is key, in particular in the study of the geological past ({\it i.~e.}~\cite{2014CliPa..10.2053P,BrunettiVerard2015,Ferreira2018GRL}). The study of the number and positions of climate attractors as defined herein can be applied to paleoclimate studies, in particular to geological periods where climatic shifts caused dramatic extinction events, like 
near the Permian-Triassic boundary. Adding the present  analysis to the usual comparison with geological records of paleoclimate indicators, our confidence in retrieving the evolution of the climate of the Earth in deep past will be much enhanced. 

%

\begin{acknowledgements}
We acknowledge the financial support from the Swiss
National Science Foundation (Sinergia Project CRSII5\_180253).
The computations were performed at University of Geneva on the Baobab and Climdal3
clusters.
We are grateful to David Nagy, Carmelo E. Mileto and Enzo Putti-Garcia for running some of the simulations. 
We acknowledge an anonymous reviewer and Valerio Lucarini for  helpful comments and suggestions. M.~B. thanks Martin Beniston for inspiring discussions. 
\end{acknowledgements}

\bibliographystyle{spbasic} 
      

\end{document}